\documentclass[fleqn,usenatbib,useAMS]{mnras}
\usepackage{newtxtext,newtxmath}

\usepackage{graphicx}	
\usepackage{amsmath}	
\usepackage{amssymb}	
\usepackage{multicol}        
\usepackage{bm}		
\usepackage{pdflscape}	

\usepackage{tikz}
\usetikzlibrary{arrows,automata,calc,positioning}
\usepackage[percent]{overpic}

\newenvironment{rcases}
  {\left.\begin{aligned}}
  {\end{aligned}\right\rbrace}

\newcommand{\Ms}{ \mathrm{M}_\odot }
\newcommand{\MM}{ \mathrm{M} }
\newcommand{\LL}{ \mathrm{L} }
\newcommand{\cMpc}{ \mathrm{cMpc} }
\newcommand{\LE}{ \mathrm{L}_{\mathrm{E}} } 
\newcommand{\lE}{ \lambda_{\mathrm{E}} } 
\newcommand{\LEM}{ \mathrm{L}_{\mathrm{E},\MM} } 
\newcommand{\lEM}{ \lambda_{\mathrm{E},\MM} } 
\newcommand{\mmax}{ \mathrm{max} }

\newcommand{\TwentyMpcBar}{
    \linethickness{3pt}    
    \put(77,8) {\textcolor{black}{\bf 20 Mpc}}
    \put(75,5){\color{black}\line(1,0){20}} }
\newcommand{\SixMpcBar}{
    \linethickness{3pt}    
    \put(77,8) {\textcolor{black}{\bf 6 Mpc}}
    \put(75,5){\color{black}\line(1,0){20}} }
\newcommand{\TwoMpcBar}{
    \linethickness{3pt}    
    \put(77,8) {\textcolor{black}{\bf 2 Mpc}}
    \put(75,5){\color{black}\line(1,0){20}} }

\newcommand{\lp}{\lambda_{\mathrm{max}\sigma^2}} 
\newcommand{\lc}{\lambda_{\mathrm{cut}}} 

\title[Setting the Stage: Structures from Gaussian Random Fields]{Setting the Stage: Structures from Gaussian Random Fields}

\author[T. Sawala et al.]{Till Sawala,$^{1}$\thanks{E-mail: till.sawala@helsinki.fi}, Adrian~Jenkins$^{2}$, Stuart McAlpine$^{1}$, Jens Jasche$^{3},$
\newauthor Guilhem Lavaux$^{4}$, Peter~H.~Johansson$^{1}$,  Carlos~S.~Frenk$^{2}$ \\
$^{1}$Department of Physics, Gustaf H\"allstr\"omin katu 2, University of Helsinki, Finland\\
$^{2}$Institute for Computational Cosmology, Durham University, South Road, Durham DH1 3LE, United Kingdom \\
$^{3}$The Oskar Klein Centre, Department of Physics, Stockholm University, Albanova University Center, 106 91 Stockholm, Sweden \\
$^{4}$CNRS \& Sorbonne Universit\'e, UMR7095, Institut d'Astrophysique de Paris, 75014 Paris, France
}

\date{Accepted XXX. Received YYY; in original form ZZZ}
\pubyear{2019}

\begin{document}
\label{firstpage}
\pagerange{\pageref{firstpage}--\pageref{lastpage}}
\maketitle

\begin{abstract}
We study structure formation in a set of cosmological simulations to uncover the scales in the initial density field that gave rise to the formation of present-day structures. Our simulations share a common primordial power spectrum (here $\Lambda$CDM), but the introduction of hierarchical variations of the phase information allows us to systematically study the scales that determine the formation of structure at later times.
We consider the variance in $z=0$ statistics such as the matter power spectrum and halo mass function. We also define a criterion for the {\it existence} of individual haloes across simulations, and determine what scales in the initial density field contain sufficient information for the non-linear formation of unique haloes. We study how the characteristics of individual haloes such as the mass and concentration, as well as the position and velocity, are affected by variations on different scales, and give scaling relations for haloes of different mass.
Finally, we use the example of a cluster-mass halo to show how our hierarchical parametrisation of the initial density field can be used to create variants of particular objects. With properties such as mass, concentration, kinematics and substructure of haloes set on distinct and well-determined scales, and its unique ability to introduce variations localised in real space, our method is a powerful tool to study structure formation in cosmological simulations.
\end{abstract}

\begin{keywords}
methods: numerical -- galaxies: formation -- cosmology: theory, dark matter, large-scale structure of the Universe\end{keywords}



\section{Introduction}
In the standard cosmological model galaxies and dark matter haloes originate from random, adiabatic density fluctuations in the Big Bang, magnified by inflation, and amplified under the force of gravity in competition with cosmic expansion. If, as inflation predicts~\citep[e.g.][]{Linde-2005}, and observations indicate ~\citep[e.g.][]{Bouchet-1993, Nusser-1995, Planck-2016, Planck-2019}, the primordial density field is Gaussian, and statistically homogeneous and isotropic, its late-time power spectrum and the distribution of haloes that form in a sufficiently large volume are fully determined by the laws of physics, and the universal cosmological parameters \citep[e.g.][]{Peebles-1980, Bardeen-1986}.

Due to the random nature of the initial density field, a comparison between theoretical predictions and observations is usually done on a population level, rather than for individual objects. Constraints on the halo mass function or the matter power spectrum require large surveys, and in the case of simulations, similarly large volumes to sufficiently sample the underlying distributions \citep[e.g.][]{Springel-2005-millennium, Klypin-2011, Angulo-2012}. While scales much smaller than the survey or simulation size are sampled many times, scales represented by the largest haloes, or the largest modes of the density field, are sampled much more sparsely. This problem of ``cosmic variance'' \citep[see, e.g.][]{Colombi-2000} is not just a question of size, however. Any particular object, observed with enough detail, is unlikely to have a closely matching counterpart in a finite simulation volume, and the search for the initial conditions that give rise to the formation of particular observed structures remains an ongoing challenge \cite[e.g.][]{Hoffman-1991, Bistolas-1998, Yepes-2013, Doumler-2013,  Jasche-2013, Hoffman-2015, Jasche-2015, Lavaux-2016, Carlesi-2016}.

While the statistics of the halo population are fully determined by the initial power spectrum, the formation of particular objects and their characteristics also depend on the particular phase information. In a simple, monolithic collapse model  \citep[e.g.][]{Peebles-1980}, the formation of a structure would be governed only by modes with wavelengths at or above the scale of the Lagrangian region from which the structure originated. N-body simulations \citep[e.g.][]{Davis-1985} have shown, however, that structure formation in $\Lambda$CDM is ``bottom up'', and large structures form in part through the merger of smaller ones, causing even smaller-scale modes to affect the formation as well as the properties of more massive haloes.

The fact that distinct constituents of the halo population originate from independent scales in the initial density field has been pointed out, for example, by \cite{Aragon-Calvo-2016}, who used simulations with shared large-scale modes in order to create an ensemble of simulations. Notably, and analogous to our Sections~\ref{sec:results:population} and~\ref{sec:results:powerspectra}, they show how a set of semi-independent realisations with some shared phase information can overcome the statistical limitations created by single sample.

In this paper, we systematically investigate which scales in the initial density field are responsible for the formation of haloes at later times. Starting from the density field of the $100^3$cMpc$^3$~{\sc eagle} simulation volume\footnote{Throughout this paper, we use physical units of mass and density, and comoving physical units of length, unless otherwise specified.} \citep{Schaye-2014}, we generate a sequence of simulations that systematically introduce random variations of the white noise field on increasingly larger scales. At each scale, we compare the resulting simulations in terms of the evolved density field, the population of haloes, the existence of individual haloes, and the variation in halo properties.

In particular, we uncover the scales in the initial density field that contain the information responsible for the non-linear formation of individual haloes. We also examine how the properties of haloes of different masses change when the density field is perturbed on scales below this {\it existence} scale. Finally, we demonstrate how our method of introducing hierarchical, random perturbations to existing density fields can be used deliberately to create variations of simulations and simulated objects, including with variations in mass, concentration and kinematics. This makes it a particularly powerful tool for future zoom and constrained simulations, allowing to efficiently explore the parameter space of possible initial conditions that give rise to the formation of haloes with particular observed properties. We will explore the full potential of this method in a forthcoming paper.

This paper is organised as follows. In Section~\ref{sec:methods:ICs}, we describe our method for parametrising the phase information in an octree basis, and the {\sc Panphasia} white noise field. In
Section~\ref{sec:methods:simulations}, we outline the set-up of the simulations used in this paper, our way of identifying haloes, and of matching objects across simulations. Section~\ref{sec:results} presents the global results of our simulations, in terms of the density field and its power spectrum, in Section~\ref{sec:results:density}, and the halo population, in Section~\ref{sec:results:population}. In Section~\ref{sec:results:individual}, we discuss the formation, and variation in properties, of individual objects. We present a definition of the identity of particular haloes across simulations in Section~\ref{sec:results:individual:identity}, which allows us, in Section~\ref{sec:results:individual:existence}, to study the scales in the initial density field that determine the existence of particular haloes. In Section~\ref{sec:results:properties}, we study the variation of halo properties: mass (Section~\ref{sec:results:properties:mass}), concentration (Section~\ref{sec:results:properties:concentration}), position (Section~\ref{sec:results:properties:displacements}) and velocity (Section~\ref{sec:results:properties:velocity}). We present variations of a particular, cluster mass halo in Section~\ref{sec:results:example}, and conclude with a summary and an outlook to future work in Section~\ref{sec:summary}.

\section{Methods}\label{sec:methods}
The results in this paper are based on cosmological "dark matter only" N-body simulations, i.e. both baryons and dark matter are subsumed into a single type of simulation particle, and evolved only under the effect of gravity. In this section, we describe the creation of our initial conditions, the set-up of the simulations, and the identification of structures. Additional information about the parametrisation of the primordial density field, and the {\sc Panphasia} white noise field that is used in its construction, can be found in \cite{Jenkins-2013} and \cite{Jenkins-Booth-2013}.

\subsection{Initial conditions}\label{sec:methods:ICs}
A natural way to describe the primordial Gaussian density or displacement field in a cosmological simulation of a cubic region with periodic boundary conditions is a Fourier representation, introduced for cosmological simulations by \cite{Efstathiou-1985}, and employed by many subsequent initial condition generators \citep[e.g.][]{Katz-1994, Bertschinger-2001, Springel-2005-millennium, Jenkins-2010, Hahn-2011, Oleary-2012}.

In the standard $\Lambda$CDM model, it is assumed that the initial density fluctuations after inflation are Gaussian. The statistical properties of the overdensity field, $\delta(\mathbf{x})$, and its Fourier transform, $\delta(\mathbf{k})$, for a volume, $V$, are then completely defined by the one dimensional linear power spectrum, $P_{\rm lin}(k)$:

\begin{equation}
    P_{\rm lin}(k) = \frac{1}{V}\langle | \delta_k |^2 \rangle.
\end{equation}

To create a set of $\Lambda$CDM initial conditions, it is necessary to specify the cosmological parameters, the dimensions of the periodic region, the power spectrum, and the phase information, which can be encapsulated as a realisation of a Gaussian white noise field \citep{Salmon-1996}. Gaussian white noise fields are particularly convenient to work with numerically: for example, their two-point autocorrelation functions are zero for any non-zero lag. Consequently, the values for an unconstrained white noise field can be set by any high quality pseudorandom number generator.
As \cite{Salmon-1996} also pointed out, not only does this offer a simple way to produce multi-scale random fields, it is also straightforward to include linear constraints: the corresponding $\Lambda$CDM linear overdensity field for a given linear matter power spectrum is given by the convolution, in real space, of the white noise `phase information' field with a specific spherically symmetric window function, computed from a one dimensional integral over the linear matter power spectrum.

\subsubsection{Octree basis}\label{sec:methods:ICs:octree}
\cite{Jenkins-2013} introduced a way of constructing multi-scale real-space white noise fields based on an octree decomposition of a cubic period volume. In this formalism, the Gaussian white noise field is built hierarchically from a linear superposition of octree basis functions.

The individual basis functions are zero outside of the particular cubic cell they occupy, and orthogonal to each other, even when completely or partially overlapping. Consequently, an unconstrained Gaussian white noise field can be created by choosing the octree function amplitudes from a pseudorandom sequence of uncorrelated Gaussian variables of zero mean and unit variance. 

The white noise field can be refined at any location by adding information from higher (or deeper) levels of the octree. The white noise field is then specified in each cell at the highest level as an appropriate polynomial of the local Cartesian cell coordinates. The octree functions themselves are not polynomials, but an octree function occupying any parent cell is represented by eight distinct piecewise polynomials, with each one filling one of the parent's eight child cells.

It proves convenient in this paper to use these octree functions to represent the phase information in all the simulations we present. The primary rationale for this choice, however, will be made evident in subsequent papers. Our ultimate goal is to construct multi-scale initial conditions which also satisfy linear constraints, derived for example from reconstructions of the large-scale structure around the Milky Way. Here, the octree decomposition is crucial, as it allows for the introduction of localised changes to the density field, independently from the external constraints.

There are an infinite number of possible sets of orthogonal octree basis function, using different polynomial functions to represent the field within the cells of the octree. The degree to which the density information is attached to a particular level of the octree depends on the choice of basis. Throughout this paper we will use the $S_8$ octree basis functions developed by \cite{Jenkins-2013}. The $S_8$ function set have been used in making initial conditions for many cosmological simulations of the  Virgo Consortium\footnote{\url{http://virgo.dur.ac.uk/}}, from modelling galaxy clusters~\citep{Bahe-2017} via galaxy groups \citep{Sawala-2016a} and individual galaxies \citep{Grand-2017}, down to simulating the smallest CDM dark haloes at the present day \citep{Wang-2019}.

\subsubsection{Contribution to the Variance from Individual Levels}

\begin{figure}
    \includegraphics[width=\columnwidth]{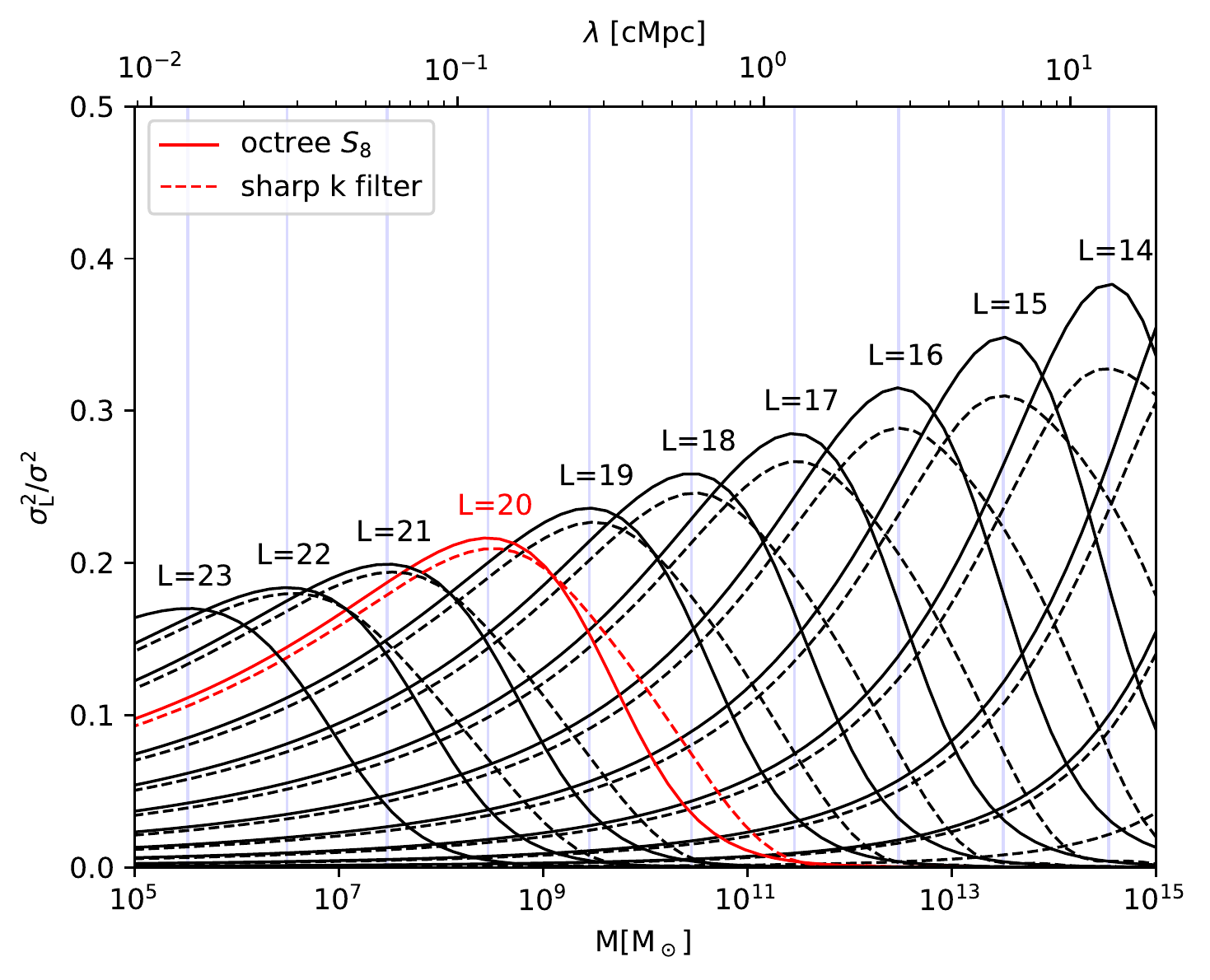}
    \caption{Relative contribution to total the variance of the density field, $\sigma^2$, by different octree levels, using the $S_8$ octree basis functions (solid lines), or a sharp k-space filter (dashed lines), as a function of wavelength, $\lambda$, top axis, and corresponding spherical top-hat mass, bottom axis. The scale at which the relative contribution from a given octree level $\LL$ is maximal defines $\lambda_{\mmax \sigma^2}(\LL)$ and  $\MM_{\mmax \sigma^2}(\LL)$, and is indicated by faint vertical lines.  \label{fig:sigma-p1}}  
\end{figure}

To illustrate the relative importance of individual layers of octree functions to the $\Lambda$CDM overdensity field, we first consider the fractional contribution of each level of octree functions to the total variance of the overdensity field when smoothed with a spherical tophat function of radius $R$, containing mass $M$, at the mean density of the universe.
This ratio is given by:
$$
\frac{\sigma^2_L}{\sigma^2} = \frac{\int {\rm d}^3\mathbf{k}\phantom{x} W^2(kR)\left(j^2_{lmn}(\mathbf{k}\Delta_L) - j^2_{lmn}(2\mathbf{k}\Delta_L)\right)P_{\rm lin}(k)}
{\int {\rm d}^3\mathbf{k}\phantom{x} W^2(kR)P_{\rm lin}(k)},
$$
where $W(kR)\equiv \left(\sin(kR)-kR\cos(kR)\right)/(kR)^3$
is the spherical top-hat window function in
$k$-space, $P_{lin}(k)$ is the $\Lambda$CDM linear power spectrum, $\Delta_L$ is the cell size of the octree at level $L$, and the functions $j_{lmn}$ are 
given by:
\begin{equation}
\begin{aligned}
j_{lmn}({\bf k}\Delta_L ) =~& {\large \left( (2l + 1)(2m+1)(2n+1)\right)}^{1/2} \\
&  \times j_l\left(\frac{k_x\Delta_L}{2}\right)j_m\left(\frac{k_y\Delta_L}{2}\right)j_n\left(\frac{k_z\Delta_L}{2}\right).
\end{aligned}
\end{equation}

The three functions $j_{l,m,n}$, on the right hand side, are Spherical Bessel functions of the first kind, and 
$k_x,k_y,k_z$ are the Cartesian components of the wave vector $\mathbf{k}$.

The phase information for level $L$ is defined as that given by the octree functions that fully occupy octree cells at level $L-1$, and therefore contribute phase information that can be represented
as sets of disjoint polynomials filling the octree cells at level $L$.

In Fig.~\ref{fig:sigma-p1} we plot the ratio $\sigma_L^2 / \sigma^2$ against the spherical top-hat filter mass, M, and size of a spherical perturbation, $\lambda$, related via $\MM=\overline{\rho} \times 4/3 \pi r^3$, where $\overline{\rho} = \Omega_0 \times 3 \mathrm{H_0}^2 / (8 \pi G)$, and $r = \lambda/2 = \pi / k$, for a series of single octree levels, each shown by a solid line. For each level there is a range of top-hat filter masses and wavelengths where a particular level contributes most to the total variance. We call these scales $\MM_{\mmax \sigma^2}$ and $\lambda_{\mmax \sigma^2}$, respectively. 

For comparison, the dashed curves show the fractional contribution to the variance if, instead of setting octree functions to zero except in a single octree layer, we do the equivalent transformation assuming we use Fourier modes to represent the phase information. More precisely, we set all Fourier
modes to zero outside the factor of two range of $1/\sqrt{2}<k\Delta_L<\sqrt{2}$.

The peaks of the dashed and solid curves line up quite well, and the shape of the sets of curves are similar. This indicates that for quantities such as the variance of the overdensity field smoothed with a top-hat, we can establish a close correspondence between octree layers and sharp-$k$ space shells in Fourier space. We expect to get very similar results to those we present in this paper, had we chosen instead only to work with Fourier modes. 

\subsubsection{The Cut-Off Scale}

\begin{figure}
   \includegraphics[width=\columnwidth]{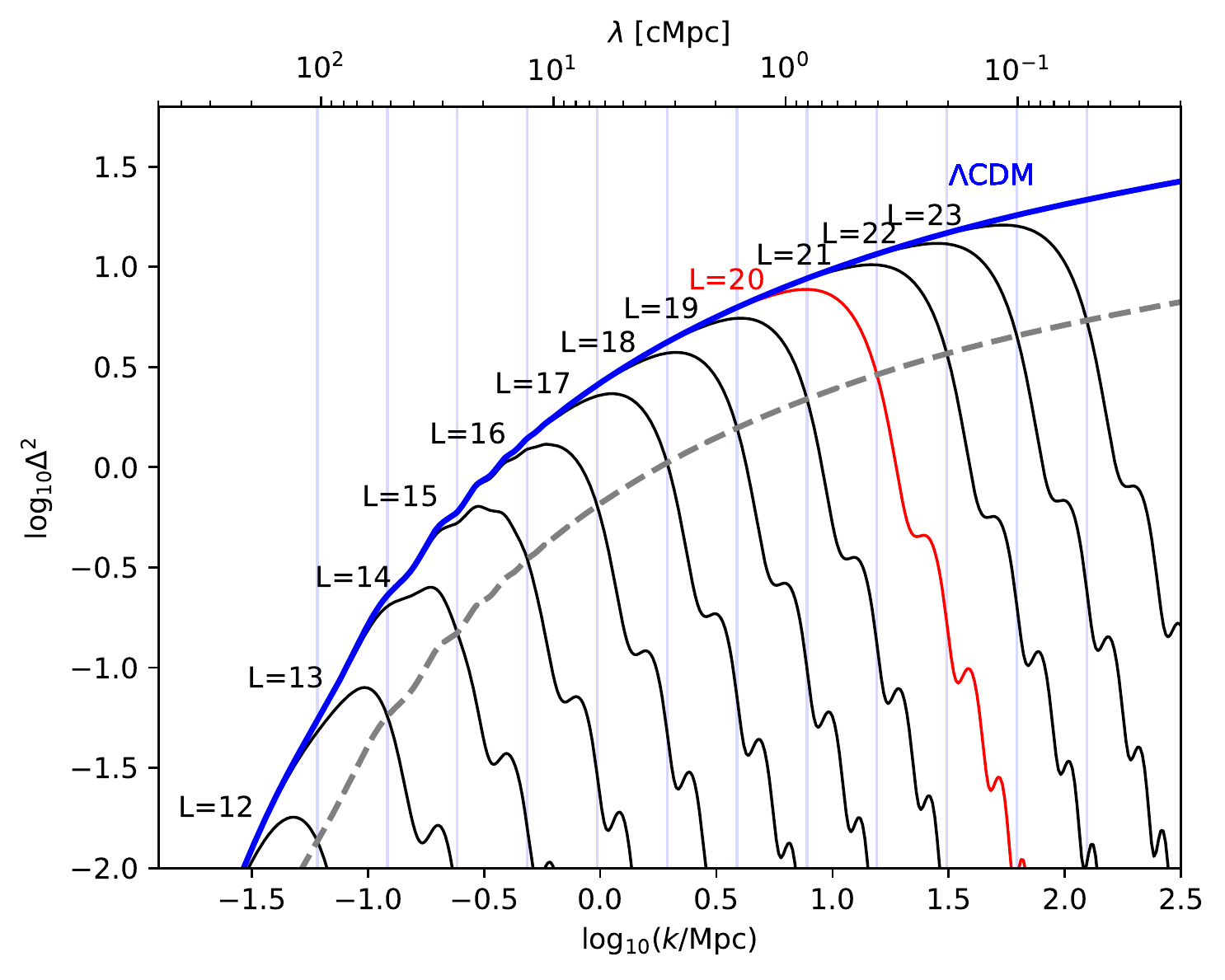}
    \caption{The reproduction of the $\Lambda$CDM dimensionless power spectrum (blue solid line) as a function of wavenumber, $k$, bottom axis, and wavelength, $\lambda$, top axis, with increasing accuracy. Individual black lines show the truncated power spectra given by Eqn.~\ref{eqn:trunc_powspec}, successively including the phase information for octree levels up to the indicated level $\LL$. The dashed grey line indicates a reduction in $\Delta^2$ by a factor of 4. Faint vertical lines mark its intercept with the truncated power spectra, allowing us to define a cut-off wavenumber, $k_\mathrm{cut}$, and cut-off scales, $\lambda_\mathrm{cut}(\LL)$ and $\MM_{\mathrm{cut}}(\LL)$ for each octree level.
    \label{fig:power-p1}}
\end{figure}

If we set to zero all the octree functions occupying cells at level $L$ and above (so that the white noise field is approximated by disjoint polynomial functions occupying the octree cells at level $L$), the resulting, truncated power spectrum has a high-$k$ cut-off of the form:
\begin{equation}
 \label{eqn:trunc_powspec}
\langle P_{lin}^L({\bf k})\rangle  = P_{lin}(k) \left(\sum_{\scriptscriptstyle l,m,n=0,1}j^2_{lmn}({\bf k}\Delta_L)\right),
\end{equation}
The high-$k$ cut-off is due to the summation term in brackets on the right-hand side of Eqn.~(\ref{eqn:trunc_powspec}) which tends to one in the limit of small $k$, and to zero for large $k$.

In Fig.~\ref{fig:power-p1} we illustrate how the full $\Lambda$CDM dimensionless power spectrum is built up from successive higher `layers' of octree functions by plotting Eqn.~\ref{eqn:trunc_powspec} for a $\Lambda$CDM power spectrum for several values of $L$. We can see that the $\Lambda$CDM linear power spectrum is approximated better and better as we add successive layers of octree function phase information. Truncating the octree representation for all cells higher than a given level produces a relatively sharp cut-off in the spherically averaged power spectrum. We can define an associated cut-off wavenumber, $k_{cut}(L)$, by determining where $P^L_{lin}(k)$ falls by a factor of two in amplitude, or four in power, below the $\Lambda$CDM linear power spectrum.

\subsubsection{Length Scales and Physical Quantities}
In Table~\ref{tab:lengthscales} we give a set of conversion factors which relate the level, $L$, to physical quantities. The first column gives the octree level. Level 23 is the deepest level and hence the smallest scale at which we sample the Panphasia field for this project. The second column lists the side length of the octree cells themselves. As the 100~cMpc side length of the simulation volume is represented by 12 cells at level 12, the length-scale at level $L$ is $(100/3)2^{14-L}$~cMpc. 

Column three gives $\lp(\LL)$, the wavelength of a perturbation for which the contribution from level $\LL$ to the fractional variance $\sigma^2_L/\sigma^2$ in the $S_8$-parametrisation is maximal, as shown in Fig.~\ref{fig:sigma-p1}. Column four gives $\MM_{\mmax \sigma^2}$, the equivalent spherical top-hat mass. Column five gives $\lc(\LL)$, the cut-off scale, i.e. the small-scale limit at which the power due to the truncation at level $\LL$ falls to a quarter of the value of the full, linear $\Lambda$CDM power spectrum, and $\MM_{\mathrm{cut}}$, in column 6, is the mass of an equivalent spherical top-hat.

While the size of octree cell is of interest, this length-scale on its own is not particularly revealing. This is because, as described above, the physical length-scales affected by the octree functions result from a combination both of the cell size and the functional forms of the octree functions themselves. In general, basis choices that use high-order polynomials affect smaller physical length-scales in units of the octree cell size.

As shown in Fig.~\ref{fig:sigma-p1}, for most scales of interest, the contribution from any one level to the variance is less than 1/3, and any object contains phase information from multiple levels. Comparing the red lines in Fig.~\ref{fig:sigma-p1} and Fig.~\ref{fig:power-p1}, we see that, while the octree functions at $\LL=20$ contribute most at $\lp = 0.12$~cMpc, the power already falls to 1/4 at $\lc = 0.41$~cMpc if the power spectrum is truncated at $\LL=20$. This is due to the fact that the information from $\LL=20$ already contributes significantly to the full $\Lambda$CDM power spectrum at this scale, in addition to smaller contributions from $\LL>20$. While $\lp$ thus locates the peak contribution from a single level, for the effect of variations at and above a certain scale on structure formation, $\lc$ proves the most useful quantity. In Section~\ref{sec:results:individual:existence}, we also define a new length scale based on the existence of unique haloes across variations at a given level.

\begin{table}
	\begin{center}
	    \caption{Correspondence between octree levels and physical scales.}
	    \label{tab:lengthscales}
	    \begin{tabular}{lccccc} 
		    \hline
    		L & $\Delta_{\mathrm{cell}}$  & $\lambda_{\mmax \sigma^2}$ &
    		$\mathrm{log}_{10}\left(\frac{M_{\mmax \sigma^2}}{\Ms}\right)$ & $\lambda_{\mathrm{cut}}$ & $\mathrm{log}_{10}\left(\frac{M_\mathrm{cut}}{\Ms}\right)$
    		\\
    		\hline
    		& [Mpc] & [Mpc] & & [Mpc] &  \\ 
	    	\hline
		    23 & $0.065$ & 0.014 & 5.6 & 0.051 & 7.3 \\
    		22 & $0.13$ & 0.027 & 6.5 & 0.101 & 8.2 \\
	    	21 & $0.26$ & 0.061 & 7.5 & 0.202 & 9.1 \\
		    20 & $0.52$ & 0.12 & 8.4 & 0.405 & 10.0 \\
		    19 & $1.04$ & 0.27 & 9.5 & 0.81 & 10.9 \\
		    18 & $2.08$ & 0.61 & 10.5 & 1.62 & 11.8 \\
		    17 & $4.17$ & 1.22 & 11.4 & 3.24 & 12.7 \\
		    16 & $8.33$ & 2.73 & 12.5 & 6.48 & 13.6 \\
		    15 & $16.67$ & 6.11 & 13.5 & 12.97 & 14.5 \\
		    14 & $33.33$ & 13.68 & 14.6 & 25.94 & 15.4 \\
		    13 & $66.67$ & 30.63 & 15.6 & 51.88 & 16.3 \\
		    12 & $133.33$ & 61.11 & 16.5 & 103.78 & 17.2 \\
		    \hline
	    \end{tabular}
	\end{center}
	{\small
	$\LL$: octree level, $\Delta_\mathrm{cell}$: size of the octree cell at level $L$, $\lambda_{\mmax \sigma^2}$: characteristic wavelength, where the fractional contribution to the variance from level $\LL$ is maximal, $\MM_{\mmax \sigma^2}$: mass of a spherical top hat of diameter $\lambda_{\mmax \sigma^2}$, $\lambda_{\mathrm{cut}}$: wavelength at which power is reduced by four when power at levels $\LL$ and above are set to zero, $M_\mathrm{cut}$: mass of a spherical top-hat of diameter $\lambda_{\mathrm{cut}}$.}
\end{table}

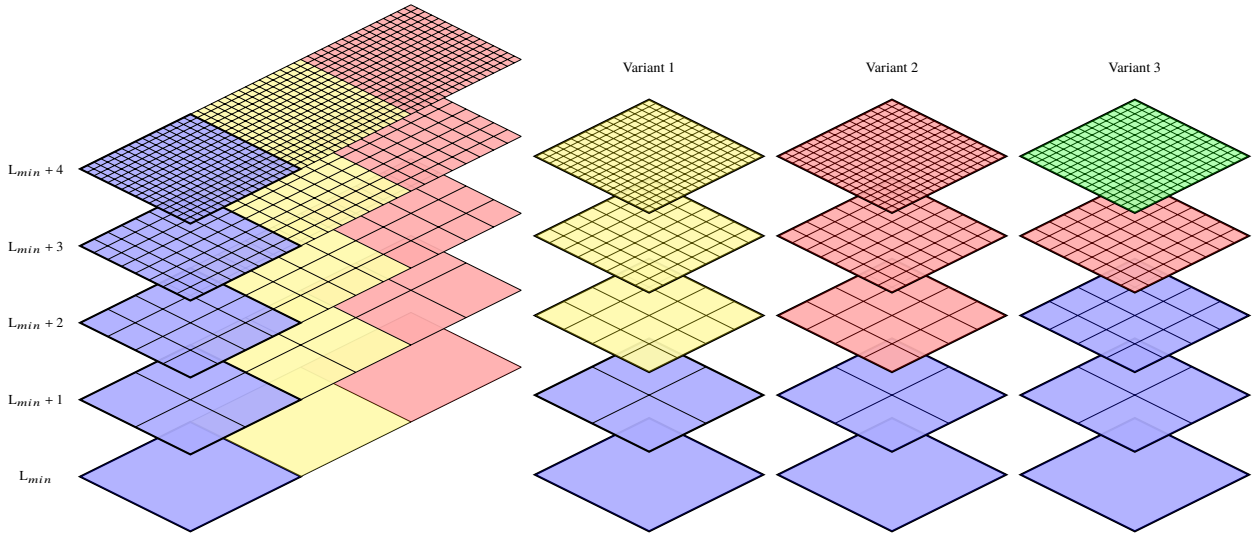
\begin{figure*}

\resizebox{!}{7cm}{
\begin{tikzpicture}[scale=.9,every node/.style={minimum size=1cm},on grid]

    \draw
    (-3.5,3.) node {$\LL_{min}$}
    (-3.5,4.75) node {$\LL_{min}+1$}
    (-3.5,6.5) node {$\LL_{min}+2$}
    (-3.5,8.25) node {$\LL_{min}+3$}
    (-3.5,10.) node {$\LL_{min}+4$};

    \begin{scope}[yshift=50,every node/.append style={yslant=0.5,xslant=-1,},yslant=0.5,xslant=-1]
        \draw[step=2.5, black] (0,0) grid (7.5,2.5);
        \draw[black,very thick] (0,0) rectangle (2.5,2.5);
        \fill[blue, fill opacity = 0.3] (0,0) rectangle (2.5,2.5);
        \fill[yellow, fill opacity = 0.3] (2.5,0) rectangle (5,2.5);
        \fill[red, fill opacity = 0.3] (5,0) rectangle (7.5,2.5);
    \end{scope}
    	
    \begin{scope}[yshift=100,every node/.append style={yslant=0.5,xslant=-1},yslant=0.5,xslant=-1]
    	\fill[white,fill opacity=.9] (0,0) rectangle (7.5,2.5);
        \fill[blue, fill opacity = 0.3] (0,0) rectangle (2.5,2.5);
        \fill[yellow, fill opacity = 0.3] (2.5,0) rectangle (5,2.5);
        \fill[red, fill opacity = 0.3] (5,0) rectangle (7.5,2.5);
        \draw[black,very thick] (0,0) rectangle (2.5,2.5);
        \draw[step=1.25] (0,0) grid (7.5,2.5);
    \end{scope}
    
    \begin{scope}[yshift=150,every node/.append style={yslant=0.5,xslant=-1},yslant=0.5,xslant=-1]https://www.overleaf.com/project/5cc7efcbdc86b619ccf1faa6
       	\fill[white,fill opacity=.9] (0,0) rectangle (7.5,2.5);
	    \fill[blue, fill opacity = 0.3] (0,0) rectangle (2.5,2.5);
        \fill[yellow, fill opacity = 0.3] (2.5,0) rectangle (5,2.5);
        \fill[red, fill opacity = 0.3] (5,0) rectangle (7.5,2.5);
    	\draw[black,very thick] (0,0) rectangle (2.5,2.5);
        \draw[step=.625] (0,0) grid (7.5,2.5);
    \end{scope}  
    
    \begin{scope}[yshift=200,every node/.append style={yslant=0.5,xslant=-1},yslant=0.5,xslant=-1]
    	\fill[white,fill opacity=.9] (0,0) rectangle (7.5,2.5);
        \fill[blue, fill opacity = 0.3] (0,0) rectangle (2.5,2.5);
        \fill[yellow, fill opacity = 0.3] (2.5,0) rectangle (5,2.5);
        \fill[red, fill opacity = 0.3] (5,0) rectangle (7.5,2.5);
        \draw[black,very thick] (0,0) rectangle (2.5,2.5);
        \draw[step=.3125] (0,0) grid (7.5,2.5);
    \end{scope}
    
    \begin{scope}[yshift=250,every node/.append style={yslant=0.5,xslant=-1},yslant=0.5,xslant=-1]
    	\fill[white,fill opacity=.9] (0,0) rectangle (7.5,2.5);
        \fill[blue, fill opacity = 0.3] (0,0) rectangle (2.5,2.5);
        \fill[yellow, fill opacity = 0.3] (2.5,0) rectangle (5,2.5);
        \fill[red, fill opacity = 0.3] (5,0) rectangle (7.5,2.5);
        \draw[black,very thick] (0,0) rectangle (2.5,2.5);
        \draw[step=.15625] (0,0) grid (7.5,2.5);
    \end{scope}
\end{tikzpicture}
} 
\resizebox{!}{6.5cm}{
\begin{tikzpicture}[scale=.9,every node/.style={minimum size=1cm},on grid]
    
     \draw
    (0.,12.) node {Variant 1};
    
    \begin{scope}[yshift=50,every node/.append style={yslant=0.5,xslant=-1,},yslant=0.5,xslant=-1]
        \fill[white,fill opacity=.9] (0,0) rectangle (2.5,2.5);
        \draw[step=2.5, black] (0,0) grid (2.5,2.5);
        \draw[black,very thick] (0,0) rectangle (2.5,2.5);
        \fill[blue, fill opacity = 0.3] (0,0) rectangle (2.5,2.5);
    \end{scope}
    	
    \begin{scope}[yshift=100,every node/.append style={yslant=0.5,xslant=-1},yslant=0.5,xslant=-1]
    	\fill[white,fill opacity=.9] (0,0) rectangle (2.5,2.5);
        \fill[blue, fill opacity = 0.3] (0,0) rectangle (2.5,2.5);
        \draw[black,very thick] (0,0) rectangle (2.5,2.5);
        \draw[step=1.25] (0,0) grid (2.5,2.5);
    \end{scope}
    	
    \begin{scope}[yshift=150,every node/.append style={yslant=0.5,xslant=-1},yslant=0.5,xslant=-1]
    	\fill[white,fill opacity=.9] (0,0) rectangle (2.5,2.5);
        \draw[black,very thick] (0,0) rectangle (2.5,2.5);
        \draw[step=0.625, black] (0,0) grid (2.5,2.5);
        \fill[yellow, fill opacity = 0.3] (0,0) rectangle (2.5,2.5);
    \end{scope}  
    
    \begin{scope}[yshift=200,every node/.append style={yslant=0.5,xslant=-1},yslant=0.5,xslant=-1]
    	\fill[white,fill opacity=.9] (0,0) rectangle (2.5,2.5);
        \draw[black,very thick] (0,0) rectangle (2.5,2.5);
        \draw[step=0.3125, black] (0,0) grid (2.5,2.5);
        \fill[yellow, fill opacity = 0.3] (0,0) rectangle (2.5,2.5);
    \end{scope}
    
    \begin{scope}[yshift=250,every node/.append style={yslant=0.5,xslant=-1},yslant=0.5,xslant=-1]
    	\fill[white,fill opacity=.9] (0,0) rectangle (2.5,2.5);
        \draw[black,very thick] (0,0) rectangle (2.5,2.5);
        \draw[step=0.15625, black] (0,0) grid (2.5,2.5);
        \fill[yellow, fill opacity = 0.3] (0,0) rectangle (2.5,2.5);
    \end{scope}
\end{tikzpicture}
} 
\resizebox{!}{6.5cm}{
\begin{tikzpicture}[scale=.9,every node/.style={minimum size=1cm},on grid]
    
     \draw
    (0.,12.) node {Variant 2};
    
    \begin{scope}[yshift=50,every node/.append style={yslant=0.5,xslant=-1,},yslant=0.5,xslant=-1]
        \fill[white,fill opacity=.9] (0,0) rectangle (2.5,2.5);
        \draw[step=2.5, black] (0,0) grid (2.5,2.5);
        \draw[black,very thick] (0,0) rectangle (2.5,2.5);
        \fill[blue, fill opacity = 0.3] (0,0) rectangle (2.5,2.5);
    \end{scope}
    	
    \begin{scope}[yshift=100,every node/.append style={yslant=0.5,xslant=-1},yslant=0.5,xslant=-1]
    	\fill[white,fill opacity=.9] (0,0) rectangle (2.5,2.5);
        \draw[black,very thick] (0,0) rectangle (2.5,2.5);
        \draw[step=1.25, black] (0,0) grid (2.5,2.5);
        \fill[blue, fill opacity = 0.3] (0,0) rectangle (2.5,2.5);
    \end{scope}
    
    \begin{scope}[yshift=150,every node/.append style={yslant=0.5,xslant=-1},yslant=0.5,xslant=-1]
    	\fill[white,fill opacity=.9] (0,0) rectangle (2.5,2.5);
        \draw[black,very thick] (0,0) rectangle (2.5,2.5);
        \draw[step=0.625, black] (0,0) grid (2.5,2.5);
        \fill[red, fill opacity = 0.3] (0,0) rectangle (2.5,2.5);
    \end{scope}  
    
    \begin{scope}[yshift=200,every node/.append style={yslant=0.5,xslant=-1},yslant=0.5,xslant=-1]
    	\fill[white,fill opacity=.9] (0,0) rectangle (2.5,2.5);
        \draw[black,very thick] (0,0) rectangle (2.5,2.5);
        \draw[step=0.3125, black] (0,0) grid (2.5,2.5);
        \fill[red, fill opacity = 0.3] (0,0) rectangle (2.5, 2.5);
    \end{scope}
    
    \begin{scope}[yshift=250,every node/.append style={yslant=0.5,xslant=-1},yslant=0.5,xslant=-1]
    	\fill[white,fill opacity=.9] (0,0) rectangle (2.5,2.5);
        \draw[black,very thick] (0,0) rectangle (2.5,2.5);
        \draw[step=0.15625, black] (0,0) grid (2.5,2.5);
        \fill[red, fill opacity = 0.3] (0,0) rectangle (2.5, 2.5);
    \end{scope}
\end{tikzpicture}
} 
\resizebox{!}{6.5cm}{
\begin{tikzpicture}[scale=.9,every node/.style={minimum size=1cm},on grid]
     \draw
    (0.,12.) node {Variant 3};
    
    \begin{scope}[yshift=50,every node/.append style={yslant=0.5,xslant=-1,},yslant=0.5,xslant=-1]
        \fill[white,fill opacity=.9] (0,0) rectangle (2.5,2.5);
        \draw[step=2.5, black] (0,0) grid (2.5,2.5);
        \draw[black,very thick] (0,0) rectangle (2.5,2.5);
        \fill[blue, fill opacity = 0.3] (0,0) rectangle (2.5,2.5);
    \end{scope}
    	
    \begin{scope}[yshift=100,every node/.append style={yslant=0.5,xslant=-1},yslant=0.5,xslant=-1]
    	\fill[white,fill opacity=.9] (0,0) rectangle (2.5,2.5);
        \draw[black,very thick] (0,0) rectangle (2.5,2.5);
        \draw[step=1.25, black] (0,0) grid (2.5,2.5);
        \fill[blue, fill opacity = 0.3] (0,0) rectangle (2.5,2.5);
    \end{scope}
    
    \begin{scope}[yshift=150,every node/.append style={yslant=0.5,xslant=-1},yslant=0.5,xslant=-1]
    	\fill[white,fill opacity=.9] (0,0) rectangle (2.5,2.5);
        \draw[black,very thick] (0,0) rectangle (2.5,2.5);
        \draw[step=0.625, black] (0,0) grid (2.5,2.5);
        \fill[blue, fill opacity = 0.3] (0,0) rectangle (2.5,2.5);
    \end{scope}  
    
    \begin{scope}[yshift=200,every node/.append style={yslant=0.5,xslant=-1},yslant=0.5,xslant=-1]
    	\fill[white,fill opacity=.9] (0,0) rectangle (2.5,2.5);
        \draw[black,very thick] (0,0) rectangle (2.5,2.5);
        \draw[step=0.3125, black] (0,0) grid (2.5,2.5);
        \fill[red, fill opacity = 0.3] (0,0) rectangle (2.5, 2.5);
    \end{scope}

    \begin{scope}[yshift=250,every node/.append style={yslant=0.5,xslant=-1},yslant=0.5,xslant=-1]
    	\fill[white,fill opacity=.9] (0,0) rectangle (2.5,2.5);
        \draw[black,very thick] (0,0) rectangle (2.5,2.5);
        \draw[step=0.15625, black] (0,0) grid (2.5,2.5);
        \fill[green, fill opacity = 0.3] (0,0) rectangle (2.5, 2.5);
    \end{scope}
\end{tikzpicture}
} 

\caption{Illustration of displacements in the {\sc Panphasia} white noise field. In all panels the region of {\sc Panphasia} represented in the simulation volume is indicated by thick lines. In the left panel, adjacent regions of {\sc Panphasia}, displaced by 1 and 2 in one dimension are shown in red and yellow, respectively. The second and third panels  (Variants 1 and 2) illustrate shifts at level L$_{min + 2}$ by 1 and 2, respectively. The fourth panel, Variant 3, represents a shift by 1 at level L$_{min + 3}$ and another arbitrary shift at level L$_{min + 4}$. Any combination of shifts is possible, and all integer shifts result in independent white noise fields on the respective levels.     \label{fig:shifts}}

\end{figure*}

\subsubsection{Panphasia}
In addition to defining the $S_8$ octree basis functions, \cite{Jenkins-2013} also defined a
single extremely large `public' realisation of a Gaussian white noise field, called `Panphasia'. The Panphasia field is an octree with 50 levels, more than enough to encompass all the phase information of all existing cosmological simulations. By design the phase information in Panphasia can be computed rapidly at
any location of the field and at any depth in the octree. We will take our phase information for this paper from the Panphasia field.  

All the simulations for this paper are of a $100^3$~cMpc$^3$ volume. We define a reference
set of phase information which is the phase information used for the
Eagle Project flagship $100^3$~cMpc$^3$ volume \citep{Schaye-2014}.  The phase
information for this simulation occupies a very small region of the entire
Panphasia field.  The reference simulation phase information
comes from a cubic patch of dimension $12^3$
at the twelfth level octree (so at this level the whole Panphasia field consists
of $2^{12} = 2048$ cells on a side). We will use the symbol, $L$, usually with
a subscript to denote the level of phase information in the Panphasia field.
The zero point for the octree levels is arbitrary and simply follows as a consequence
of a choice for the reference phases made by 
\cite{Schaye-2014}.  

\subsubsection{Variations as Shifts in {\sc Panphasia}}
As illustrated in Fig.~\ref{fig:shifts}, we can conveniently introduce random variations to the initial density field as coordinate shifts in {\sc Panphasia}. The left panel shows three adjacent sub-volumes of {\sc Panphasia}, in blue, yellow and red, respectively. Each sub-volume contains completely independent phase information, i.e. completely independent regions of the white noise field, for the same simulation volume. The next three panels, to the right, show several possible variants. Assuming that the blue {\sc Panphasia} region contains the full phase information of the Reference simulation from levels $\LL_{\mathrm{min}}$ to $\LL_\mmax$, variants~1 and~2 differ from the Reference simulation, and from each other, at levels $\LL_{\mathrm{min+2}}$ and above. Variant~3 shares levels $\LL_{\mathrm{min}}$ to $\LL_{\mathrm{min+2}}$ with the Reference simulation, levels $\LL_{\mathrm{min+3}}$ with the second variant, and differs from all other simulations at level $\LL_{\mathrm{min+4}}$.

Provided that there is significant large-scale power, initial conditions with different low-level phase information result in the formation of different objects, independent of shared high-level phase information. The Reference simulation, and the first and second variants, share the same amount of phase information ($\LL_{min}$ to $\LL_{min+2}$), so it is expected that, statistically, the structures formed in each will be equally similar to one another. Since the levels of {\sc Panphasia} are completely independent, the choice of "Reference" among the three is arbitrary. If two simulations share the same large-scale phase information, the statistical similarity in the structures formed depends on the smallest scale down to which the phase information is shared. While Variants~2 and~3 also share phase information at level $\LL_{min+4}$, the fact that they differ at $\LL_{min+3}$ means that they have no additional similarity. Out of all illustrated volumes, the Reference simulation and Variant~3 share phase information down to the smallest scale, so it is expected that they will have the greatest similarity in the structures formed.

We label a set of simulations that differ from the phase information of the Reference simulation from level $\LL$ to $\LL_\mmax$ as variants VL, and identify individual volumes that employ a shift by $i$ from level $\LL$ to $\LL_\mmax$ as VL$_i$. For example, V16 is the set of variants that differ from the Reference simulation at level $\LL=16$ and above, and V$18_5$ is the individual variant that differs from the Reference simulation at level $\LL=18$ and above by a shift by 5. Variants with multiple shifts, e.g. by $i$ from level L to level $\MM-1$ and by $j$ from level $\MM$ to level $\LL_\mmax$, where $\LL < \MM \leq \LL_\mmax$, are labelled  VL$_i/\MM_j$, etc. Hence, V18$_{2}/$21$_5$ is the individual variant that employs a shift by 2 at levels 18 to 20, and a shift by 5 at levels 21 and above.

\subsection{Simulations and structure finding}\label{sec:methods:simulations}
All simulations presented here assume a $\Lambda$CDM cosmology, with parameters $h=0.6777$, $n_s=0.9611$, $\sigma_8=0.8288$, $\Omega_0=0.307$, $\Omega_b=0.0483$, and $\Omega_\Lambda=0.693$. They are set up in a volume of $100^3$ cMpc$^3$, using $N=384^3$ $(5.3 \times 10^7)$ particles, giving a particle mass of $7.4\times 10^8 \Ms$, a mean interparticle separation of 260 ckpc, and a comoving softening length of 13 ckpc throughout.

We use the IC\_Gen initial conditions code and the methods described in the papers \cite{Jenkins-2010,Jenkins-2013} to make second-order Lagrangian perturbation theory initial conditions for a starting redshift of 127. We used a $1536^3$ Fourier mesh to generated all of the initial conditions. The Nyquist frequency of this Fourier mesh is a factor of four smaller than the particle Nyquist frequency. We set all Fourier modes to zero if the magnitude of their wavevector equals or exceeds the one-dimensional particle Nyquist frequency.
 
The simulations are run using {\sc P-Gadget-3}, a TreePM code based on the publicly available code {\sc Gadget-2} \citep{Springel-2005-gadget}. In total, we have performed 469 simulations: one simulation with phase information identical to the 100 cMpc {\sc Eagle} volume of \cite{Schaye-2014}, hereafter called "Reference", and 39 "variant" simulations, for each of the 12 levels from 12 to 23. At every level, there are thus 40 simulations including the Reference simulation, which are equidistant in their white noise fields.

Overdensities and self-bound structures are identified using the FoF \citep{Davis-1985} and {\sc Subfind} \citep{Springel-2001-subfind} algorithms, respectively. Throughout this paper, unless otherwise mentioned, the term "halo" refers to a self-bound structure, as identified by {\sc Subfind}, and we limit some of our analysis to only central haloes, i.e. the most massive self-bound structures within their FoF groups.

\begin{figure*}
    \begin{overpic}[width=0.33\textwidth,trim={0 100 0 100},clip]{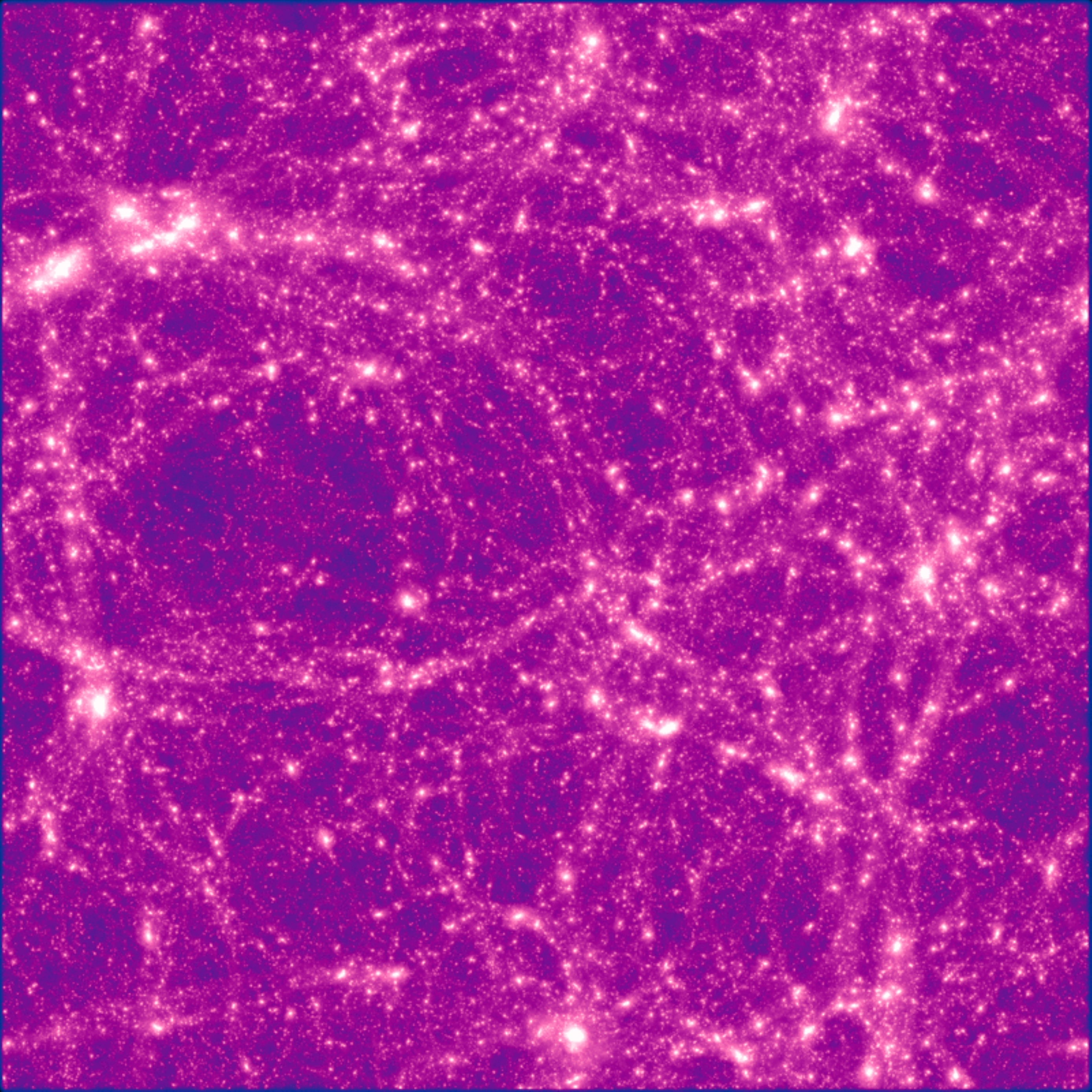}
    \put(5,5) {\textcolor{white}{\bf Reference}} 
    \linethickness{3pt}
    \put(77,8) {\textcolor{white}{\bf 20 Mpc}}
    \put(75,5){\color{white}\line(1,0){20}}
    \end{overpic} 
    \begin{overpic}[width=0.33\textwidth, trim={0 100 0 100},clip]{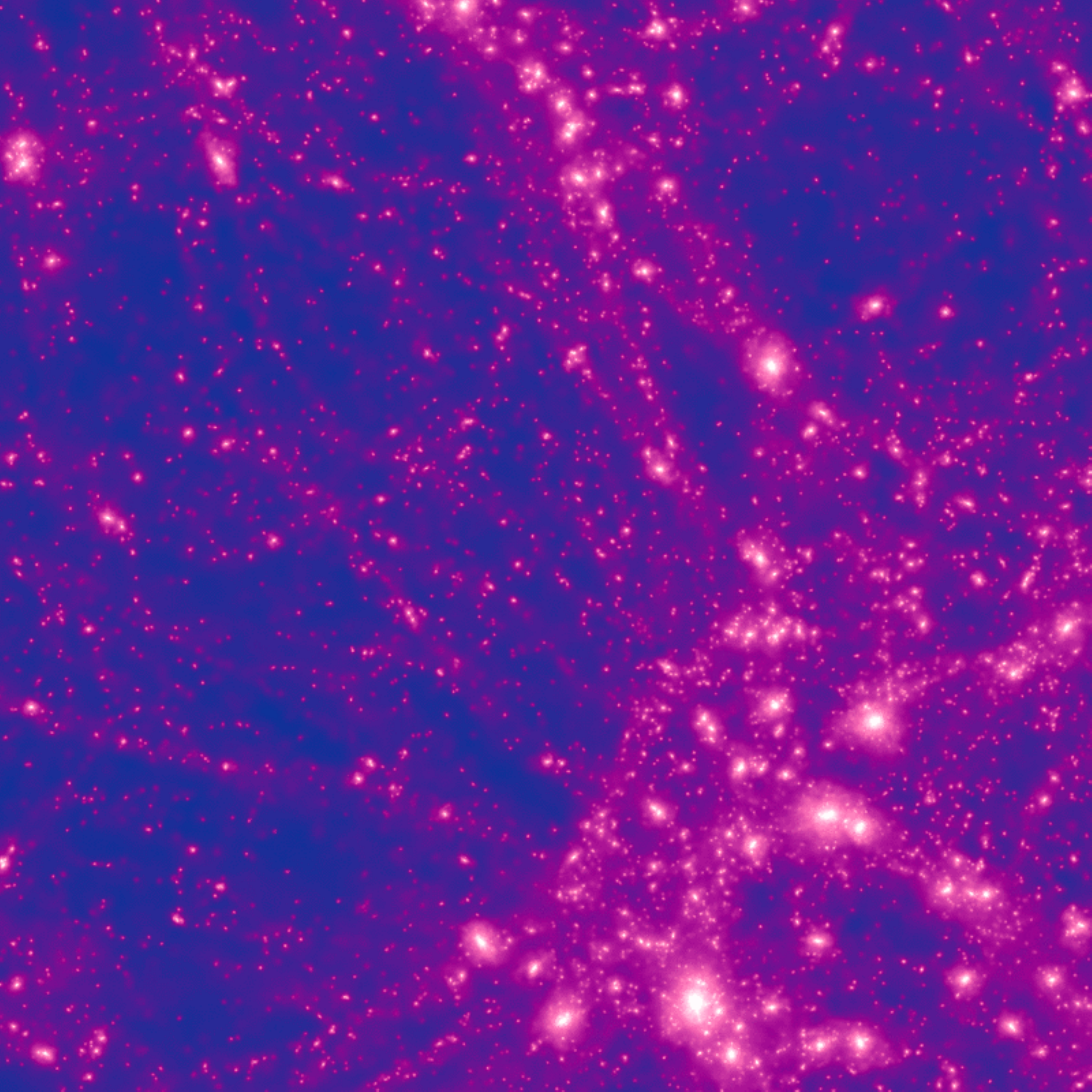}
    \put(5,5) {\textcolor{white}{\bf Reference}} 
    \linethickness{3pt}
    \put(77,8) {\textcolor{white}{\bf 6 Mpc}}
    \put(75,5){\color{white}\line(1,0){18}}
    \end{overpic} 
    \begin{overpic}[width=0.33\textwidth, trim={0 100 0 100},clip]{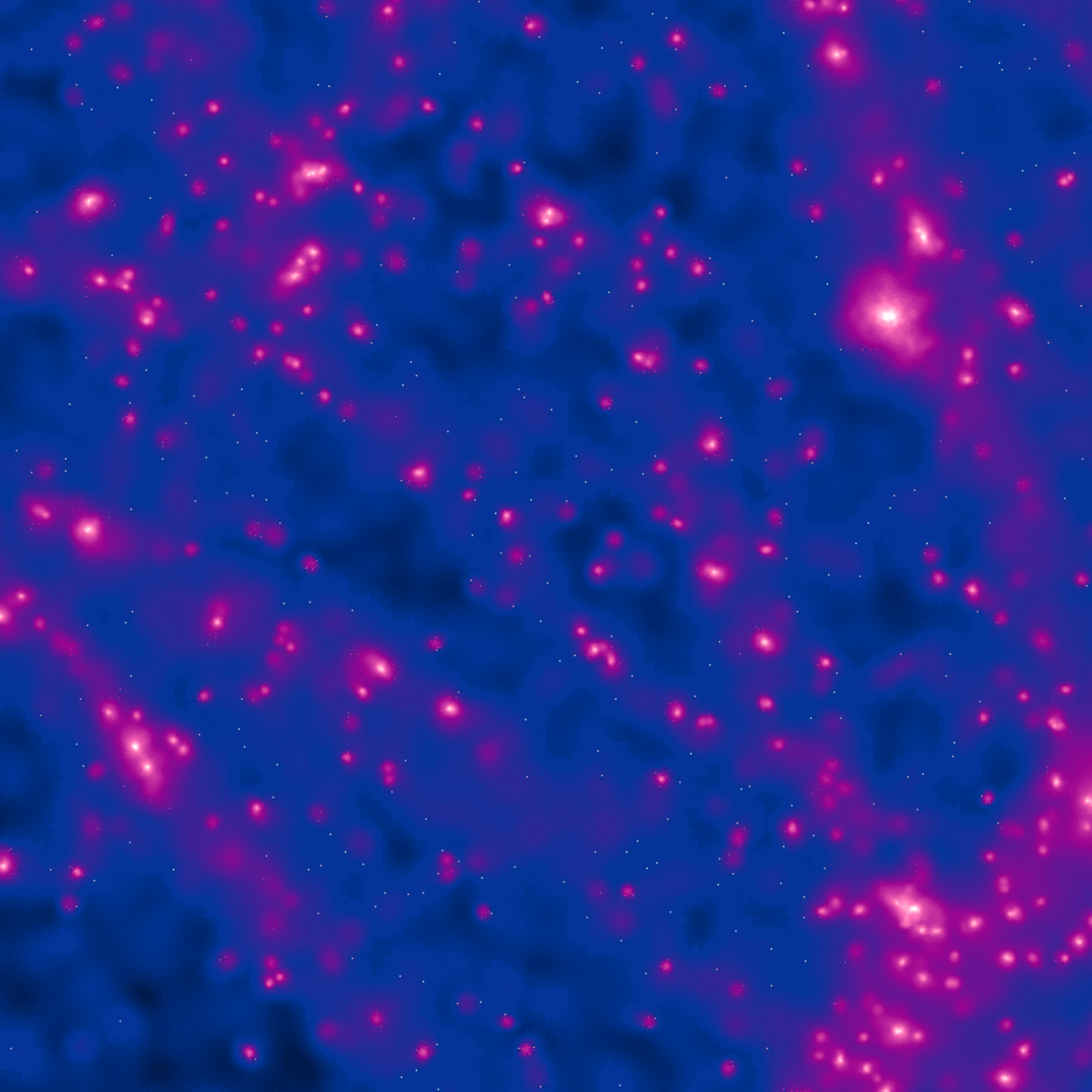}
    \put(5,5) {\textcolor{white}{\bf Reference}} 
    \linethickness{3pt}
    \put(78,8) {\textcolor{white}{\bf 2 Mpc}}
    \put(75,5){\color{white}\line(1,0){20}}
    \end{overpic} \\
    \vspace{1mm}
    \begin{overpic}[width=0.33\textwidth, trim={0 100 0 100},clip]{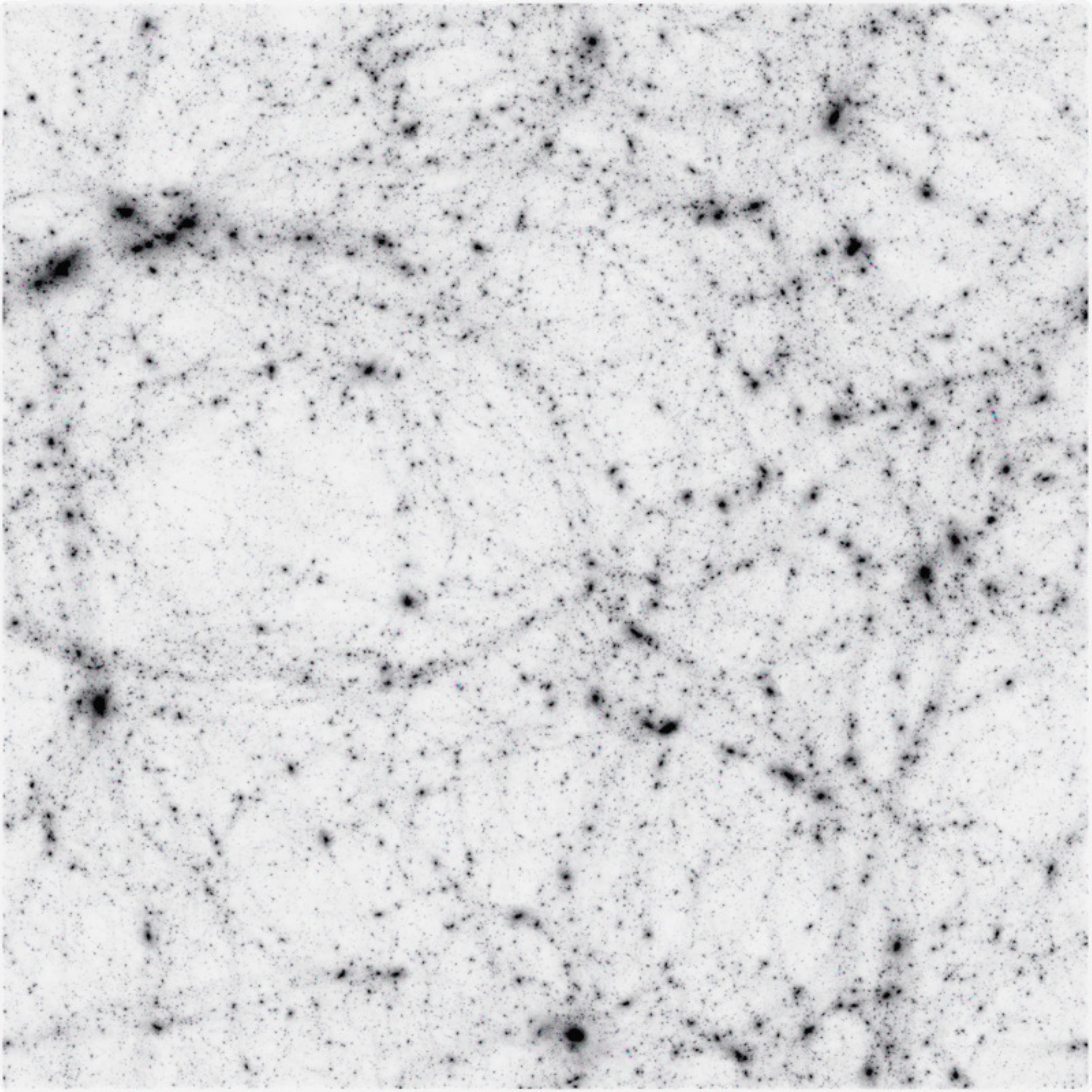}
    \put(5,5) {\textcolor{black}{\bf V22$_{1}$ }} 
    \TwentyMpcBar
    \end{overpic} 
    \begin{overpic}[width=0.33\textwidth, trim={0 100 0 100},clip]{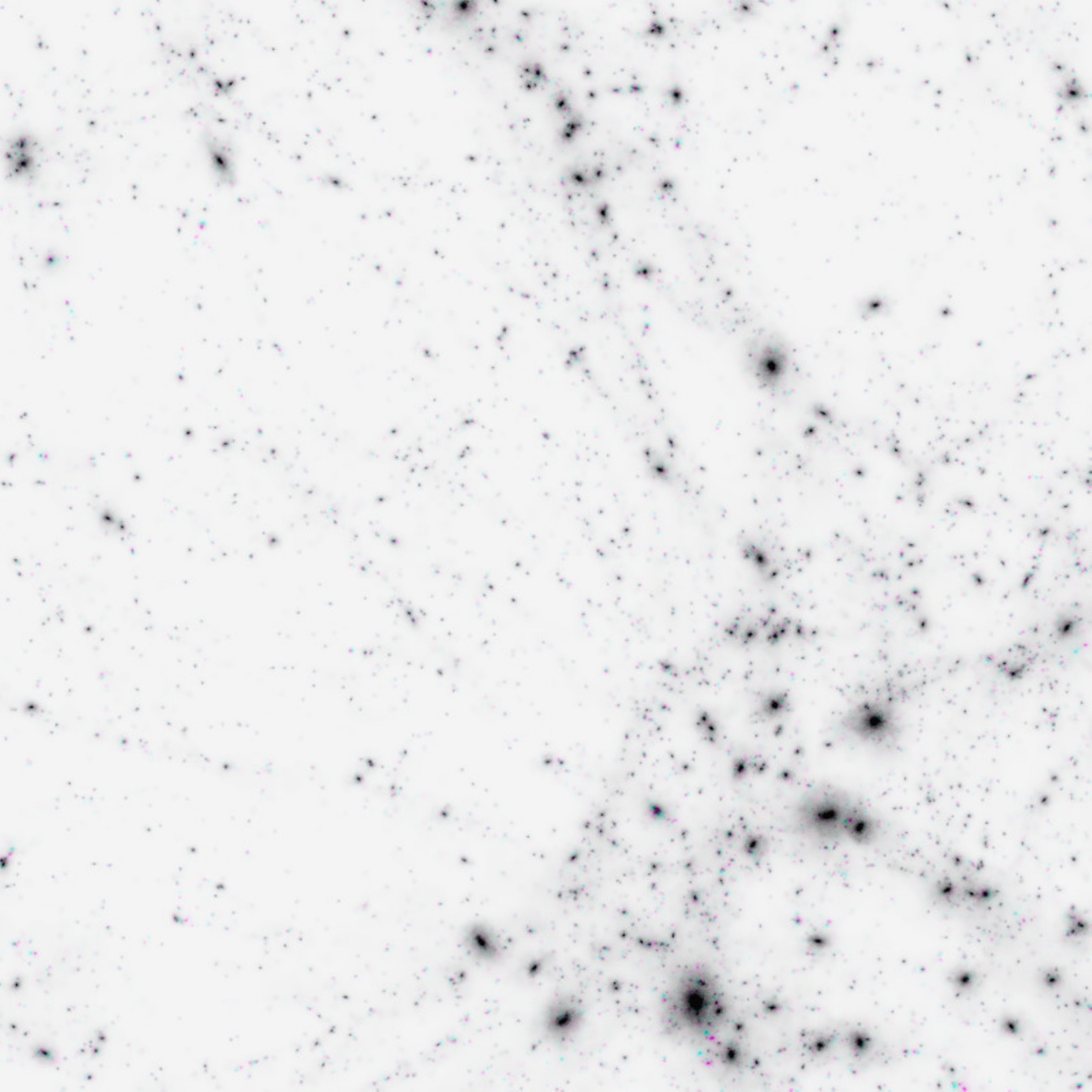}
    \put(5,5) {\textcolor{black}{\bf V22$_{1}$}}
    \SixMpcBar
    \end{overpic} 
    \begin{overpic}[width=0.33\textwidth, trim={0 100 0 100},clip]{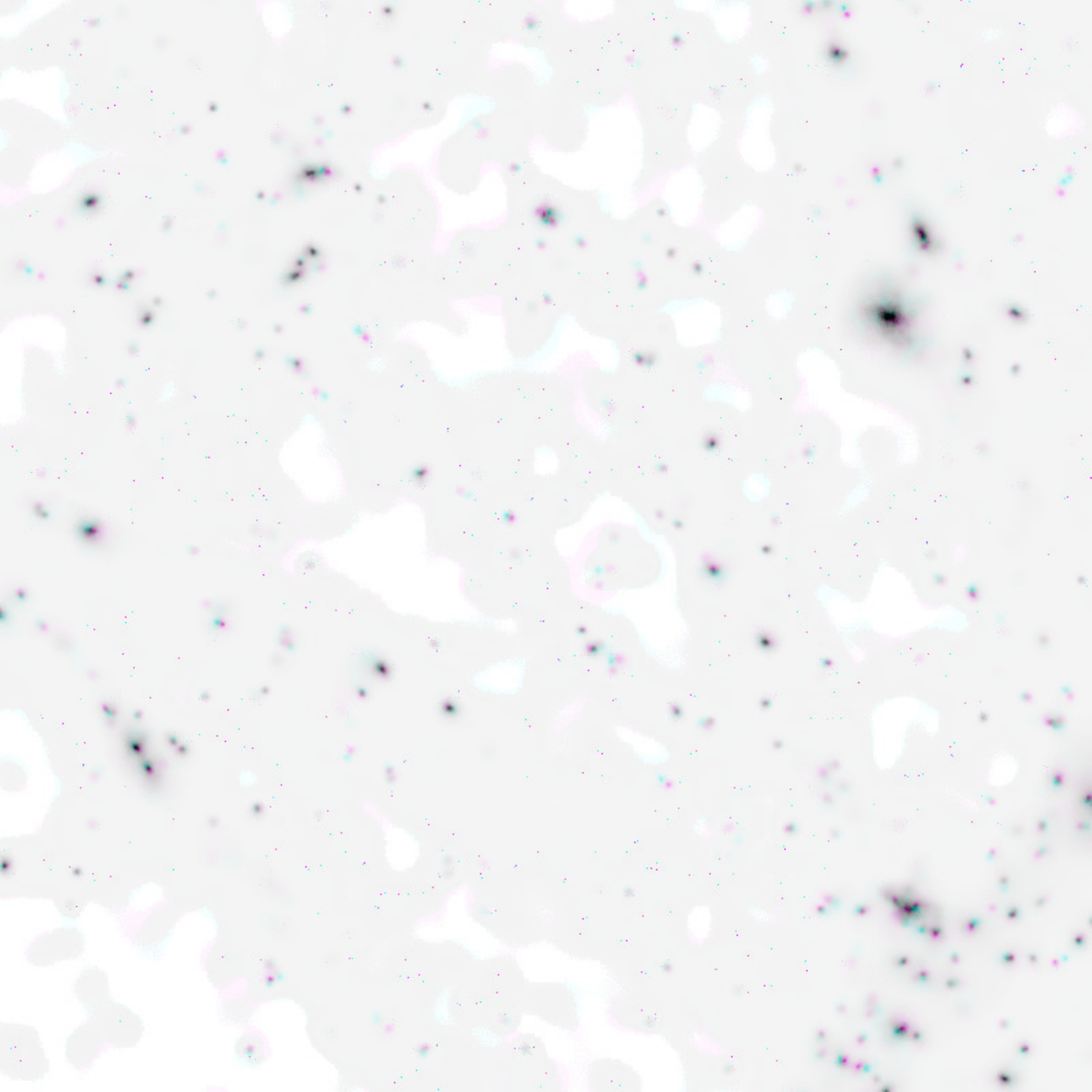}
    \put(5,5){\textcolor{black}{\bf V22$_{1}$}}
    \TwoMpcBar
    \end{overpic} \\
    \vspace{1mm}
    \begin{overpic}[width=0.33\textwidth, trim={0 100 0 100},clip]{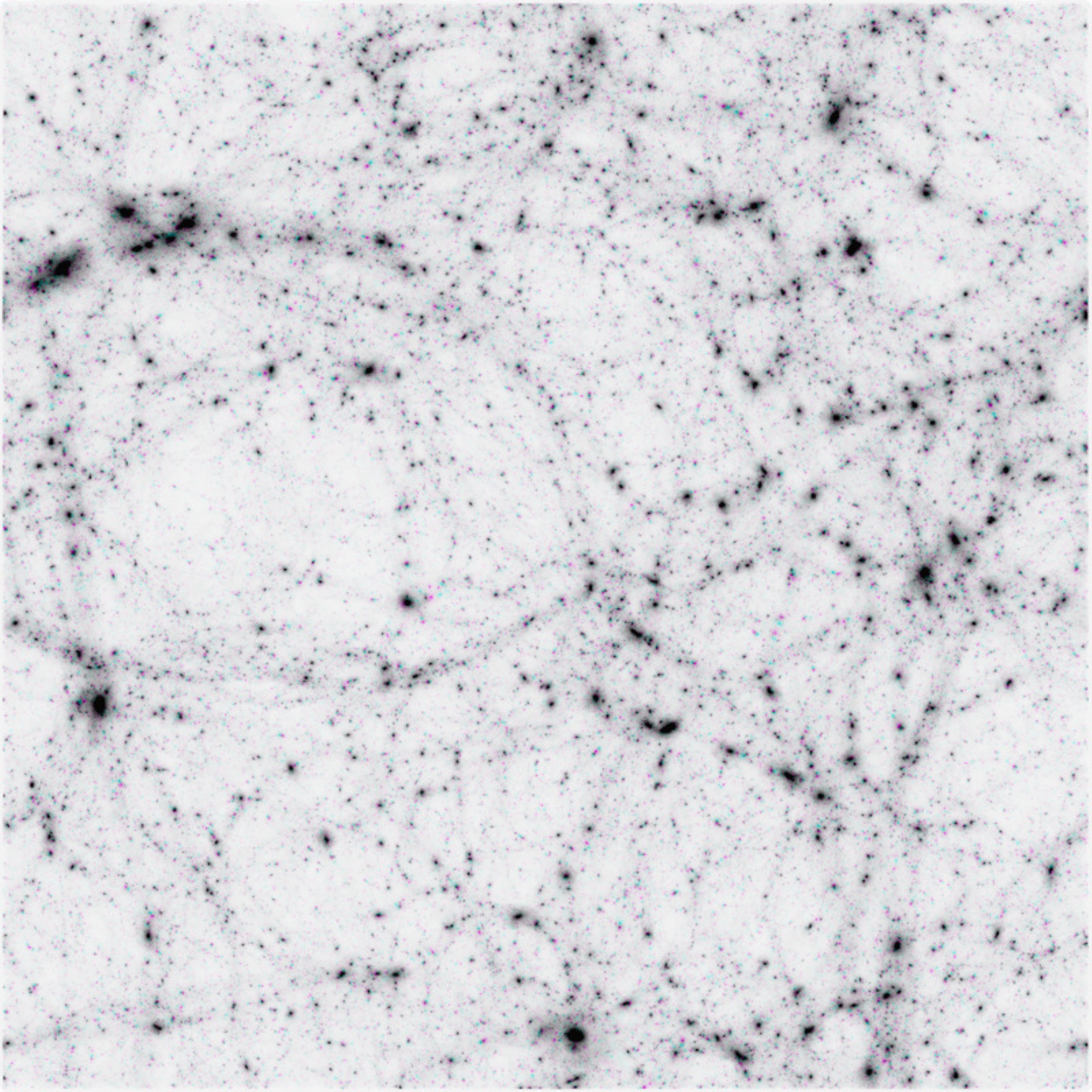}
    \put(5,5){\textcolor{black}{\bf V20$_{1}$}} 
    \TwentyMpcBar
    \end{overpic} 
    \begin{overpic}[width=0.33\textwidth, trim={0 100 0 100},clip]{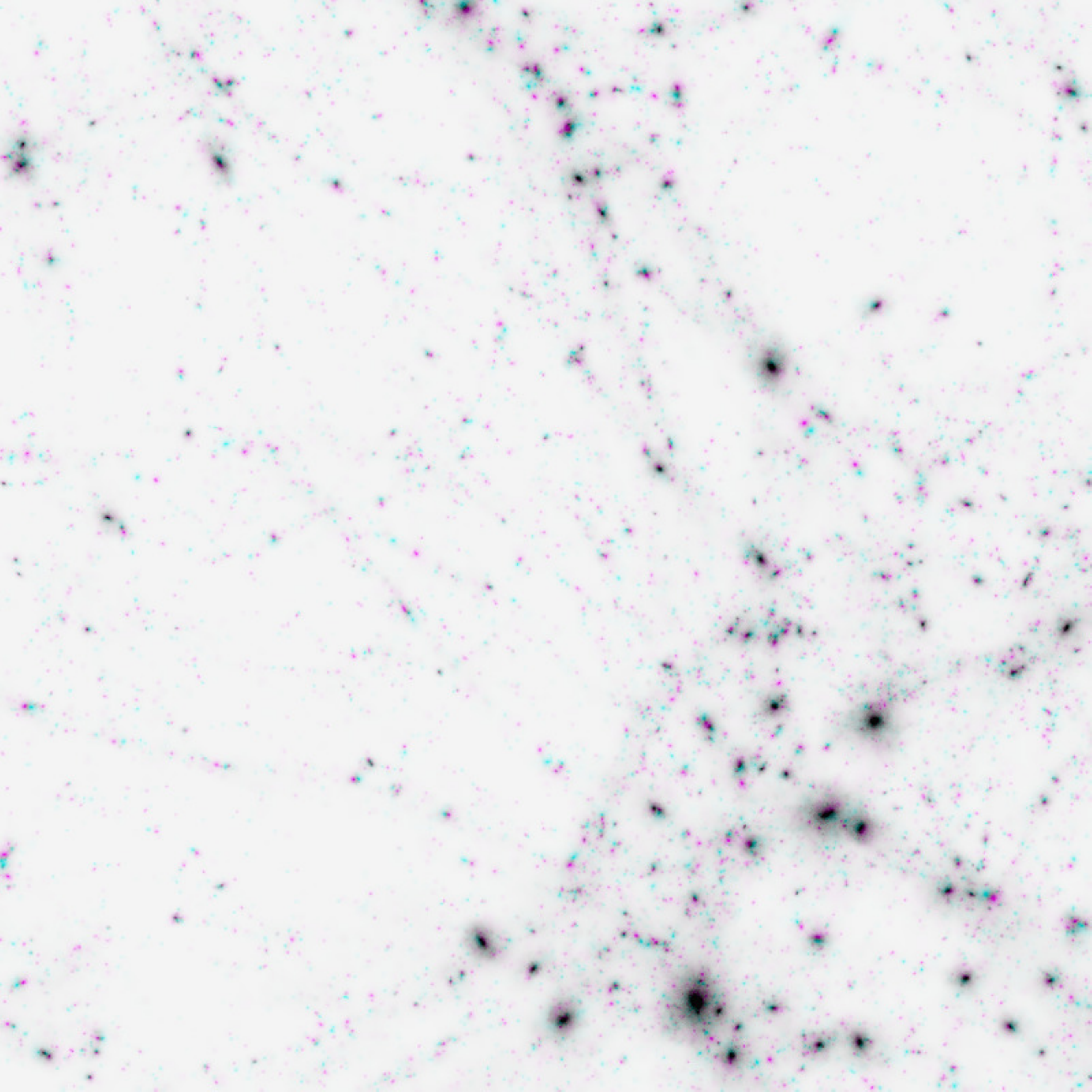}
    \put(5,5){\textcolor{black}{\bf V20$_{1}$}}
    \SixMpcBar
    \end{overpic} 
    \begin{overpic}[width=0.33\textwidth, trim={0 100 0 100},clip]{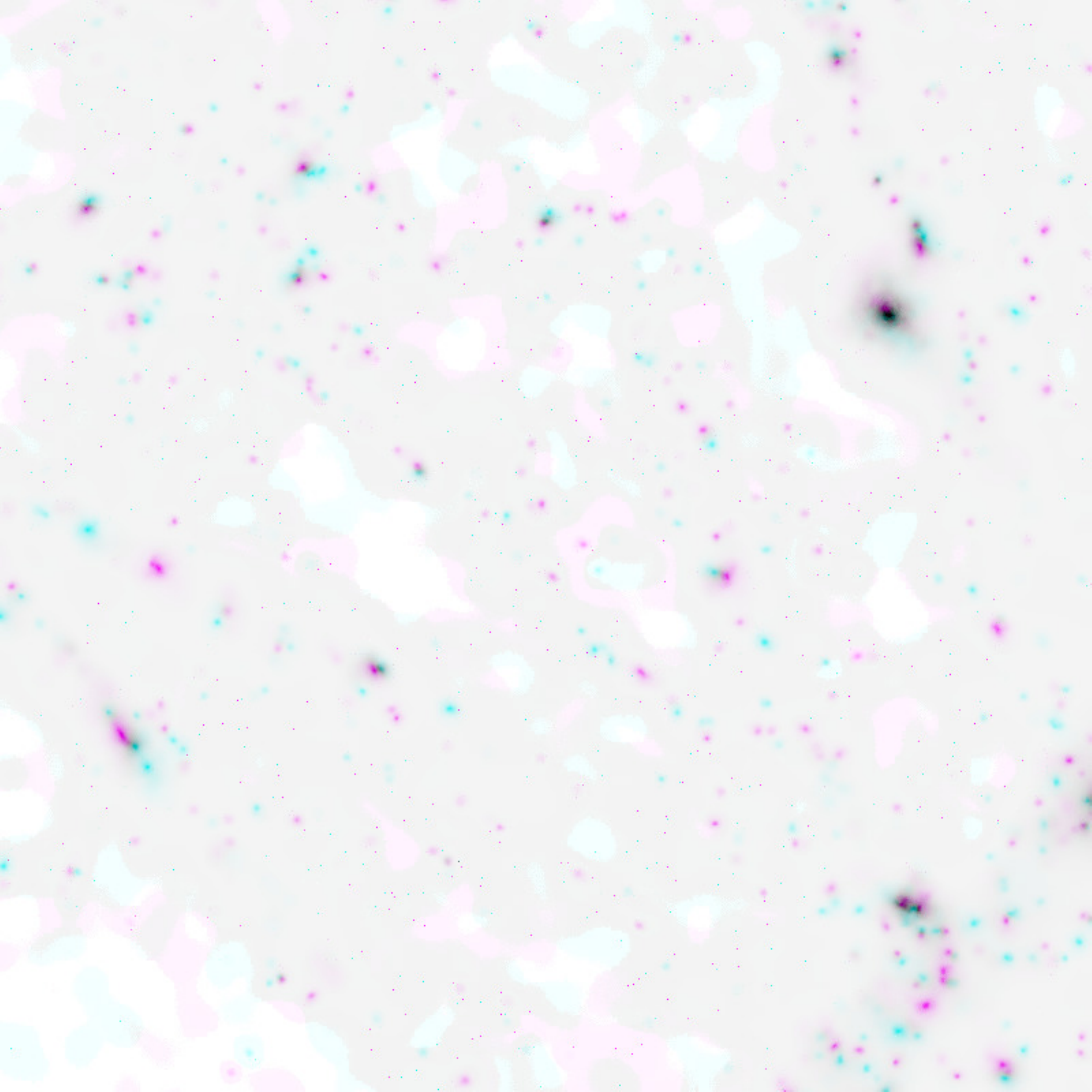}
    \put(5,5) {\textcolor{black}{\bf V20$_{1}$}}
    \TwoMpcBar
    \end{overpic} \\
    \vspace{1mm}
    \begin{overpic}[width=0.33\textwidth, trim={0 100 0 100},clip]{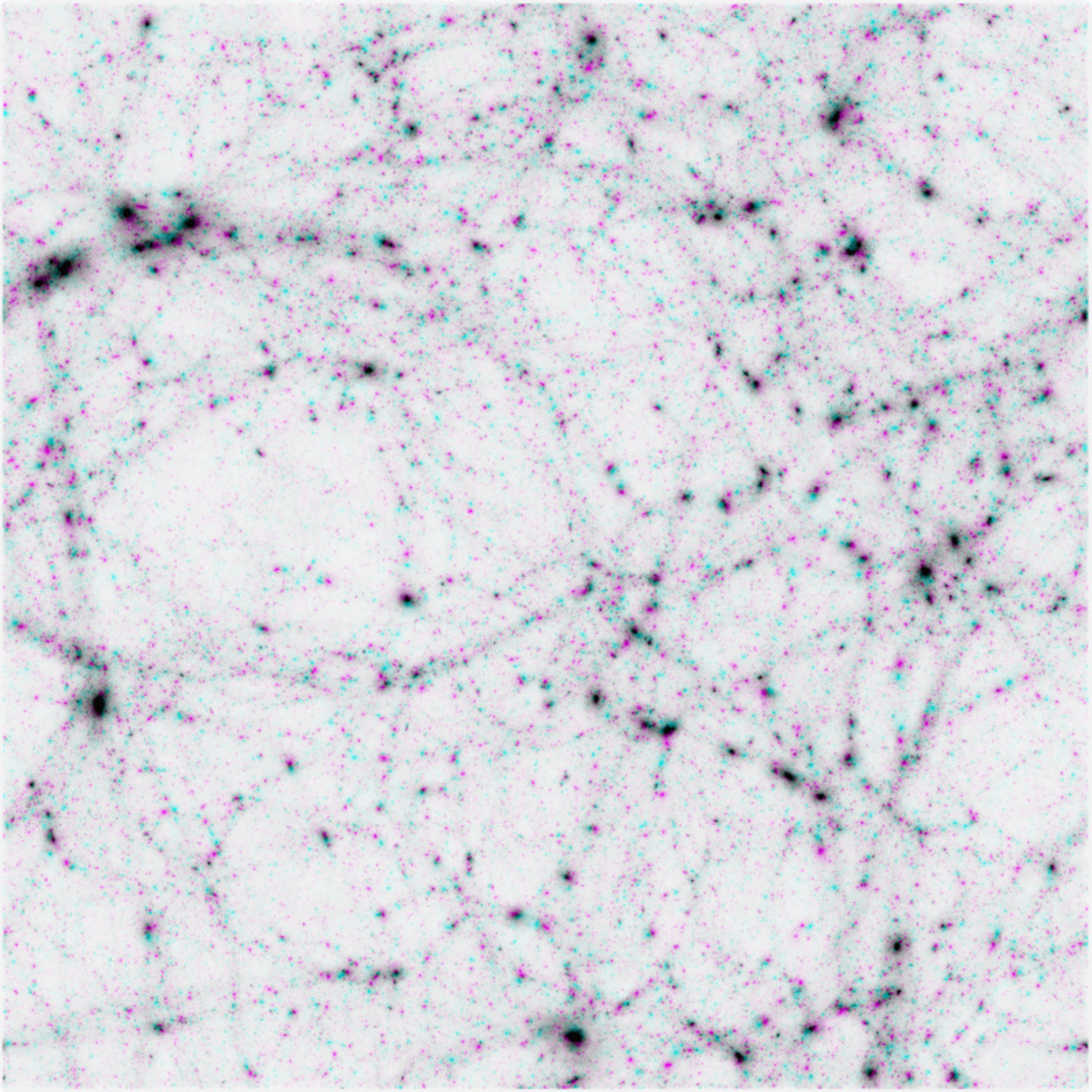}
    \put(5,5) {\textcolor{black}{\bf V18$_{1}$}}
    \TwentyMpcBar
    \end{overpic} 
    \begin{overpic}[width=0.33\textwidth, trim={0 100 0 100},clip]{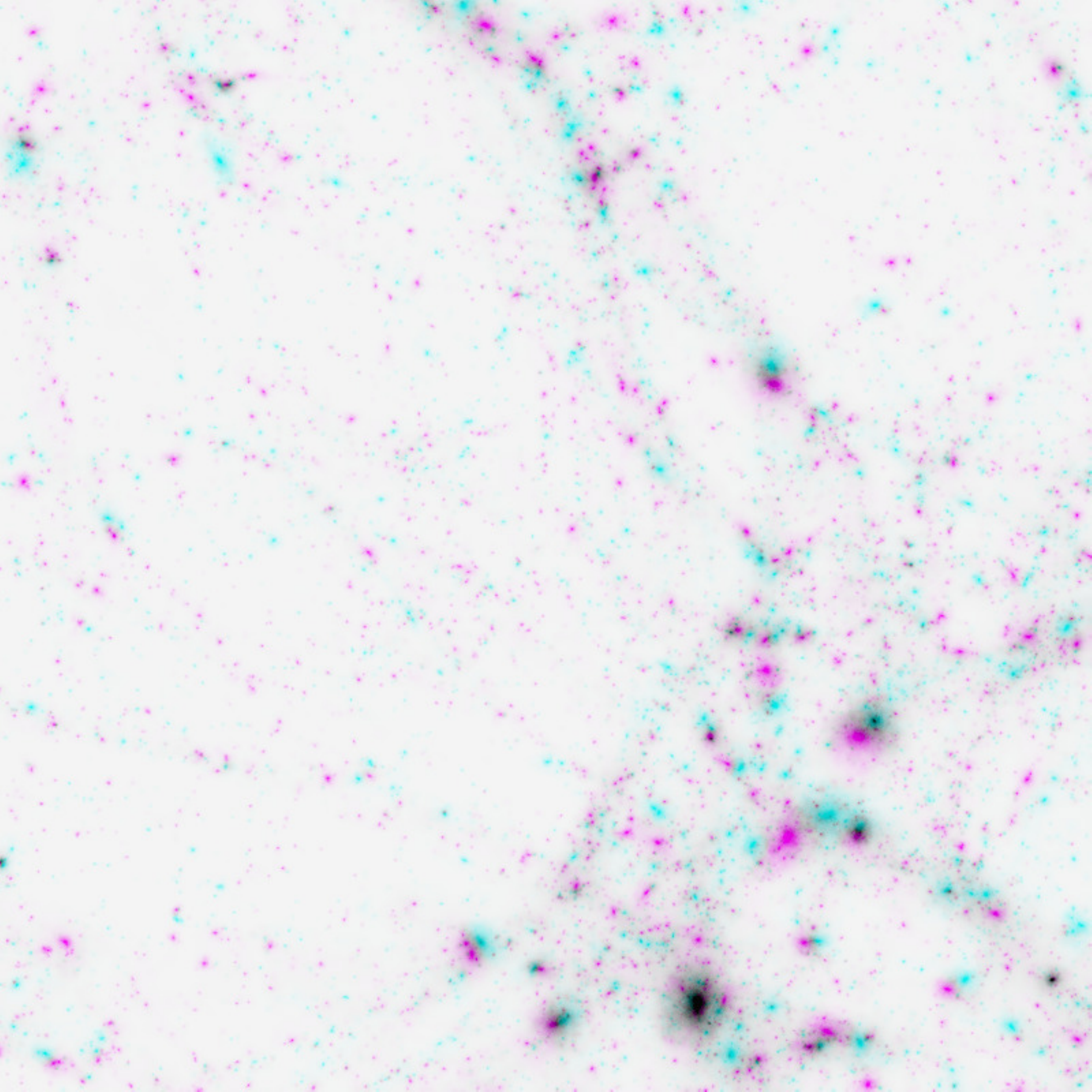}
    \put(5,5) {\textcolor{black}{\bf V18$_{1}$}} 
    \SixMpcBar
    \end{overpic} 
    \begin{overpic}[width=0.33\textwidth, trim={0 100 0 100},clip]{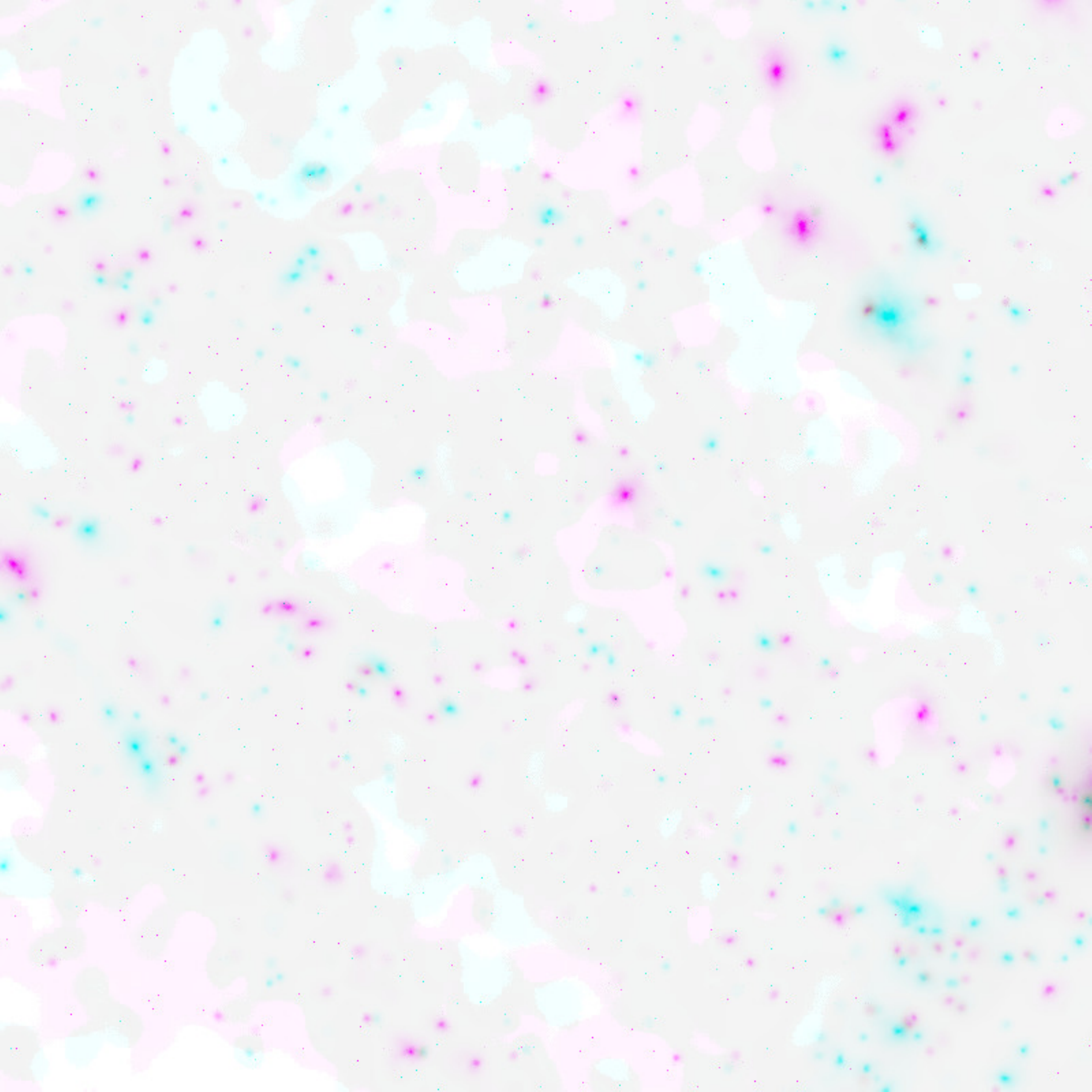}
    \put(5,5) {\textcolor{black}{\bf  V18$_{1}$}}
    \TwoMpcBar
    \end{overpic} \\
    \vspace{1mm}
    \begin{overpic}[width=0.33\textwidth, trim={0 100 0 100},clip]{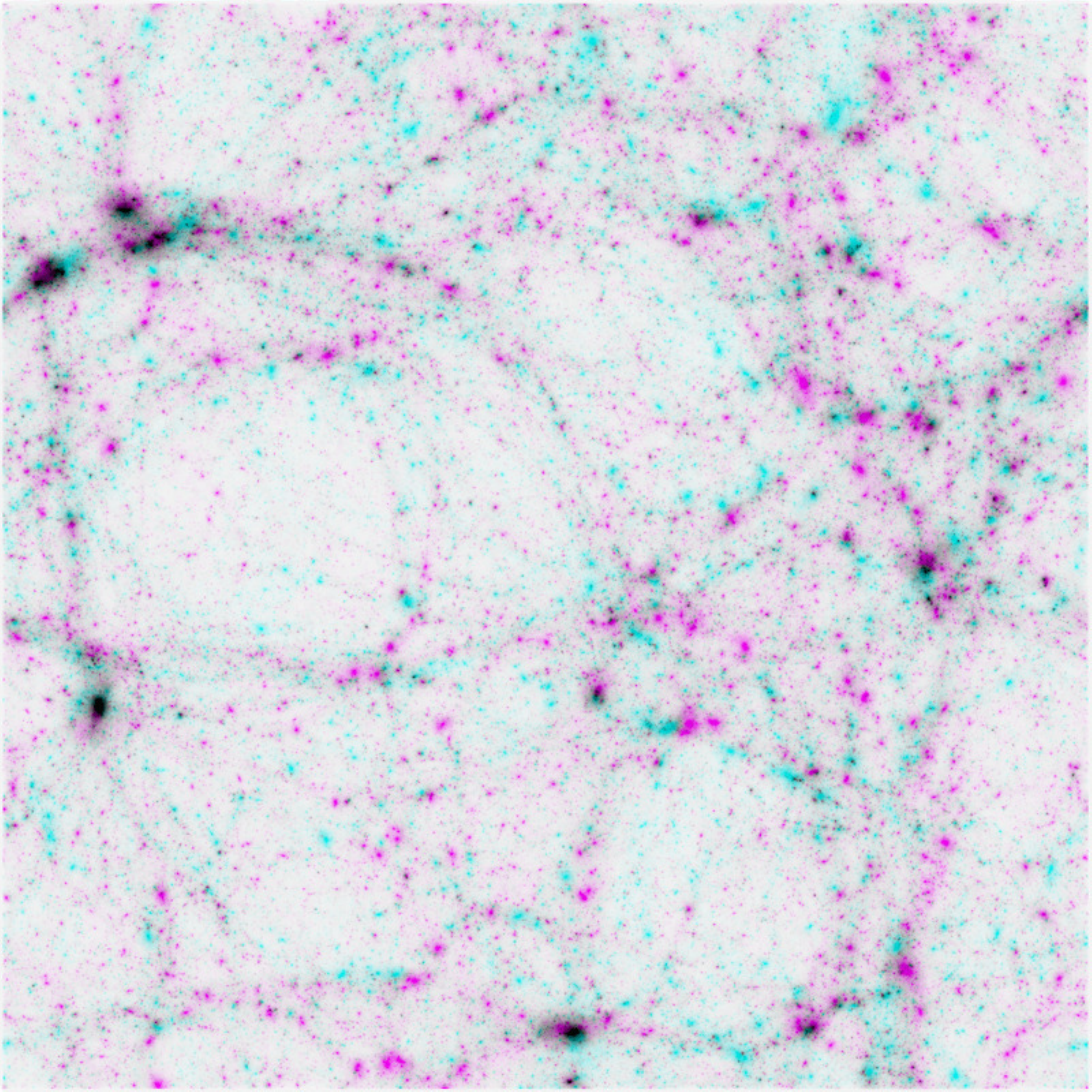}
    \put(5,5) {\textcolor{black}{\bf  V16$_{1}$}}
    \TwentyMpcBar
    \end{overpic} 
    \begin{overpic}[width=0.33\textwidth, trim={0 100 0 100},clip]{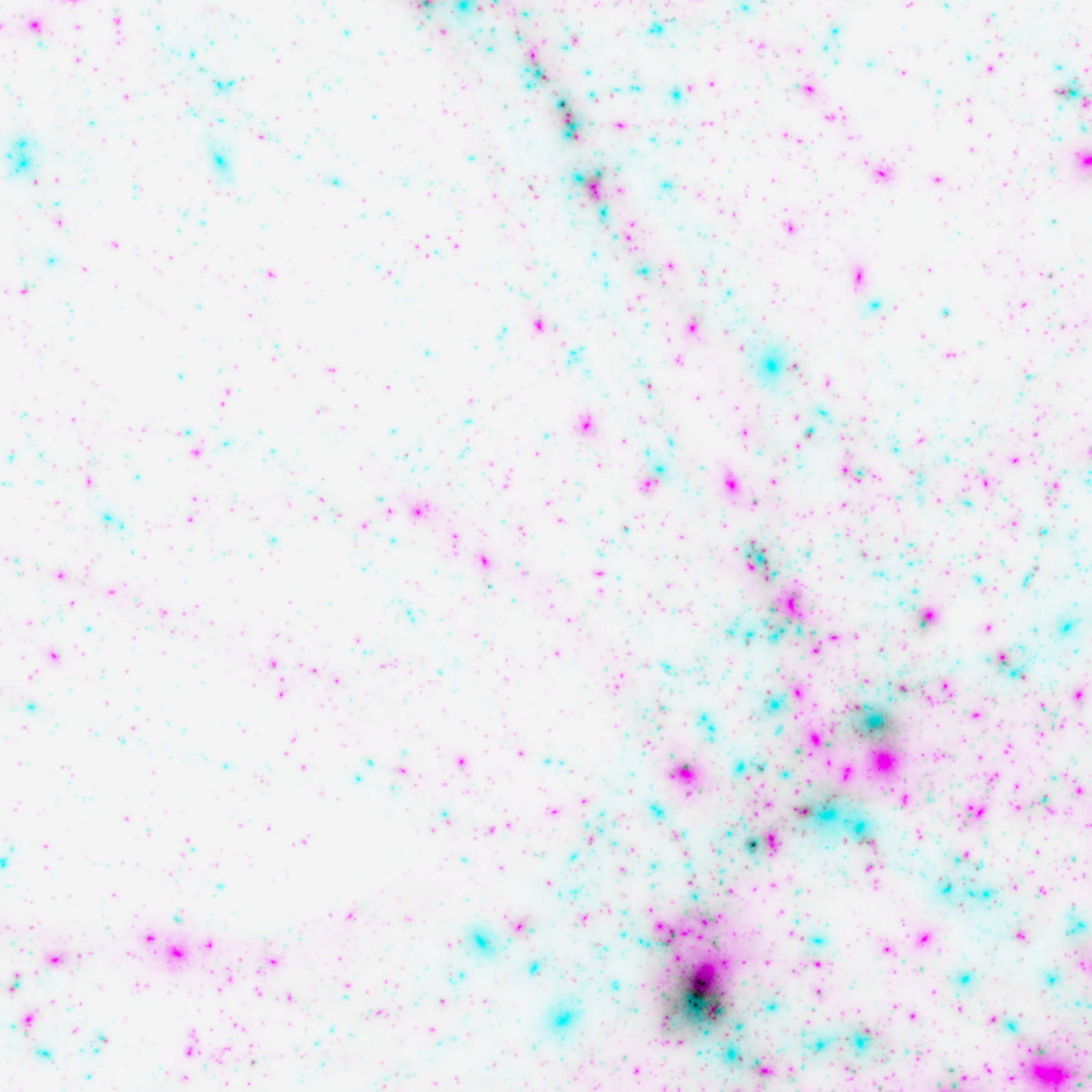}
    \put(5,5) {\textcolor{black}{\bf  V16$_{1}$}} 
    \SixMpcBar
    \end{overpic} 
    \begin{overpic}[width=0.33\textwidth, trim={0 100 0 100},clip]{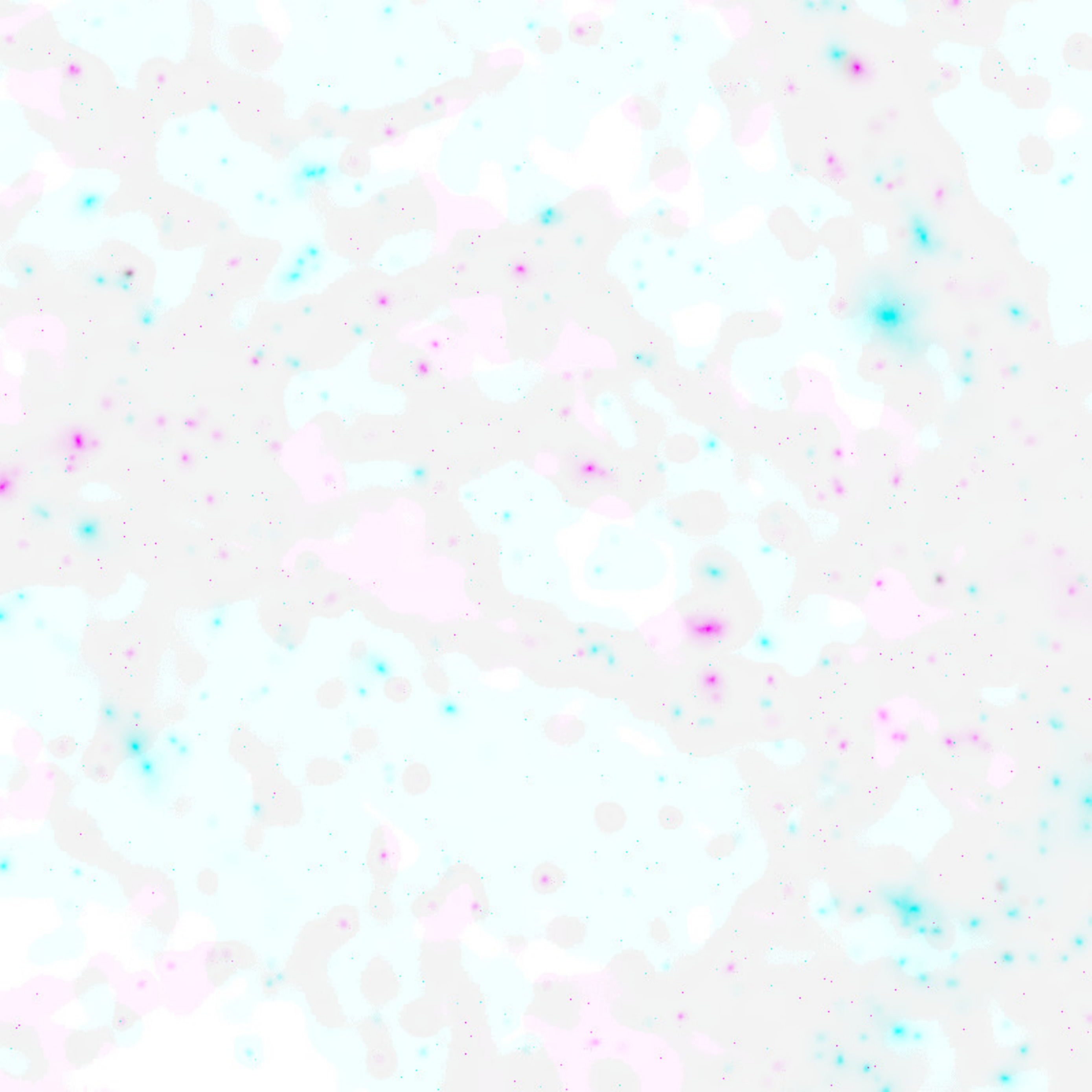}
    \put(5,5) {\textcolor{black}{\bf  V16$_{1}$}}
    \TwoMpcBar
    \end{overpic}\\
    \caption{Dark matter density at $z=0$. From left to right, columns show boxes of width and depth 100, 33 and 10 Mpc, respectively. The top row shows the simulation with the Reference phase information, while the following rows show the effect of randomising phase information at and above levels $22, 20, 18$, respectively. On each panel, shades of grey indicate high projected density in both the Reference and variant simulations, blue indicates higher density in the Reference simulation, purple indicates higher density in the variant simulations. \label{fig:structure-levels} }
\end{figure*}
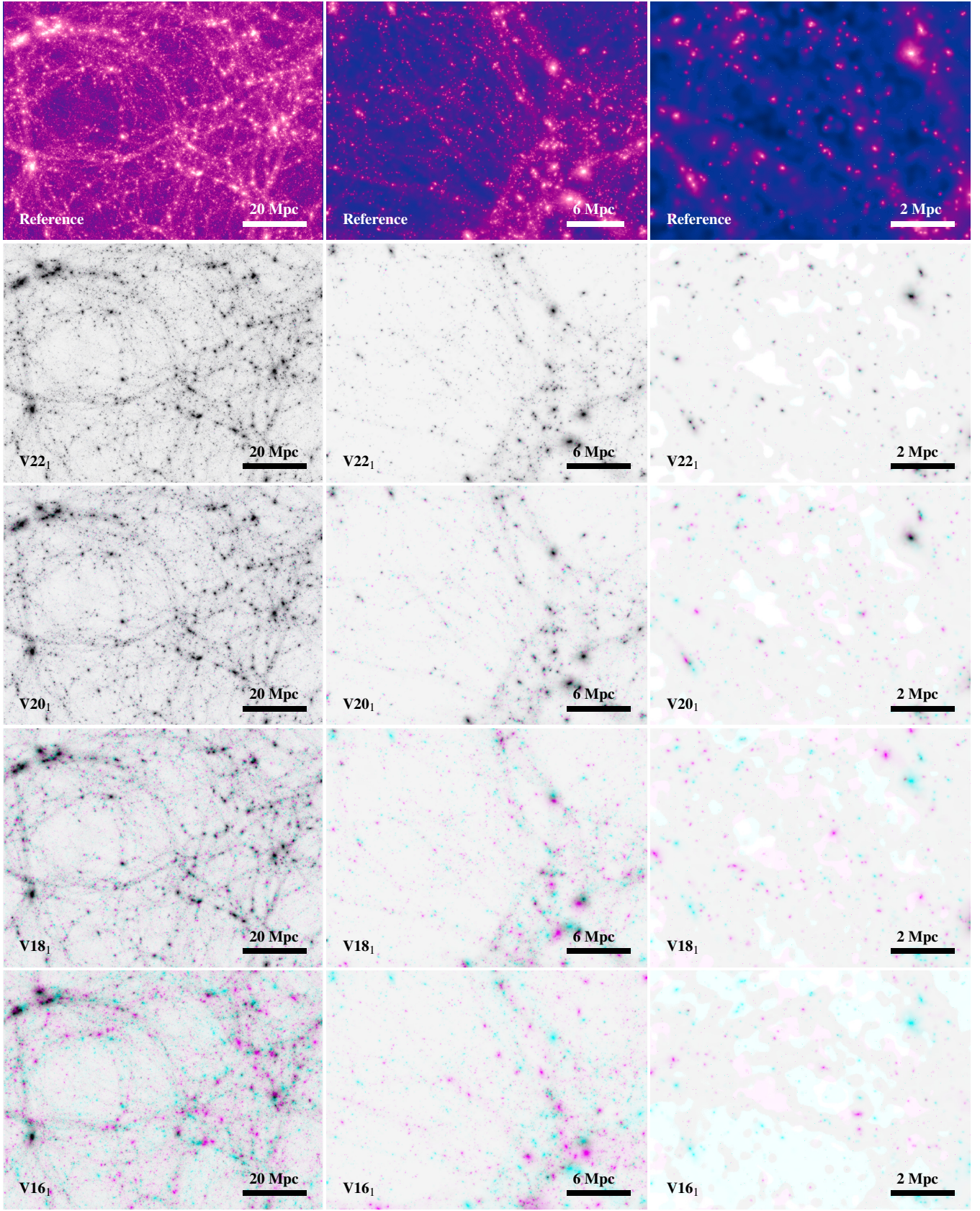

\section{Global Results} \label{sec:results}
In this section, we present the overall results of our simulations, and the effect of the varying the initial conditions on the $z=0$ density field, the matter power spectrum, and the abundance of dark matter haloes of different mass.

\subsection{Density Fields}\label{sec:results:density}
In Fig.~\ref{fig:structure-levels}, we visually compare the
structures at $z=0$ formed in the Reference simulation (top row) to
those formed in variants with independent WNFs at and above levels
22, 20, 18, and 16, respectively. From left to right, the columns show the projected dark matter density in boxes of side length 100, 33 and 10 Mpc, respectively. Shades of grey indicate similar projected density in the Reference and the variant simulations, shades of blue or purple indicate higher densities in the Reference or variant simulations, respectively.

In the right column, differences from the Reference simulation can already be perceived at level 22. While nearly all identifiable haloes can
be matched by eye, some low mass haloes appear slightly displaced, often by less than the size of the halo. At this level of variation, scales visible in the middle and left panels appear almost identical. For variants at level 20, most haloes in the right column are offset, but can still be matched across simulations by eye. Differences are also apparent in the middle column, where some lower mass haloes now show a noticeable displacement. At $\LL=18$, differences are noticeable in all three panels. In the right column, all haloes appear visibly displaced, and many low mass haloes can no longer be matched by eye, while in the left column, displacements are still mostly below the size of the identifiable haloes. At $\LL=16$ all differences are enhanced: while the right hand panel shows similar amounts of structure in both simulations, most individual objects can no longer be identified and appear at random. The middle panel still shows some correlation between the position of the more massive groups, as well as filaments, but only the largest haloes still appear in dark grey, indicating that they are displaced by less than their size.

We will examine the changes to individual, matched objects more rigorously in Section~\ref{sec:results:individual}. As an example, we will also discuss changes to a single, cluster-mass halo in more detail in Section~\ref{sec:results:example}. 

\begin{figure}
    \begin{tabular}{@{}ccc@{}}
    \begin{overpic}[width=\columnwidth]{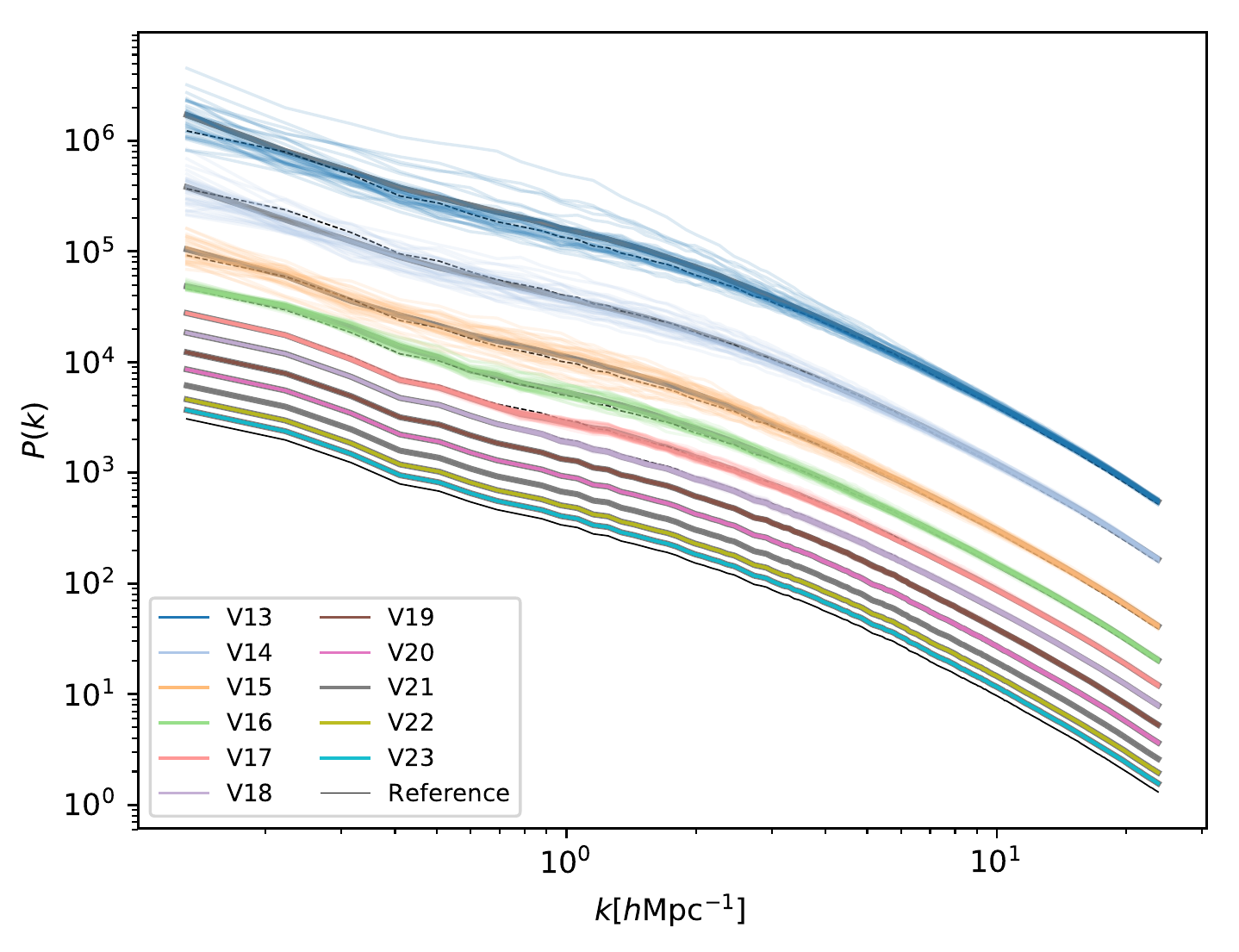}
    \end{overpic} 
    \end{tabular}
    \caption{Matter power spectra at $z=0$, for the Reference simulation (black, bottom), and 39 variants at each level from 13 to 23. For visual clarity, the sets of power spectra of the variant simulations are offset from the Reference simulation, which is repeated (in dashed) for each set. Solid grey lines show the average power spectrum over all variants at each level. Where the dashed lines are not visible, the power spectra of the variants follow that of the Reference simulation very closely. 
    }
    \label{fig:matterpower}
\end{figure}

\subsection{Power Spectrum}\label{sec:results:powerspectra}
In Fig.~\ref{fig:matterpower}, we show the matter power spectrum, $P(k)$, of our simulations measured at $z=0$. We define $$P(k) = \frac{1}{V}\langle | \delta_k | \rangle^2 $$ as the volume-averaged power spectrum, where  $\delta_k$ is given by the three-dimensional Fourier transform of the density perturbation field, $\delta({\bf x}) = \rho({\bf x}) / <\rho> - 1 $, over the simulation volume.

While all simulations are set up with an identical input power spectrum, each one only contains a finite volume, and hence each mode is sampled only a finite number of time. The black solid line in  Fig.~\ref{fig:matterpower} shows the result of the Reference simulation, and coloured lines show the results of the 39 variants at each level from $12$ to $23$, offset for visual clarity. The Reference result is also repeated as a dashed line with every set. In addition, thick grey lines show the average power spectra off all variants at a given level. It can be seen that, for small variations (e.g. for variations from level 23), all variants have a nearly identical power spectra on all scales, and follow all the peculiar features of the Reference simulation. As the scale of variations increases, differences between individual variants can be seen, but not at all scales: very small scales (large $k$) are sampled so well within each volume that differences between the variants are averaged out, while very large scales (small $k$) are not yet affected at moderate scales of variation. The large-scale limit of scatter in the matter power spectra grows with decreasing $\LL$. For larger variations, it can also be seen that the average power spectrum is much smoother than any individual power spectrum, effectively sampling a larger volume. The difference between levels 12 and 13, however, is minimal. It can be seen that individual lines for individual volumes nearly match one another. This is due to the fact that there is very little power at level 12, subject to the mean-density constraint of the simulation volume. By coincidence, the Reference simulation is quite close to this average on all scales.

\begin{figure}
    \includegraphics[width=\columnwidth]{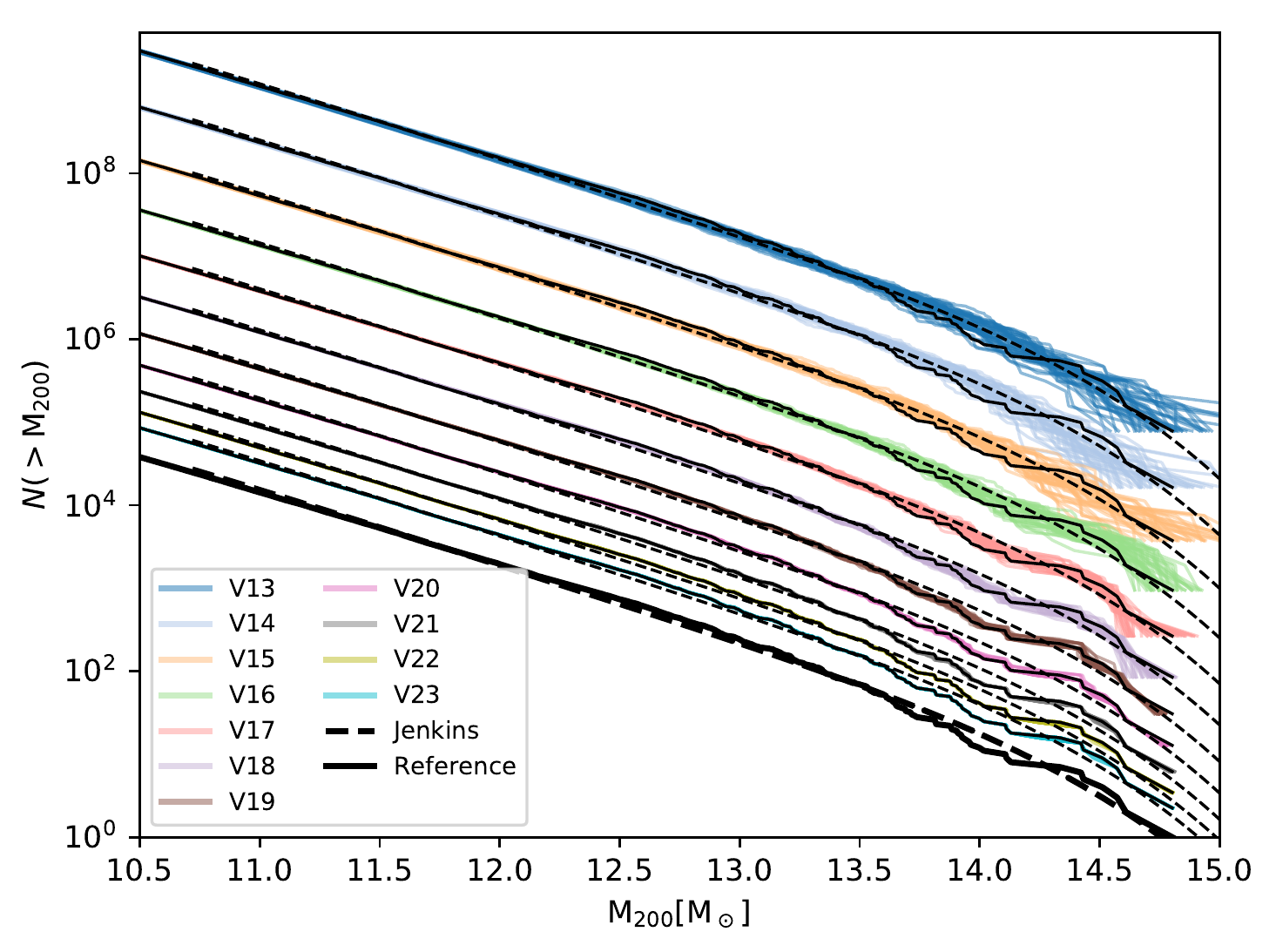}
    \caption{Halo mass function ($M_{200,\mathrm{crit}}$) at $z=0$, for the Reference simulation (thick black solid), and 39 variants at each level from 12 to 23. Also shown (thick black dashed) is the halo mass function of \protect\cite{Jenkins-2001}, calculated for the same cosmology. For visual clarity, the sets of halo mass functions of the variant simulations are offset, and the mass functions of the Reference simulation, and of \protect\cite{Jenkins-2001} are repeated with each set, as thin solid and dashed lines, respectively.  \label{fig:massfunction}}
\end{figure}

\subsection{Halo abundance} \label{sec:results:population}
In Fig.~\ref{fig:massfunction}, we compare the halo mass functions in terms of $M_{200,\mathrm{crit}}$ of our simulations to the analytic mass function of \cite{Jenkins-2001}, calculated for the same cosmology. Due to the small sample size, the Reference simulation (thick black solid line), measured in a single, $100^3$ Mpc$^3$ volume, differs slightly from the mass function of \cite{Jenkins-2001} (thick black dashed line) at high masses. Similarly to Fig.~\ref{fig:matterpower}, coloured lines show the results of the 39 variants at each level from $\LL=14$ to $\LL=23$. For visual clarity, each set of lines is offset, and the mass functions of \cite{Jenkins-2001} and the Reference simulations are repeated as thin dashed and solid black lines with each set.

\begin{table}
	\centering
	\caption{Median and standard deviation in the number haloes of a given mass across the simulations of a given level.}
	\label{tab:halo-numbers}
	\begin{tabular}{lcccc} 
		\hline
		Level & $N(10^{11})$ & $N(10^{12})$ & $N(10^{13})$ & $N(10^{14})$ \\
		\hline
		\hline
		ref & 21689  & 2602 &  343 & 21 \\
		\hline
		23 & $21723 \pm 48$ & $2613 \pm 9$ & $343 \pm 1$ & $21 \pm 0$ \\
		22 & $21683 \pm 51$ & $2597 \pm 6$ & $345 \pm 2$ & $21 \pm 0$ \\
		21 & $21676 \pm 78$ & $2613 \pm 16$ & $341 \pm 3$ & $21 \pm 0$ \\
		20 & $21704 \pm 81$ & $2629 \pm 18$ & $344 \pm 6$ & $21 \pm 1$ \\
		19 & $21790 \pm 63$ & $2698 \pm 28$ & $347 \pm 8$ & $21 \pm 1$ \\
		18 & $21833 \pm 118$ & $2662 \pm 46$ & $356 \pm 12$ & $23 \pm 2$ \\
		17 & $21825 \pm 257$ & $2703 \pm 28$ & $328 \pm 8$ & $28 \pm 2$ \\
		16 & $21711 \pm 214$ & $2674 \pm 66$ & $327 \pm 22$ & $27 \pm 3$ \\
		15 & $21403 \pm 266$ & $2655 \pm 80$ & $318 \pm 13$ & $32 \pm 5$ \\
		14 & $21705 \pm 292$ & $2732 \pm 100$ & $322 \pm 16$ & $29 \pm 5$ \\
		13 & $21554 \pm 184$ & $2677 \pm 87$ & $318 \pm 25$ & $29 \pm 5$ \\
		12 & $21523 \pm 193$ & $2682 \pm 89$ & $318 \pm 25$ & $30 \pm 6$ \\
		\hline
	\end{tabular}
\end{table}

\begin{figure}
    \includegraphics[width=\columnwidth]{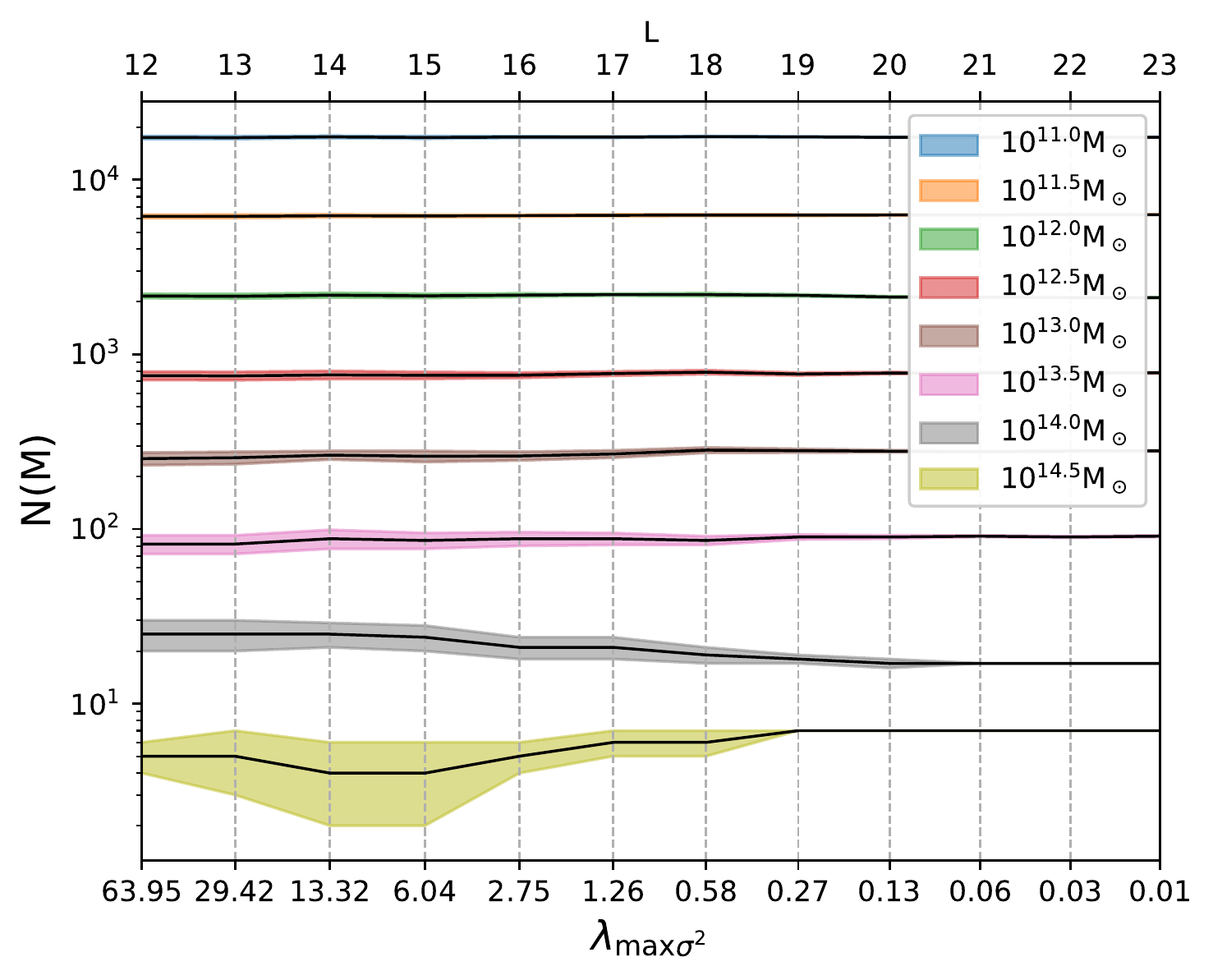}
    \caption{Median and standard deviation in the number of haloes
      for different masses, as a function of the level at which
      variations are introduced in the initial conditions at $z=0$, $\LL$, and the corresponding wavelength, $\lambda_{\mmax\sigma^2}$. The Reference simulation has the expected number of haloes up to $\sim 13.5
      \Ms$, but lies outside one $\sigma$ for both $\sim 14
      \Ms$ and $\sim 14.5 \Ms$ haloes.}
    \label{fig:number-scatter}
\end{figure}

It can be seen that, for small variations, the halo mass functions very closely follow that of the Reference simulation. The scatter continuously increases as the scale of the variations increases. Only by level 15 ($\Lambda = 25.6$ Mpc) does the scatter among the mass functions for individual volumes become large enough to erase the particular features inherited from the Reference simulation. Similar to the variation of the power spectrum, beyond level 13, there is very little additional variation.

\begin{figure*}
    \centering

\resizebox{\textwidth}{!}{

\begin{tikzpicture}[scale=1.,every node/.style={minimum size=1cm},on grid]

    \begin{scope}[yshift=0,every node/.append style={yslant=.5,xslant=0},yslant=.5,xslant=0]
            
        \fill[white,fill opacity=0.9] (0,0) rectangle (5,5);
        \draw[black,very thick] (0,0) rectangle (5,5);
        \node[
          opacity=1., 
          anchor=south west, 
          inner sep=0pt 
          ] 
          at (0,0) 
          {\scalebox{-1}[1]{\includegraphics[width=5cm]{gv_figures/L0100N0376_L24_1_z0_zoom3x.pdf}}};

    \end{scope}

    \begin{scope}[yshift=0,xshift=80,every node/.append style={yslant=.5,xslant=0},yslant=.5,xslant=0]
    	\fill[white,fill opacity=.9] (0,0) rectangle (5,5);
        \draw[black,very thick] (0,0) rectangle (5,5);
        \draw[step=0.1, gray, opacity=.5, very thin] (0,0) grid (5,5);
        
        \fill[red, opacity = .3] (1.1,2.1) circle (.55cm);
        \fill[blue, opacity = .3] (0.8,2.) circle (.48cm);
        \fill[green, opacity = .3] (1.,1.8) circle (.62cm);
        \fill[white, opacity = .5] (1,2) circle (.25cm);
        \fill[black, opacity = .5] (1,2) circle (.25cm);
        \draw[black, very thick] (1,2) circle (.25cm);
        
    \end{scope}
    
    \begin{scope}[yshift=0,xshift=160,every node/.append style={yslant=0.5,xslant=0},yslant=0.5,xslant=0]
            
        \fill[white,fill opacity=0.9] (0,0) rectangle (5,5);
        \draw[black,very thick] (0,0) rectangle (5,5);
        \node[
          opacity=.5, 
          anchor=south west, 
          inner sep=0pt 
          ] 
          at (0,0) 
          {\scalebox{-1}[1]{\includegraphics[width=5cm]{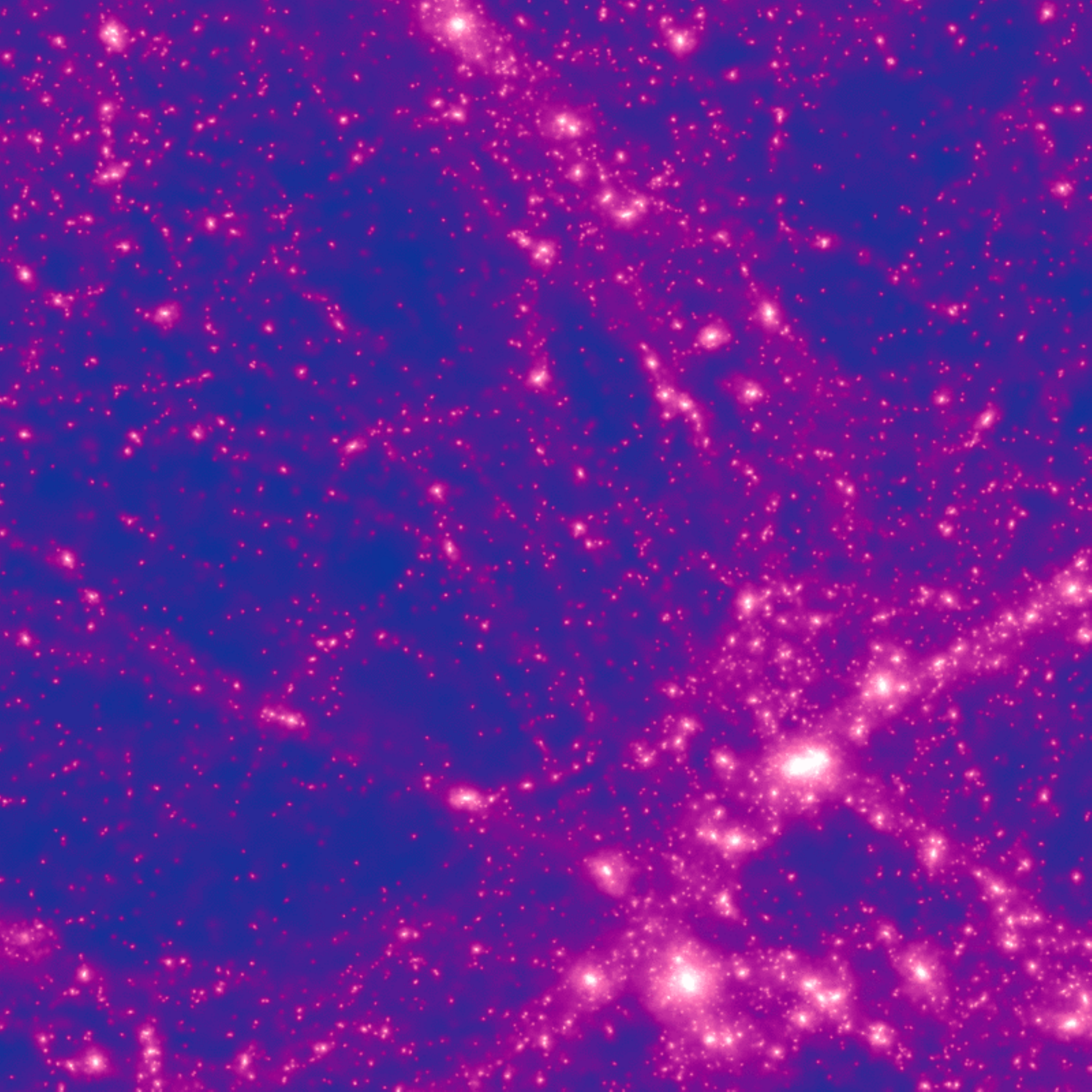}}};
        \fill[red,fill opacity=0.3] (0,0) rectangle (5,5);
    \end{scope}
    
    \begin{scope}[xshift=220,every node/.append style={yslant=.5,xslant=0},yslant=.5,xslant=0]
            
        \fill[white,fill opacity=0.9] (0,0) rectangle (5,5);
        \draw[black,very thick] (0,0) rectangle (5,5);
        \node[
          opacity=.55, 
          anchor=south west, 
          inner sep=0pt 
          ] 
          at (0,0) 
          {\scalebox{-1}[1]{\includegraphics[width=5cm]{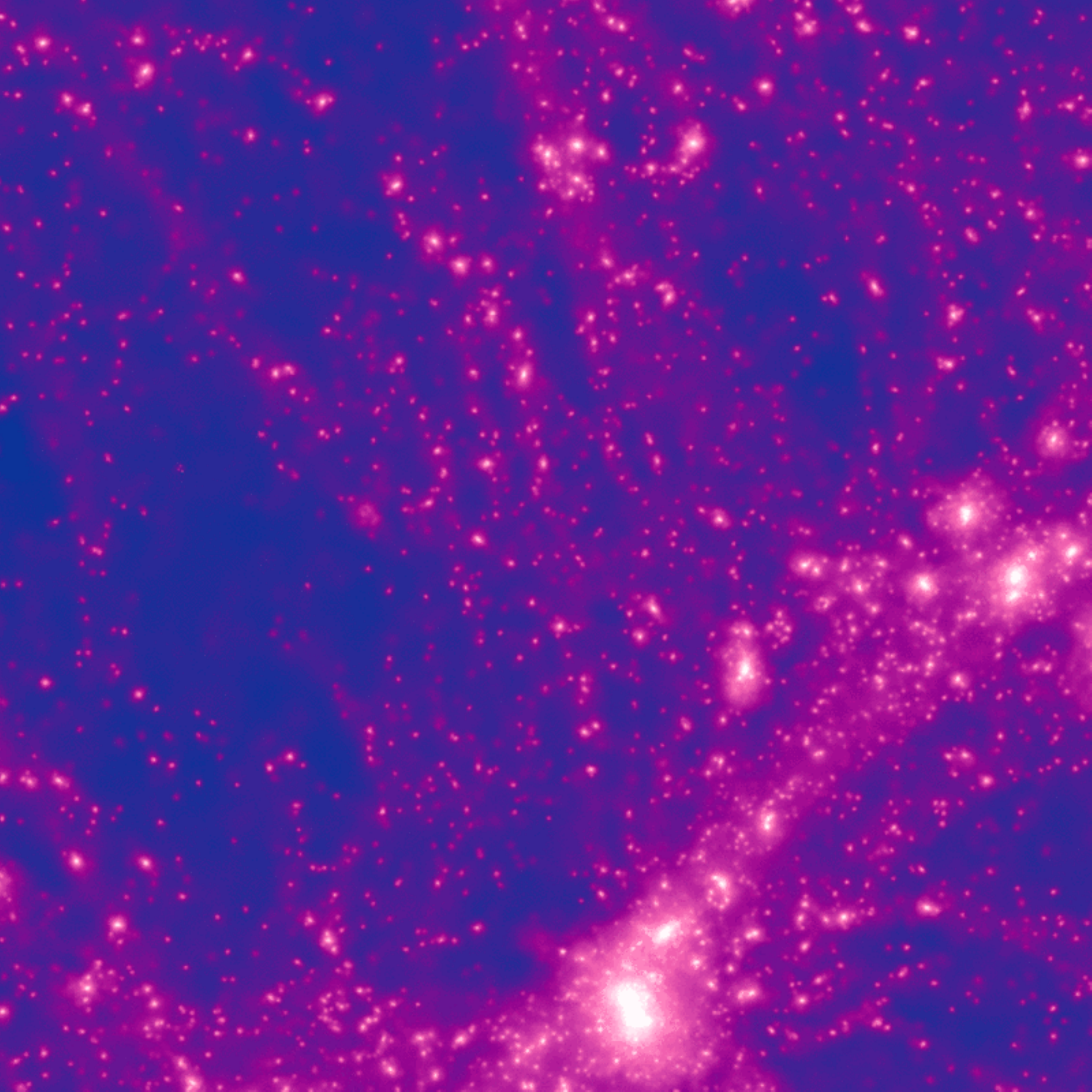}}};
        \fill[yellow,fill opacity=0.3] (0,0) rectangle (5,5);

    \end{scope}

    \begin{scope}[xshift=280,every node/.append style={yslant=0.5,xslant=0},yslant=0.5,xslant=0]
        \fill[white,fill opacity=0.9] (0,0) rectangle (5,5);
        \draw[black,very thick] (0,0) rectangle (5,5);
        \node[
          opacity=.5, 
          anchor=south west, 
          inner sep=0pt 
          ] 
          at (0,0) 
          {\scalebox{-1}[1]{\includegraphics[width=5cm]{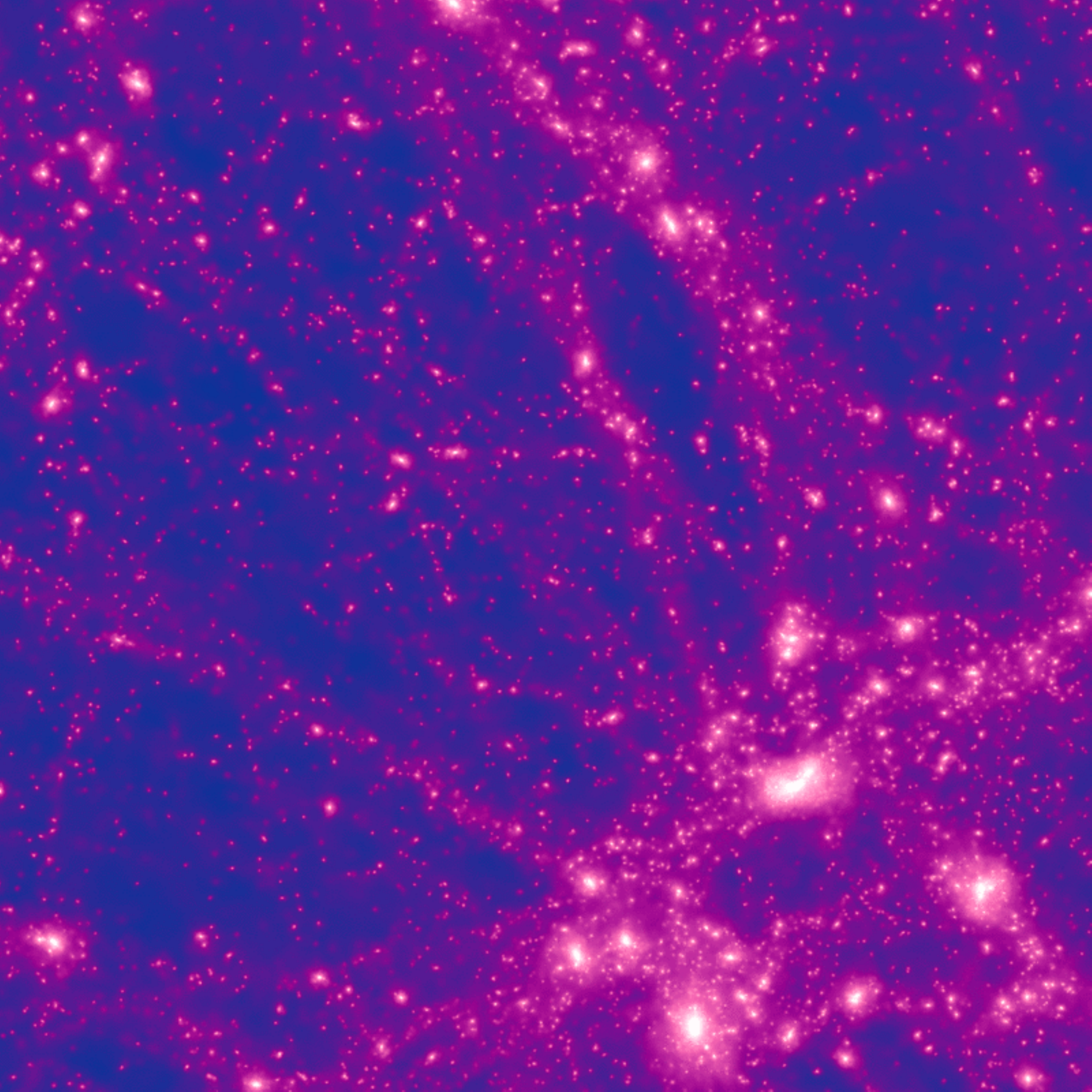}}};
        \fill[blue,fill opacity=0.3] (0,0) rectangle (5,5);

    \end{scope}

    \begin{scope}[xshift=340,every node/.append style={yslant=0.5,xslant=0},yslant=0.5,xslant=0]
        \fill[white,fill opacity=0.9] (0,0) rectangle (5,5);
        \draw[black,very thick] (0,0) rectangle (5,5);
        \node[
          opacity=.5, 
          anchor=south west, 
          inner sep=0pt 
          ] 
          at (0,0) 
          {\scalebox{-1}[1]{\includegraphics[width=5cm]{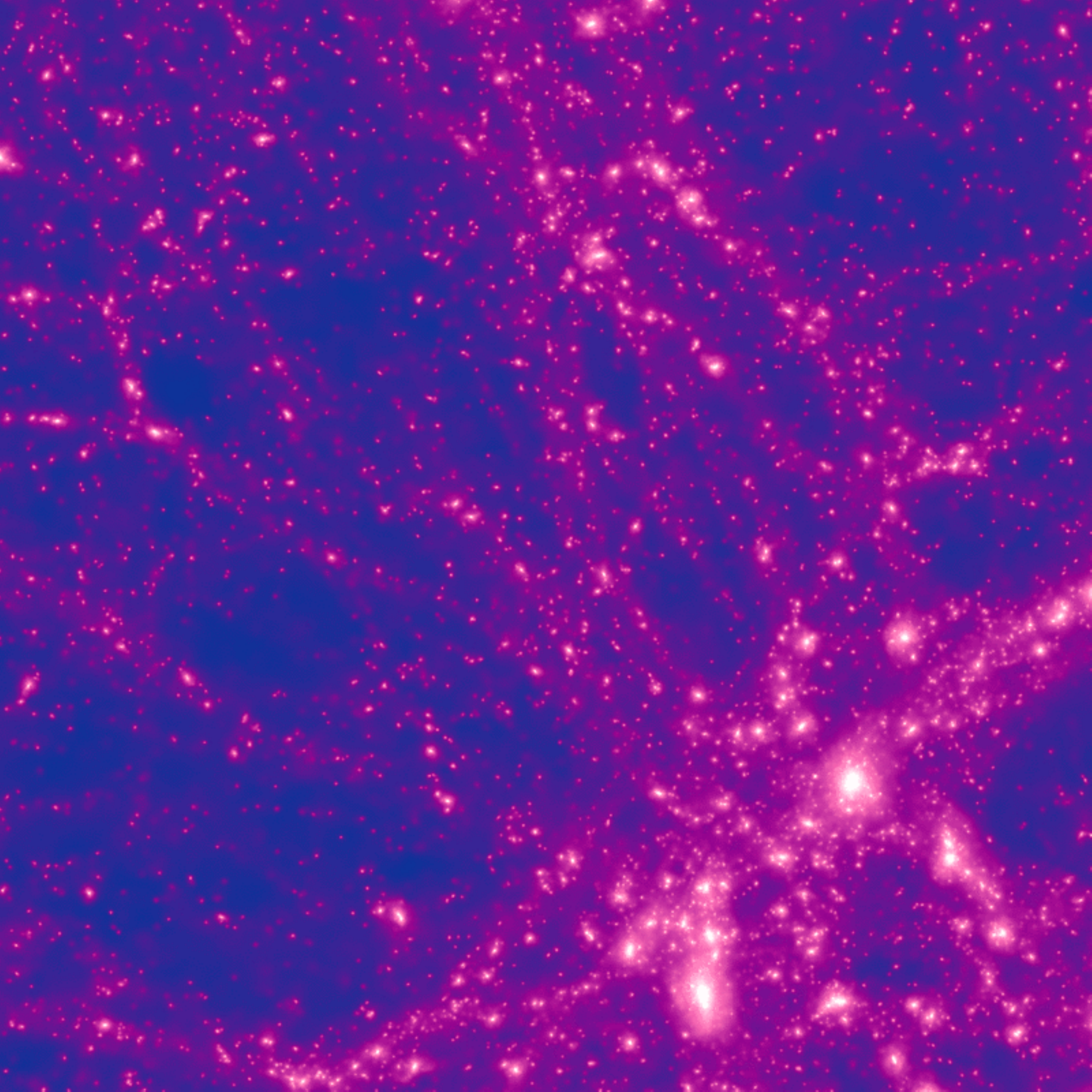}}};
        \fill[green,fill opacity=0.3] (0,0) rectangle (5,5);

    \end{scope}

    \begin{scope}[xshift=400,every node/.append style={yslant=0.5,xslant=0},yslant=0.5,xslant=0]
        \fill[white,fill opacity=0.9] (0,0) rectangle (5,5);
        \draw[black,very thick] (0,0) rectangle (5,5);
        \node[
          opacity=.5, 
          anchor=south west, 
          inner sep=0pt 
          ] 
          at (0,0) 
          {\scalebox{-1}[1]{\includegraphics[width=5cm]{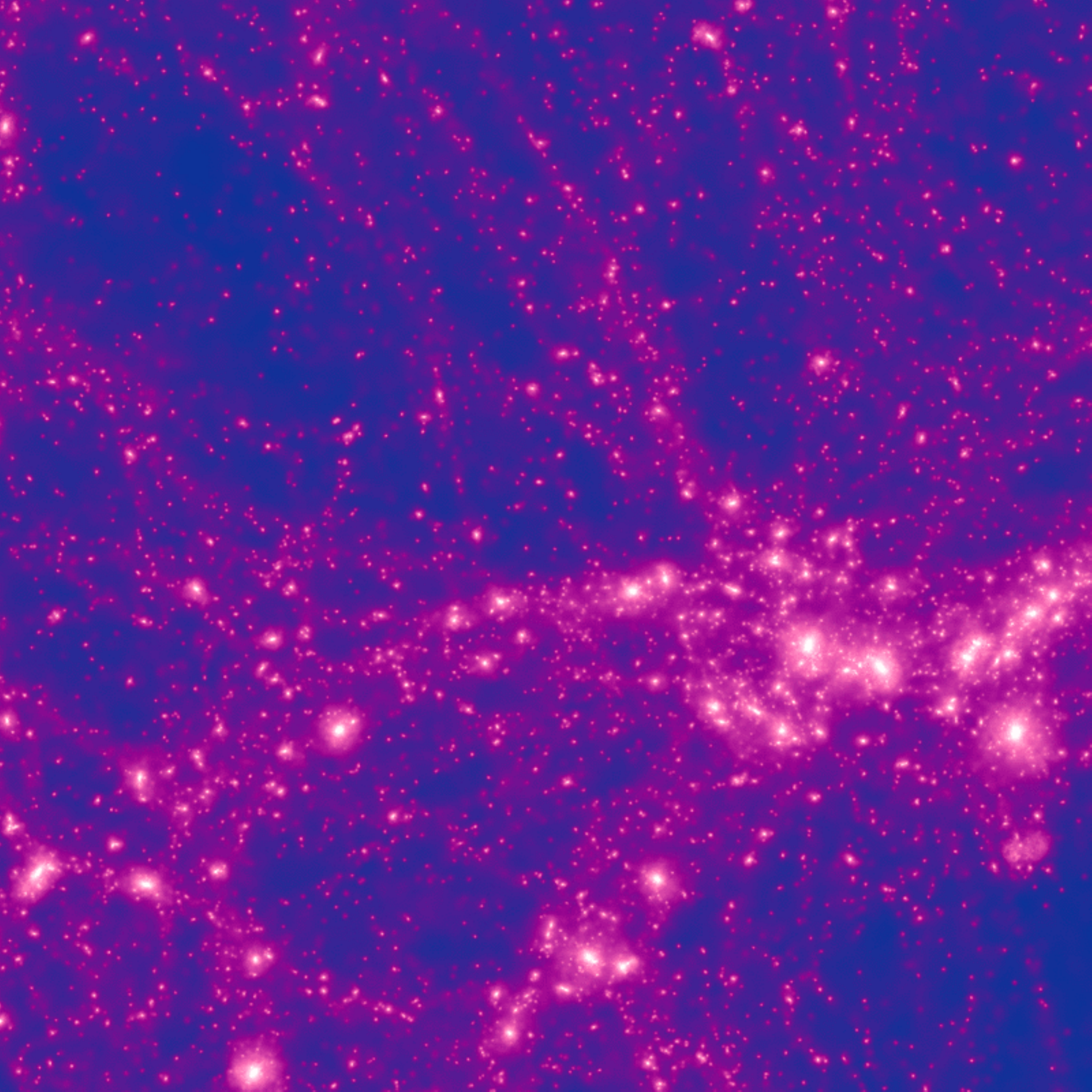}}};
         \fill[brown,fill opacity=0.3] (0,0) rectangle (5,5);

    \end{scope}


\draw[-latex, very thick] (1.3,1.9) to [out=0,in=180] (3.8,2.5);
\draw[-latex, very thick, black] (3.8,2.5) to [out=0,in=160] (6.8,2.25);
\draw[-latex, very thick, black] (3.5,2.4) to [out=0,in=160] (11.05,2.25);
\draw[-latex, very thick, black] (3.7,2.3) to [out=0,in=160] (13.,2.1);

\draw[-latex,thick] (3,8) node[right]{Reference $(z)$} to[out=180,in=90] (2,7);
\draw[-latex,thick] (7,8) node[right]{Lagrangian coordinates $(z=\infty)$} to[out=180,in=90] (6,7);
\draw[-latex,thick] (14,8) node[right]{Variants 1-5 $(z)$} to[out=180,in=90] (13,7);
\end{tikzpicture}
}  
 
\caption{Matching of haloes using particles with identical coordinates
  at $z=\infty$. The origin of the most bound particles in a given
  halo of the Reference simulation at redshift $z$ ("Reference
  particles") define a Lagrangian volume (denoted by the grey circle)
  to which all particles in haloes in the Variant simulations are
  compared. In this example, variants 1 (red), 3 (blue) and 4 (green)
  each contain a halo that includes the majority of Reference
  particles, and whose mass at redshift $z$ is within a factor of 3 of
  the halo in the Reference simulation. By contrast, variants 2
  and 5 contain no matching haloes.}
    \label{fig:match-lagrangian}
\end{figure*}

It can also be seen that the scatter among the mass functions for each set is greater at higher mass. This may appear counter-intuitive, given that (as we discuss in Section~\ref{sec:results:properties:mass}), individual, higher mass objects are {\it less} strongly affected by changes to the primordial density field at a given scale. However, the scatter in the mass function has different origins at different masses: at the high mass end, where the number of haloes is low, the scatter is due to a change in the mass of individual objects, while at the low mass end, it is due to the change in the {\it number} of independent objects, but on scale with very small sampling noise.

Table~\ref{tab:halo-numbers} gives an overview of the median number of haloes of different masses, and the associated standard deviation, across the 39 simulations for each level from $12$ to $23$. It is worth noting that, for sufficiently large haloes, or sufficiently small scale variations, the standard deviation in halo number, $\sigma_N$, is below even the value of $\sqrt{N}$ expected for a random process without any variation in the {\it bias}. For example, the average number of haloes of $10^{12}\Ms$ is $\sim 2700 \sim 52^2$, but the scatter at $\LL=20$ is only 18. This indicates that the scatter is due primarily to a change in the mass of the same haloes found across different simulations. For larger variations, the scatter typically rises above $\sqrt{N}$, which can be attributed to different bias in each volume \citep{White-1987, Cole-1989}.

Fig.~\ref{fig:number-scatter} presents the same information visually. It can be seen that, while variations in the initial density field at $\LL=19$ leads to the formation of independent $10^{11}\Ms$ haloes, their population is so well sampled in the $100$Mpc volume that their number has less than a $1\%$ scatter at any level. At the other end of the mass range most individual $10^{14}\Ms$ haloes exist across all simulations at $\LL=17$, albeit with a small change in mass. The simulation volume, however, is not large enough to sample them accurately in every volume. Averaging over all variants at levels 12 or 13, we can see that the Reference simulation contains slightly fewer than the expected $30\pm5$ haloes of mass $10^{14}\Ms$.

\section{Individual haloes} \label{sec:results:individual}
In the previous section, we have compared populations of
haloes among different simulations. We now turn our
attention to individual haloes. As we already discussed, if the
variations in the initial conditions between two simulations are small
enough, the {\it same} haloes form in both. Here, we investigate which scales in the initial density field determine the existence of unique haloes, and how the properties of individual haloes change subject to variations on smaller scales.

\begin{figure*}
\begin{tabular}{@{}ccc@{}}
    \begin{overpic}[width=0.33\textwidth]{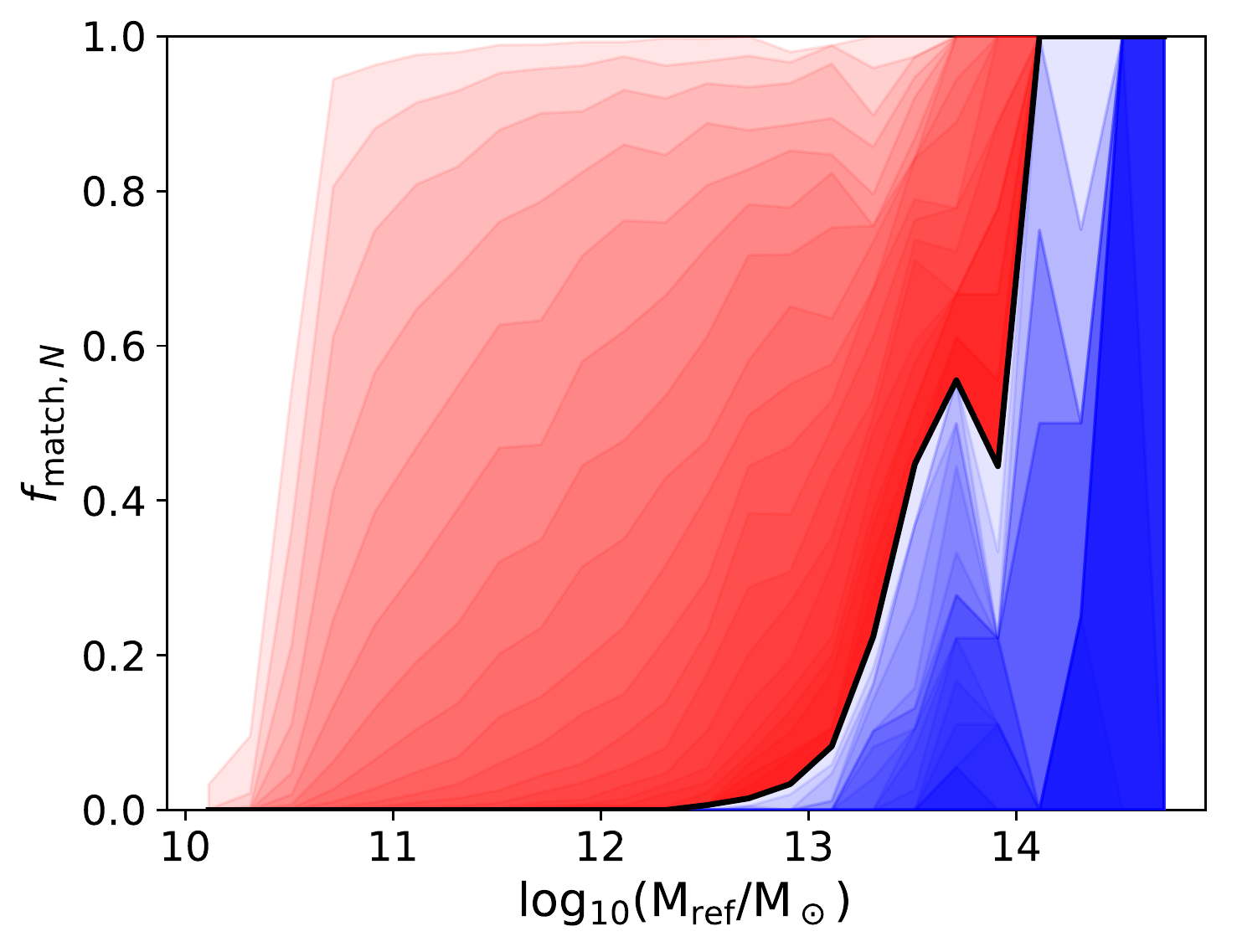}
    \put(85,12){\makebox[0pt] {\textcolor{white}{$\LL=16$}}} 
    \end{overpic} &
    \begin{overpic}[width=0.33\textwidth]{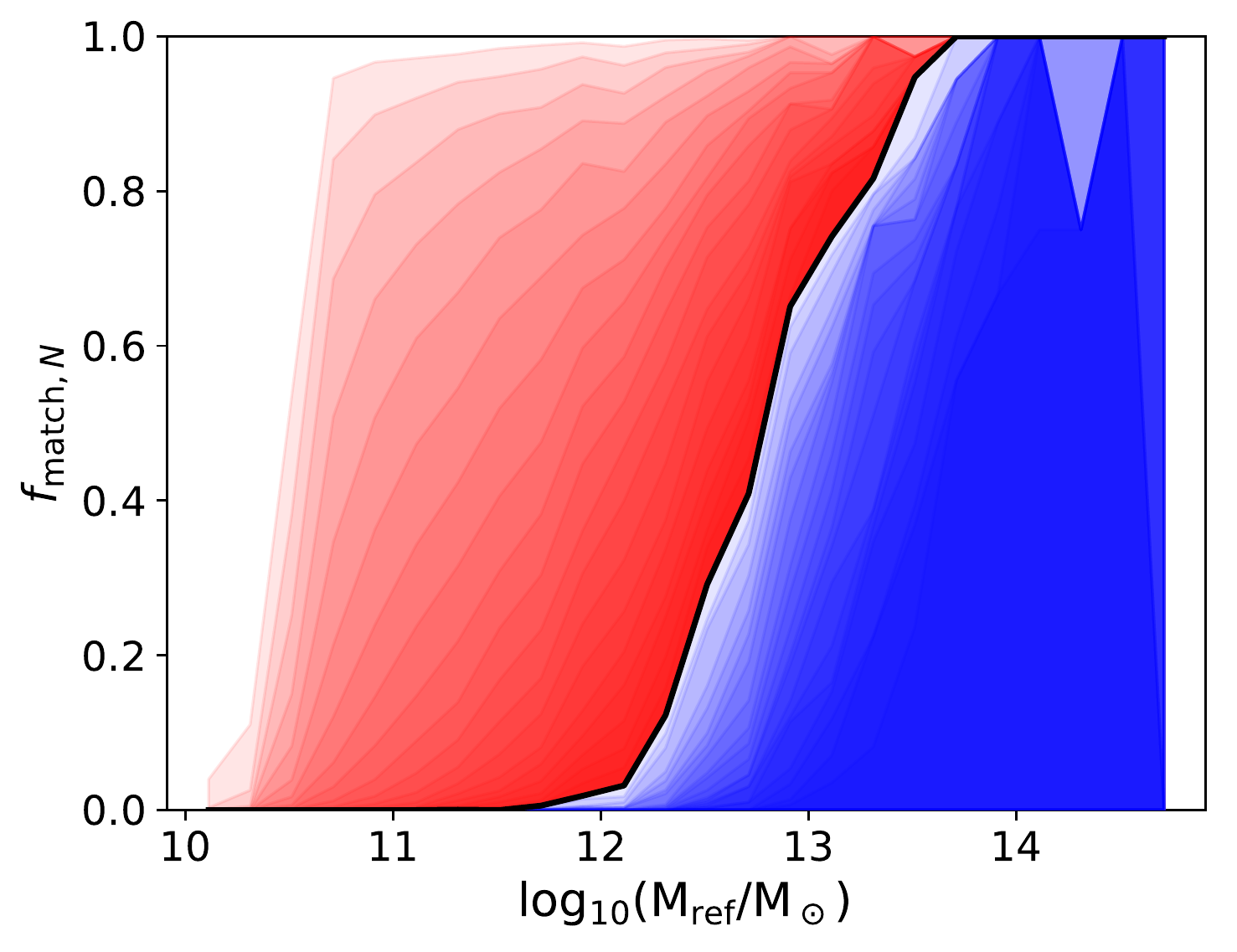}
    \put(85,12){\makebox[0pt] {\textcolor{white}{$\LL=17$}}} 
    \end{overpic} &
    \begin{overpic}[width=0.33\textwidth]{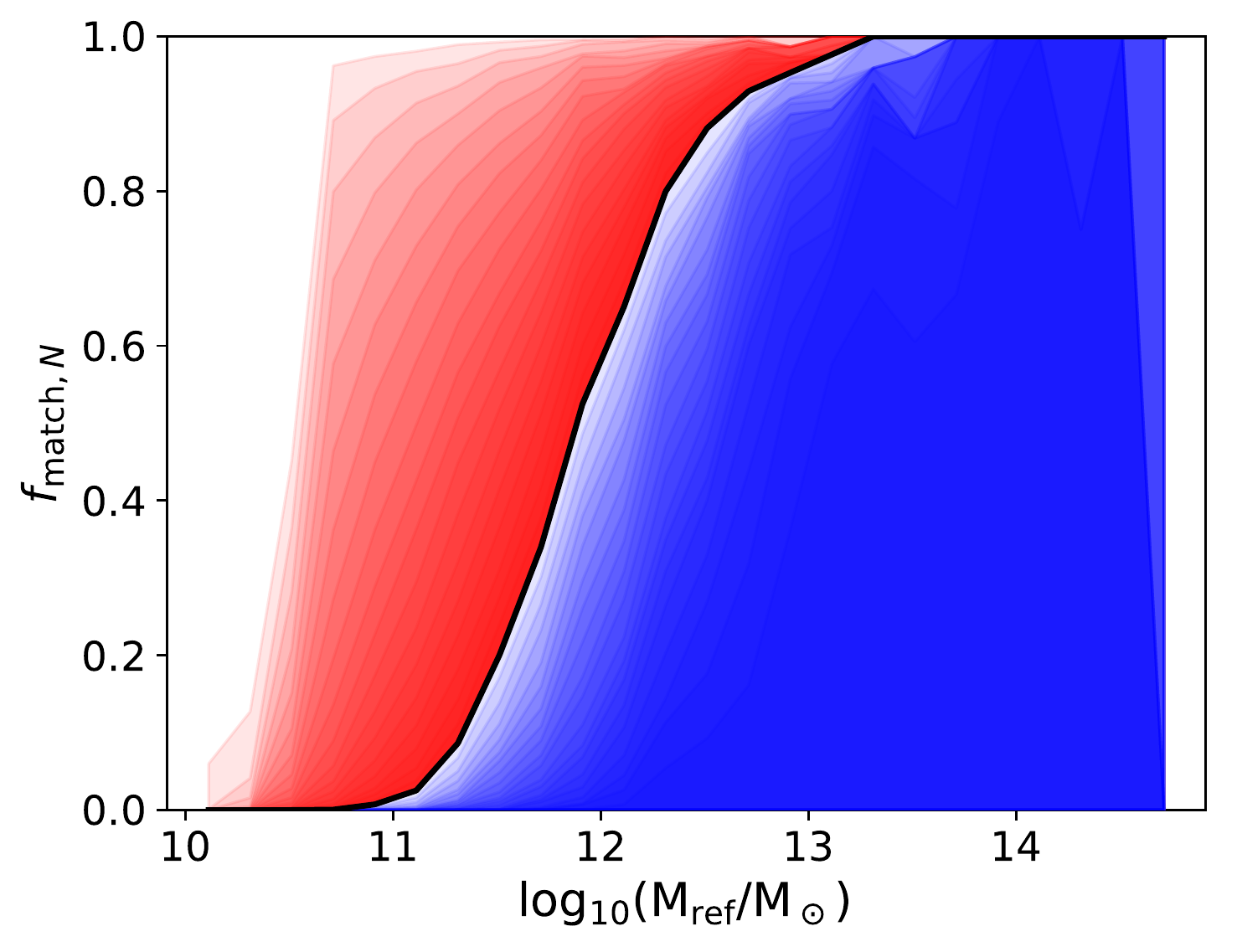}
    \put(85,12){\makebox[0pt] {\textcolor{white}{$\LL=18$}}} 
    \end{overpic} \\
    \begin{overpic}[width=0.33\textwidth]{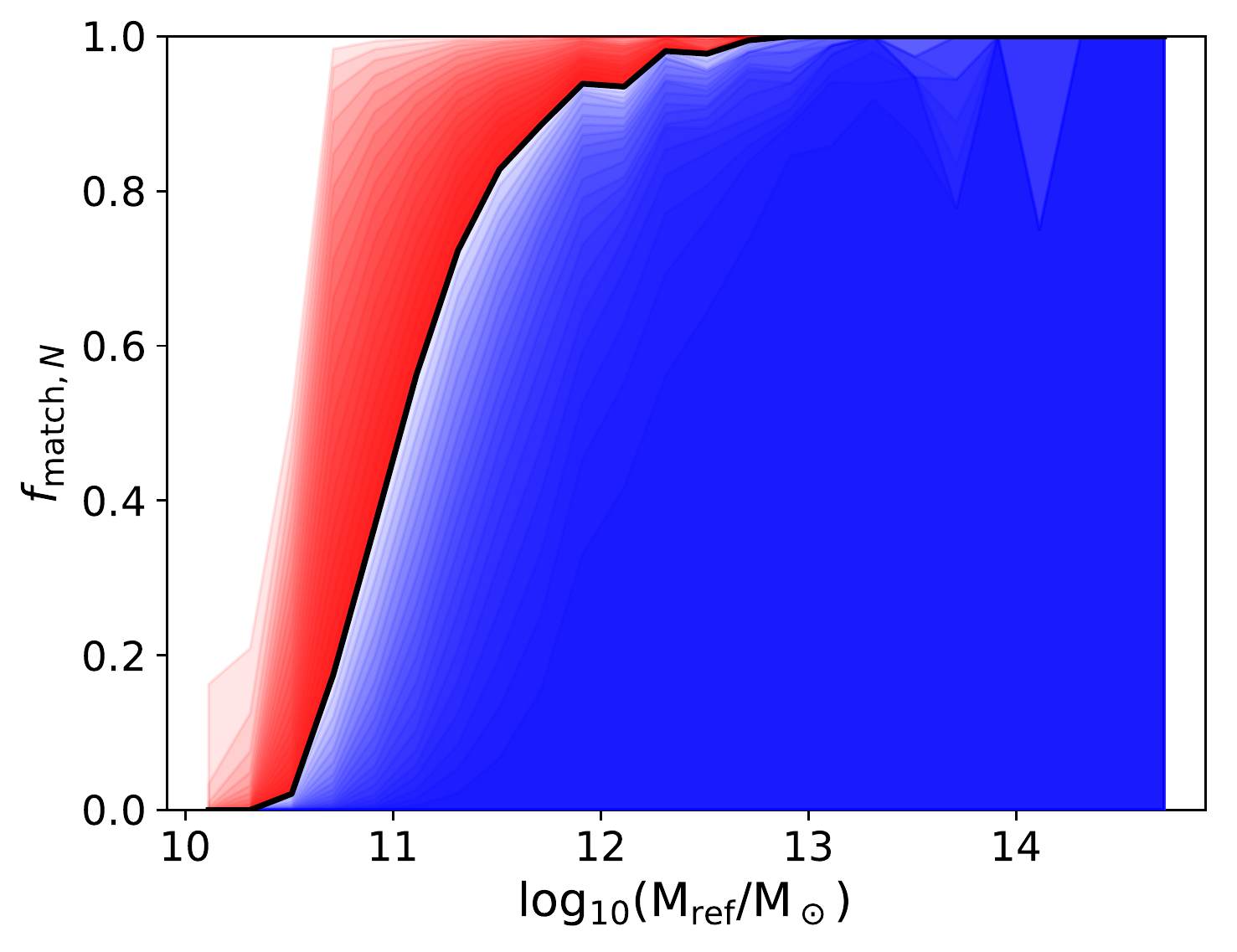}
    \put(85,12){\makebox[0pt] {\textcolor{white}{$\LL=19$}}} 
    \end{overpic} &
    \begin{overpic}[width=0.33\textwidth]{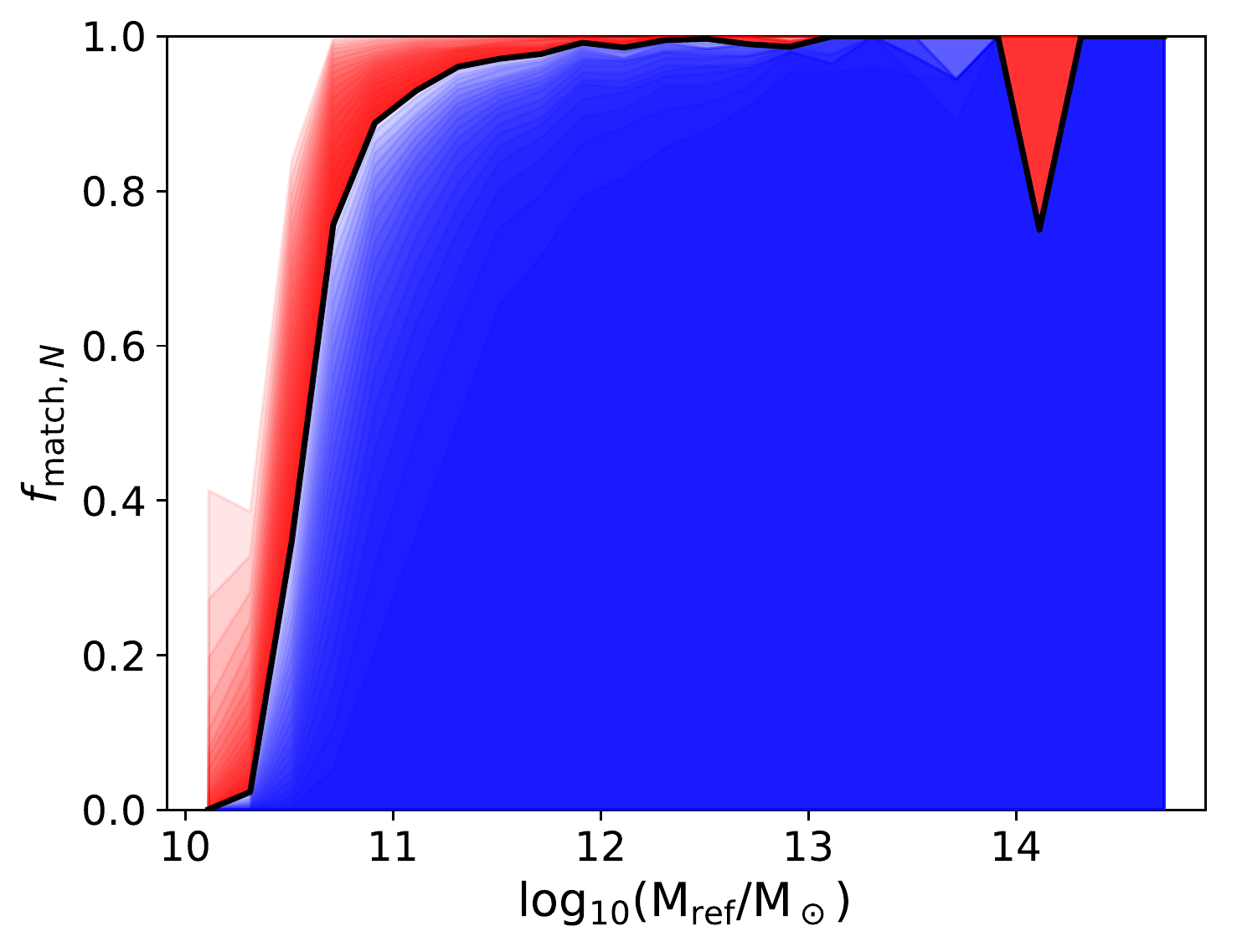}
    \put(85,12){\makebox[0pt] {\textcolor{white}{$\LL=20$}}} 
    \end{overpic} &
    \begin{overpic}[width=0.33\textwidth]{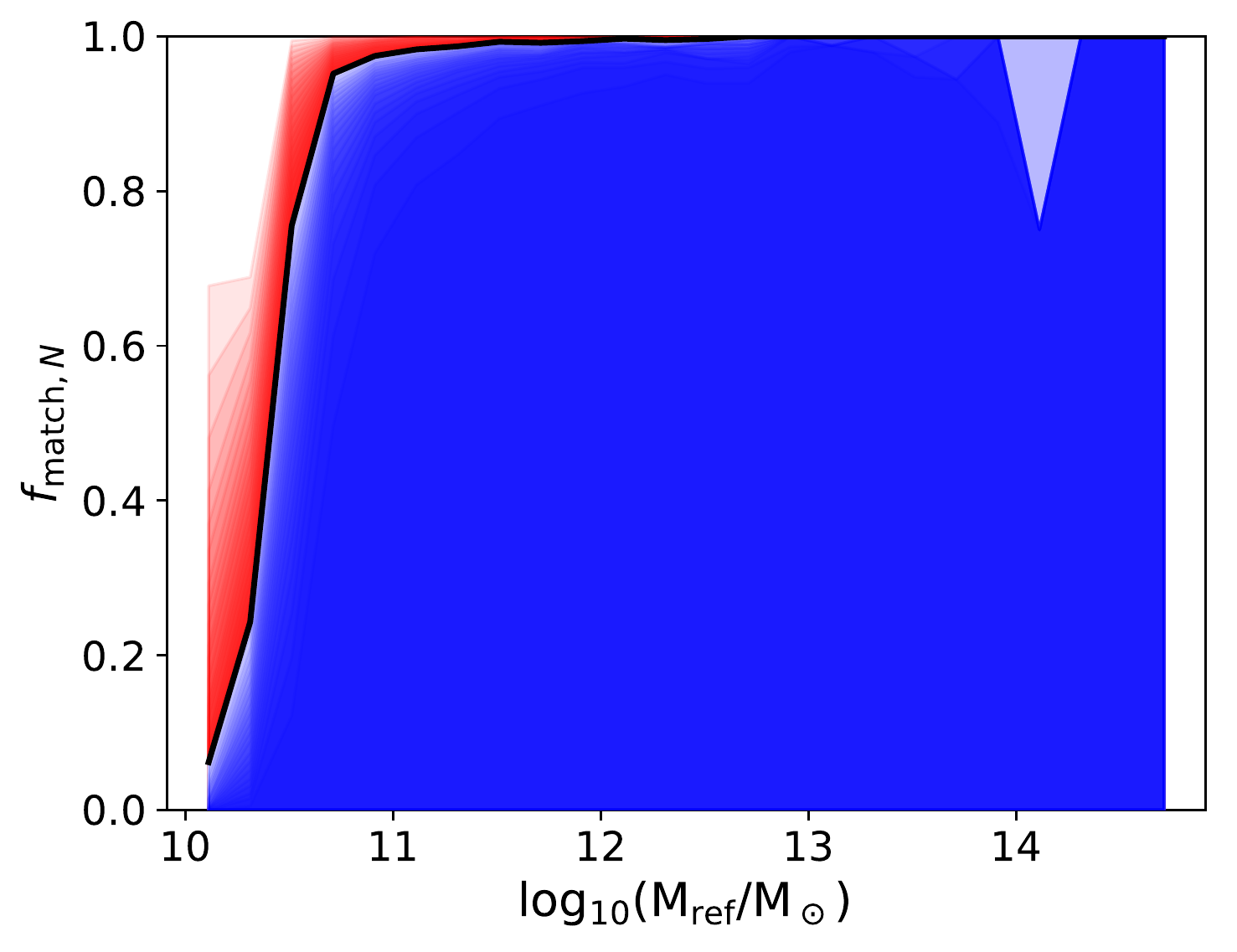}
    \put(85,12){\makebox[0pt] {\textcolor{white}{$\LL=21$}}} 
    \end{overpic} \\
  \end{tabular}
    \caption{Fraction of haloes with multiple matches, $f_{\mathrm{match},N}$, between the Reference simulation and the 39 variant simulations at each level from 16 to 21, as a function of halo mass in the Reference simulation. Shades of red and blue indicate 1 to 19, and 21 to 39 matches, respectively, while the black line indicates 20 matches. As expected, the matching rate is highest for high masses and high level (small scale) variations, and decreases towards lower masses and higher level variations. It is worth noting that even at scales that are completely uncorrelated, many haloes can be matched by chance. However, the number of {\it multiple} matches decreases sharply.}
    \label{fig:matching-rate}
\end{figure*}

In order to address these questions, we need to match haloes across simulations. Following \cite{Springel-2008}, we use the fact that all of our simulations start from identical glass files, with particles whose IDs encode their initial, unperturbed Lagrangian coordinates along a Peano-Hilbert curve. Fig.~\ref{fig:match-lagrangian} illustrates the matching procedure for a halo identified in a snapshot of the Reference simulation, represented by the leftmost slice. In a first step, we identify the halo's 50 most bound particles (or all particles, if fewer than 50), and use their IDs to determine their Lagrangian origin. This is represented by the second slice from the left, where particles identified in the previous step occupy a finite volume, indicated by the grey circle.

In the next step, we examine all haloes in the corresponding snapshot in one of the variant simulations, each represented by the five coloured slices to the right. We select haloes whose masses are within a factor of three of that in the Reference simulation. If a halo contains a large fraction of the 50 particles identified in the previous step, by definition, its Lagrangian region of origin overlaps with that of the halo in the Reference simulation. In the example shown here, the first (red) slice, third (blue) and fourth (green) slice each contain a halo that fulfils the mass criterion and contains at least $1/5$ of the particles of the halo in the Reference simulation. These haloes, which have grown to a similar mass from similar regions of origin, are considered matches. Conversely, the second (yellow) and fifth (purple) slice do not contain matching haloes in this example.

We note that this procedure is not completely symmetrical under exchange of the Reference and the variant simulations. However, we have tested that our results do not vary qualitatively when, at a given level, one of the variants is chosen as the Reference instead.

As expected, the matching rate is highest for high masses and high level (small-scale) changes in the initial density field, and decreases when the scale of changes increases relative to the size of the haloes. However, we find that even for low-mass haloes in simulations that share almost no phase information, the matching rate only falls to $\sim 15\%$. Two haloes matched under these conditions have, by coincidence, grown to similar mass from overlapping Lagrangian volumes, without their simulations sharing any relevant information. Although matched, these are not physically the same halo.

\subsection{Halo Identity Across Simulations}\label{sec:results:individual:identity}
The possibility that two similar haloes can exist in two volumes that share no phase information leaves the tantalising question: when are two haloes genuinely identical? It appears that simply asking that they consist of the same particles is not sufficient; instead, we are looking for haloes that are formed for the same physical reasons.

While we could modify our matching criteria, we cannot discriminate ab initio between a genuine match (one where the halo pair has formed because of the common phase information) and a merely coincidental one. However, if spurious matches occur purely by chance, and the rate of those matches is less than $1/2$, the probability for at least $N/2$ spurious matches to the same halo in $N$ variant simulations decreases with $N$. Conversely, if genuine matches are found with a probability above $1/2$, the probability for $N/2$ genuine matches to $N$ variants simulations increases with $N$. Consequently, for sufficiently large $N$, genuine matches have a high probability to be identified in more than half of the variants, and haloes that are matched to more than half of the variants have a high likelihood of being genuine matches.

In Fig.~\ref{fig:matching-rate}, on each panel, we show the multiple-matching rate, f$_N$, as a function of mass, for 1 to 39 possible matches between the Reference simulation and all $N=39$ variants at each level. Shades of red denote the fraction of haloes with 1 to 19 matches; shades of blue correspond to 20 to 39 matches. The thick black line shows the fraction of haloes with 20 matches. It can be seen that the fraction of haloes above $10^{10.5}\Ms$ matched at least once is over $95\%$, almost independently of mass and level. However, as expected for purely chance events, at lower masses and larger scale variations, the number of multiple matches rapidly decreases, and as expected, the fraction of haloes matched {\it at least half of the time} tends to zero for low mass haloes and low-level variations, and to unity for high masses and high level variations.

\subsection{Existence of Unique Haloes}\label{sec:results:individual:existence}
\begin{figure}
\begin{tikzpicture}[
    auto,
    node distance=1.5cm,
    semithick, bend angle=25,
    redbox/.style = {fill=red!20, rounded corners},
    yellowbox/.style = {fill=yellow!20, rounded corners},
    bluebox/.style = {fill=blue!20, rounded corners},
    greenbox/.style = {fill=green!20, rounded corners},
    brownbox/.style = {fill=brown!20, rounded corners},
    ]

    \node (BBox) [redbox, minimum width=1cm, minimum height=4.5cm] at (1.5cm, -1.5cm) {};
    \node (BBox) [yellowbox, minimum width=1cm, minimum height=4.5cm] at (3cm, -1.5cm) {};
    \node (BBox) [bluebox, minimum width=1cm, minimum height=4.5cm] at (4.5cm, -1.5cm) {};
    \node (BBox) [greenbox, minimum width=1cm, minimum height=4.5cm] at (6cm, -1.5cm) {};
    \node (BBox) [brownbox, minimum width=1cm, minimum height=4.5cm] at (7.5cm, -1.5cm) {};

    \node[state, accepting]            (ref_1)                  {$A$};
    \node[state] (ref_2) [below of=ref_1] {$B$};
    \node[state, accepting] (ref_3) [below of=ref_2] {$C$};

    \node[state]         (var1_1) [right of=ref_1]  {};
    \node[state]         (var1_2) [below of=var1_1] {};
    \node[state]         (var1_3) [below of=var1_2] {};

    \node[state]         (var2_1) [right of=var1_1] {};
    \node[state]         (var2_2) [below of=var2_1] {};
    \node[state]         (var2_3) [below of=var2_2] {};

    \node[state]         (var3_1) [right of=var2_1] {};
    \node[state]         (var3_2) [below of=var3_1] {};
    \node[state]         (var3_3) [below of=var3_2] {};

    \node[state]         (var4_1) [right of=var3_1] {};
    \node[state]         (var4_2) [below of=var4_1] {};
    \node[state]         (var4_3) [below of=var4_2] {};

    \node[state]         (var5_1) [right of=var4_1] {};
    \node[state]         (var5_2) [below of=var5_1] {};
    \node[state]         (var5_3) [below of=var5_2] {};

    \path (ref_1) edge [bend left]  node {} (var1_1);
    \path (ref_1) edge [bend left]  node {} (var2_1);
    \path (ref_1) edge [bend left]  node {} (var3_1);
    \path (ref_1) edge [bend left]  node {} (var4_1);
    \path (ref_1) edge [bend left]  node {} (var5_1);
    
    \path (ref_2) edge [bend left]  node {} (var3_2);

    \path (ref_3) edge [bend left]  node {} (var1_3);
    \path (ref_3) edge [bend left]  node {} (var3_3);
    \path (ref_3) edge [bend left]  node {} (var4_3);

	\node[rectangle] at (0cm, 1.3cm) (pdfbox) {Ref};  
	\node[rectangle] at (1.5cm, 1.3cm) (pdfbox) {Var 1};  
    \node[rectangle] at (3cm, 1.3cm) (pdfbox) {Var 2};  
	\node[rectangle] at (4.5cm, 1.3cm) (pdfbox) {Var 3};  
    \node[rectangle] at (6cm, 1.3cm) (pdfbox) {Var 4};  
    \node[rectangle] at (7.5cm, 1.3cm) (pdfbox) {Var 5};

\end{tikzpicture}
\caption{Matching by "majority vote". At each level, haloes from the Reference simulation are
  matched to haloes in all variant simulations, based on their
  constituent particles, as illustrated in Fig.~\ref{fig:match-lagrangian}. We say that a halo {\it exist at level $\LE$}
  if it can be matched to haloes in a majority of variants at that
  level. In the above example, haloes A, B and C have 5, 1 and 3 out
  of a possible 5 matches, respectively. Haloes A and C {\it exist} at
  this level, while halo B does not. }
\label{fig:existence}
\end{figure}
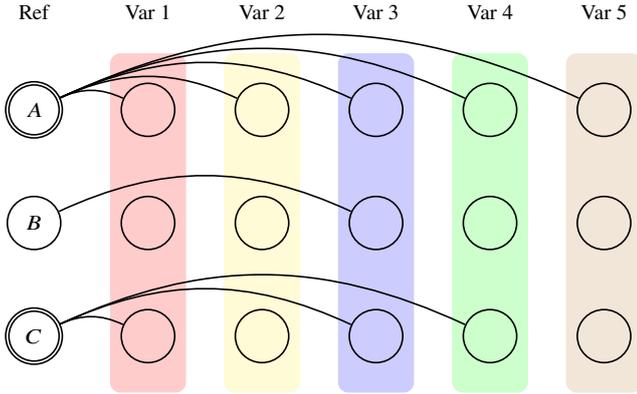

The matching by "majority vote" introduced in the previous paragraph allows us to define a new criterion for the existence of unique haloes: as illustrated in Fig.~\ref{fig:existence}, we say that a halo {\it exists} at and above level $\LE$ if it can be matched to more than half of the simulations that randomly vary the initial density fields at levels above $\LE$. In other words, a halo exists at a scale $\LE$ because, at this scale, the initial density field contains the necessary information for the formation of this particular halo.

\begin{figure}
    \begin{overpic}[width=\columnwidth]{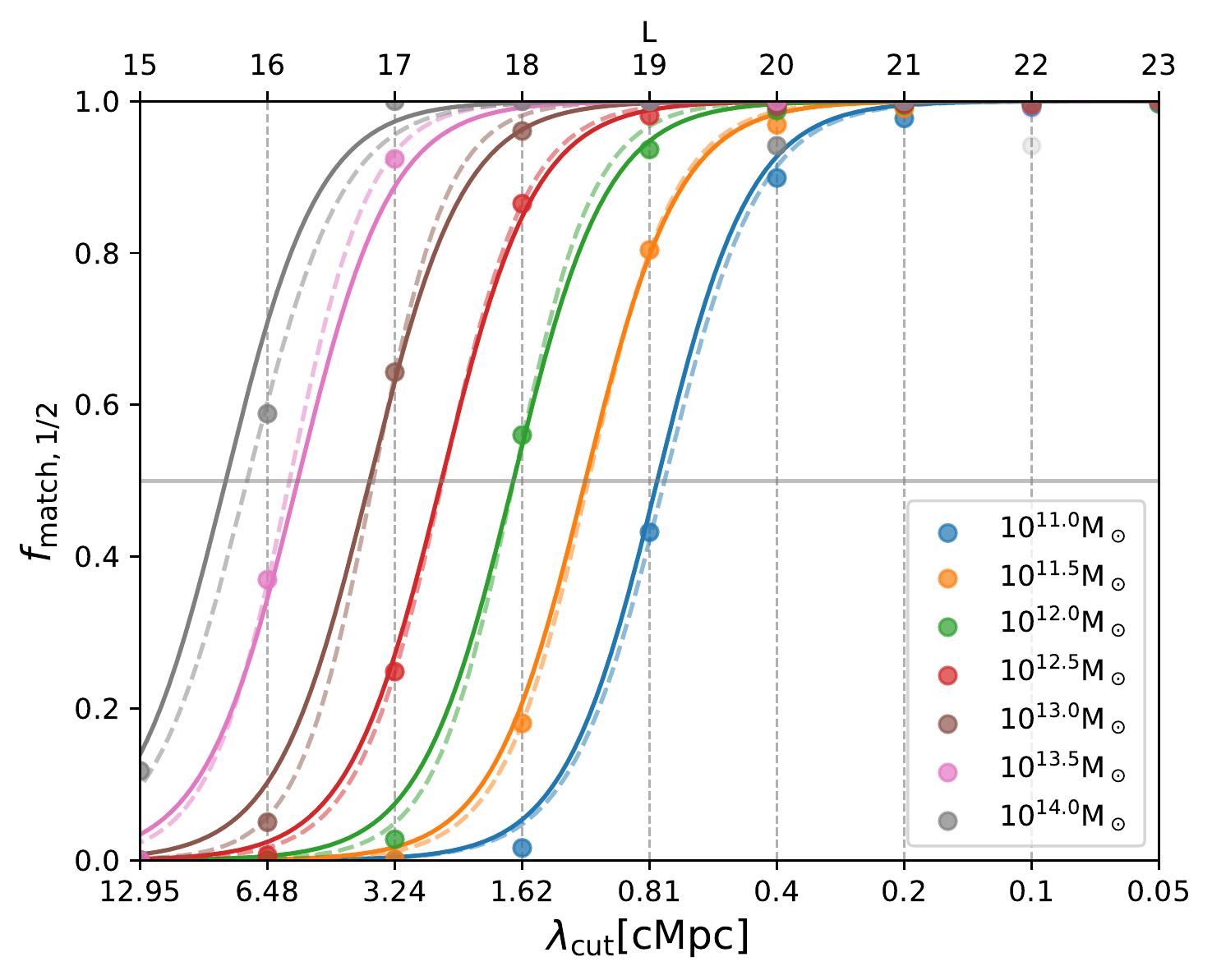}
    \end{overpic}
    \caption{$f_{\mathrm{match,1/2}}$, the fraction of haloes at $z=0$ that are matched more than half of the time across variations at a given cut-off wavelength ($\lc$, top axis), and level ($\LL$, bottom axis), for different masses. The scale at which $f_{\mathrm{match,1/2}}$ equals $1/2$ defines $\LEM$ and $\lEM$, i.e. the level and wavelength at which there is sufficient information in the primordial white noise field for the formation of unique $z=0$ haloes of a given mass, $\MM$. Circles show results measured in the simulations, dashed lines are individual fits to Eqn.~\ref{eqn:fmatch} at each mass, solid lines are fits of the same equation to the entire data set.
    }
    \label{fig:existence-levels}
\end{figure}

In Fig.~\ref{fig:existence-levels}, we show the fraction of unique haloes that can be matched to more than half of the variants, $f_\mathrm{match, 1/2}$, as a function of level, $\LL$, and of the corresponding cut-off wavelength, $\lc(\LL)$. Circles show the results measured in our simulations, dashed lines of corresponding colours show two-parameter logistic fits of the form
\begin{equation} \label{eqn:fmatch}
f_{\mathrm{match,1/2}}(\lc,\MM) = \frac{1}{ 1 + e^{-a_\MM (\lc - \lEM)}},
\end{equation}
where $a_\MM$ and $\lEM$ are free parameters fit separately at each mass, over the domain indicated by the extent of the dashed lines.

For each mass, we find a similar behaviour, with values of $a_\MM$ in the range $8-10$. Moreover, we find a regular scaling of $\mathrm{log}(\lEM) \propto ~\mathrm{log}(\MM)$. Consequently, the solid lines show a global fit to Eqn.~\ref{eqn:fmatch}, with $a_\MM = a = 9$, and
\begin{equation}\label{eqn:existence}
\mathrm{log_{10}}(\lEM / \mathrm{cMpc})  = 0.34~\mathrm{log}_{10}(\MM/\Ms) - 3.85.
\end{equation}
Dashed lines give a closer match to the individual halo mass ranges for which they are fitted independently with two free parameters each. However, the solid lines give a close fit to the entire data set with only three free parameters, $a$ and the two coefficients in Eqn.~\ref{eqn:existence} that determine $\lEM$.

The phase information of the initial density field has to be defined at least down to a scale where $\lc = \lEM$ in order to not only create the same number of haloes of a given mass at $z=0$, but to specify the formation of unique haloes. We find that this scale, $\lE$, is $\sim 30\%$ smaller than $\lc$ for haloes of the same mass. As examples, we find that unique haloes of $10^{12}$ and $10^{14}\Ms$ {\it exist} at $\lE=1.7$ and $8.1~\cMpc$, respectively, while the cut-off scales for $10^{12}\Ms$ and $10^{14}\Ms$ are $\lc = 2.4$ and $11.6~\cMpc$, respectively.

\begin{figure*}
\begin{tabular}{@{}ccc@{}}
    \begin{overpic}[width=0.33\textwidth]{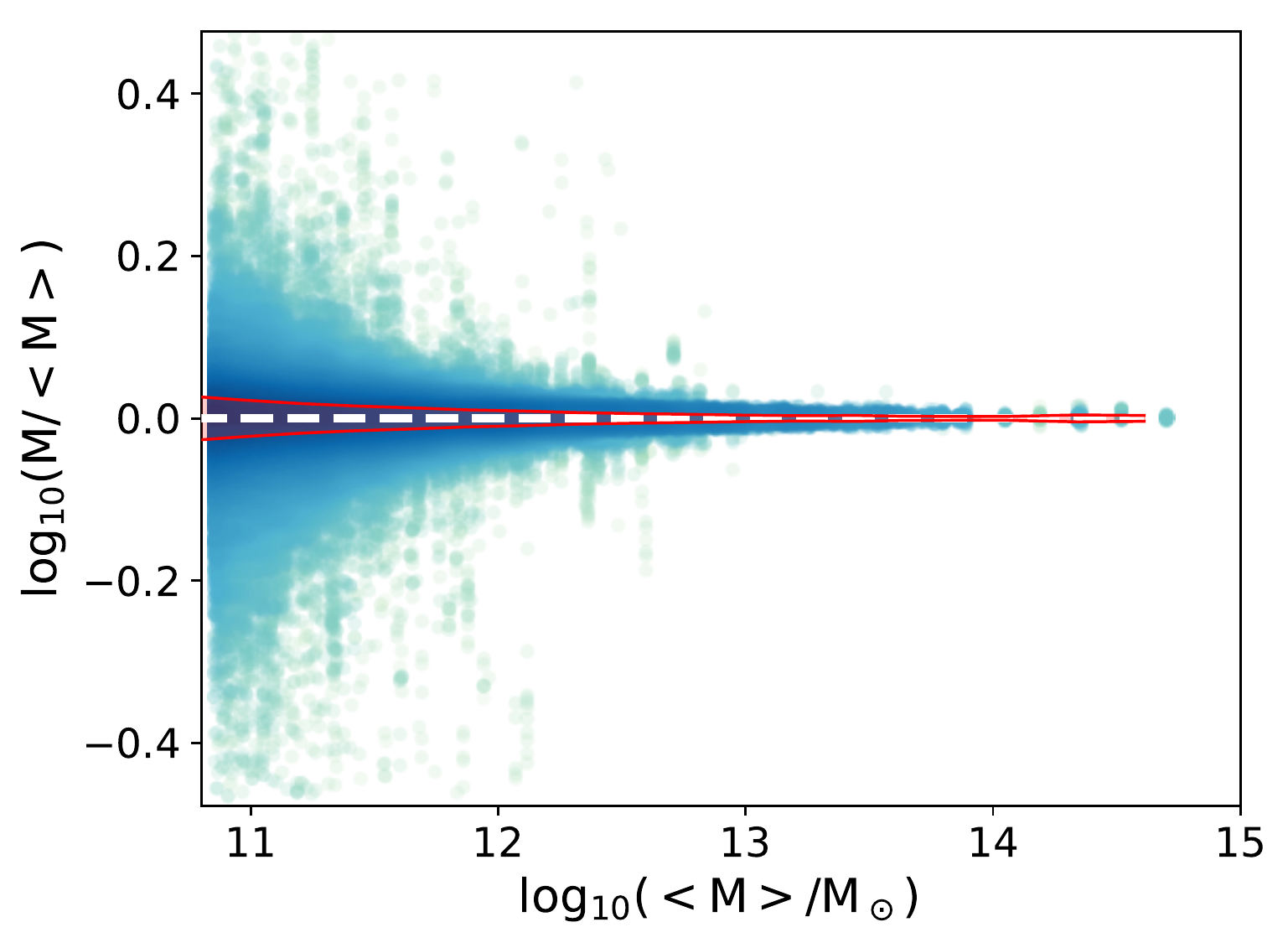}
    \put(55,65){\makebox[0pt] {\textcolor{black}{$\LL=23$}}} 
    \end{overpic} &
    \begin{overpic}[width=0.33\textwidth]{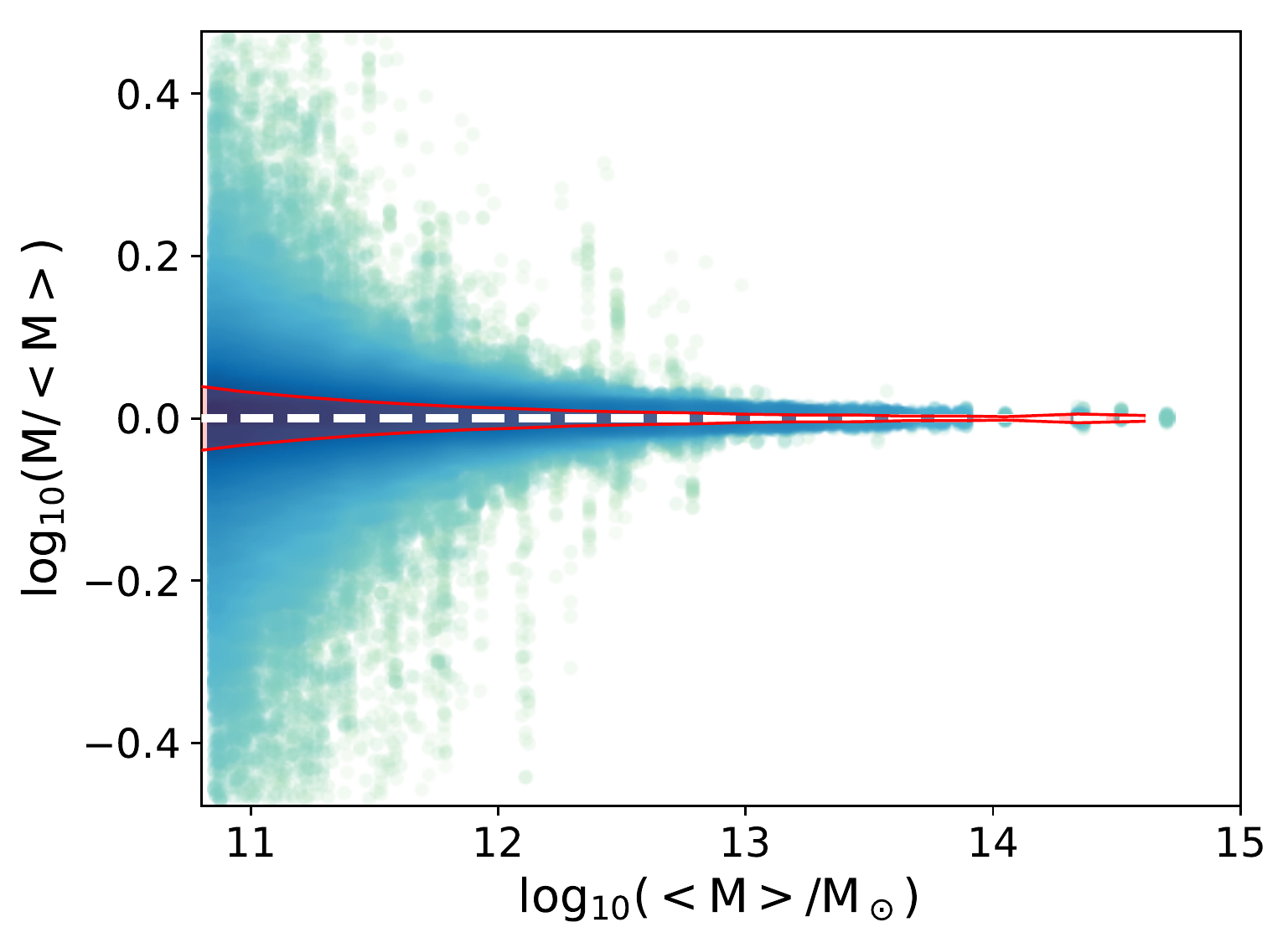}
    \put(55,65){\makebox[0pt] {\textcolor{black}{$\LL=22$}}} 
    \end{overpic} &
    \begin{overpic}[width=0.33\textwidth]{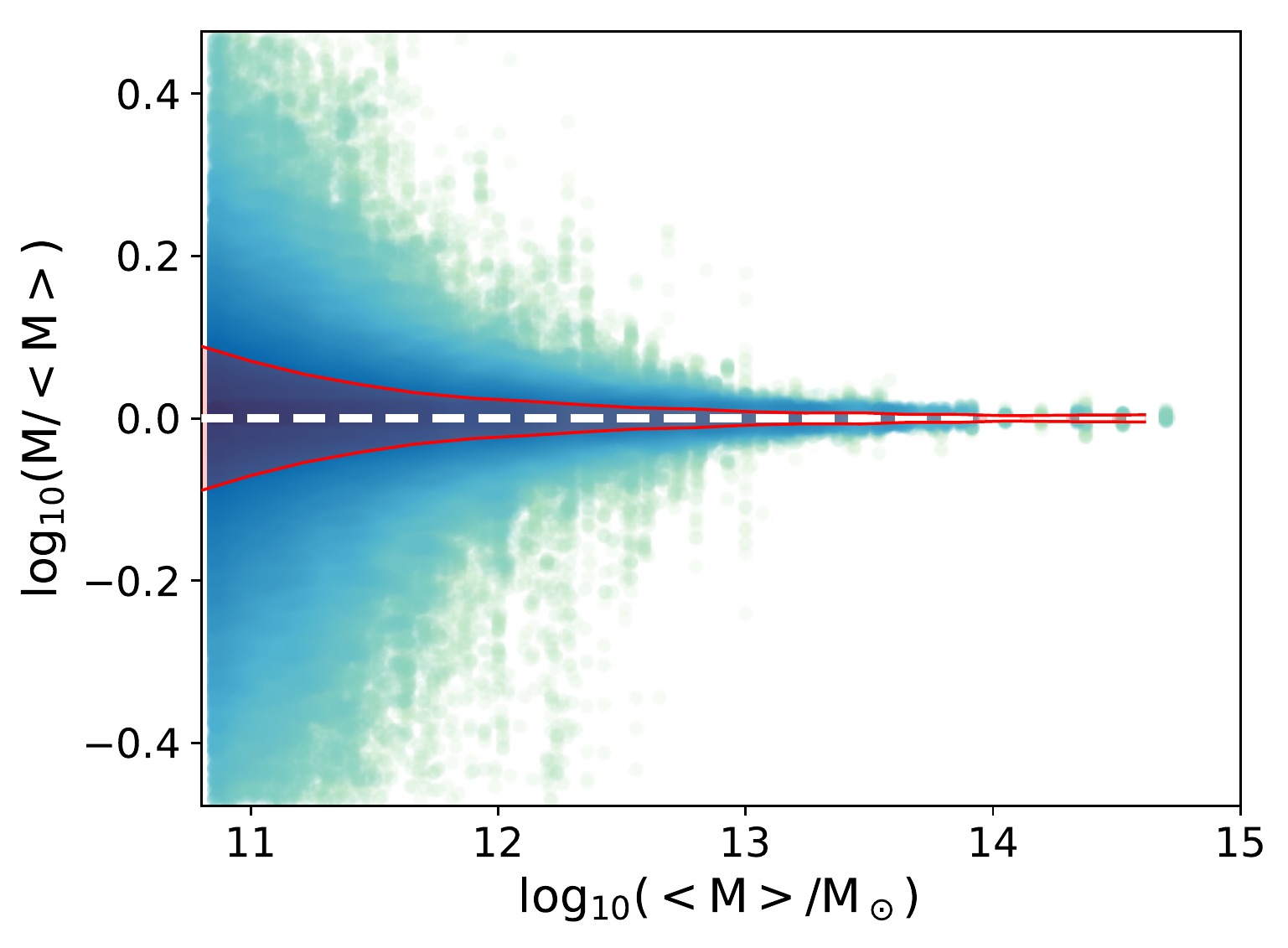}
    \put(55,65){\makebox[0pt] {\textcolor{black}{$\LL=21$}}} 
    \end{overpic} \\
    \begin{overpic}[width=0.33\textwidth]{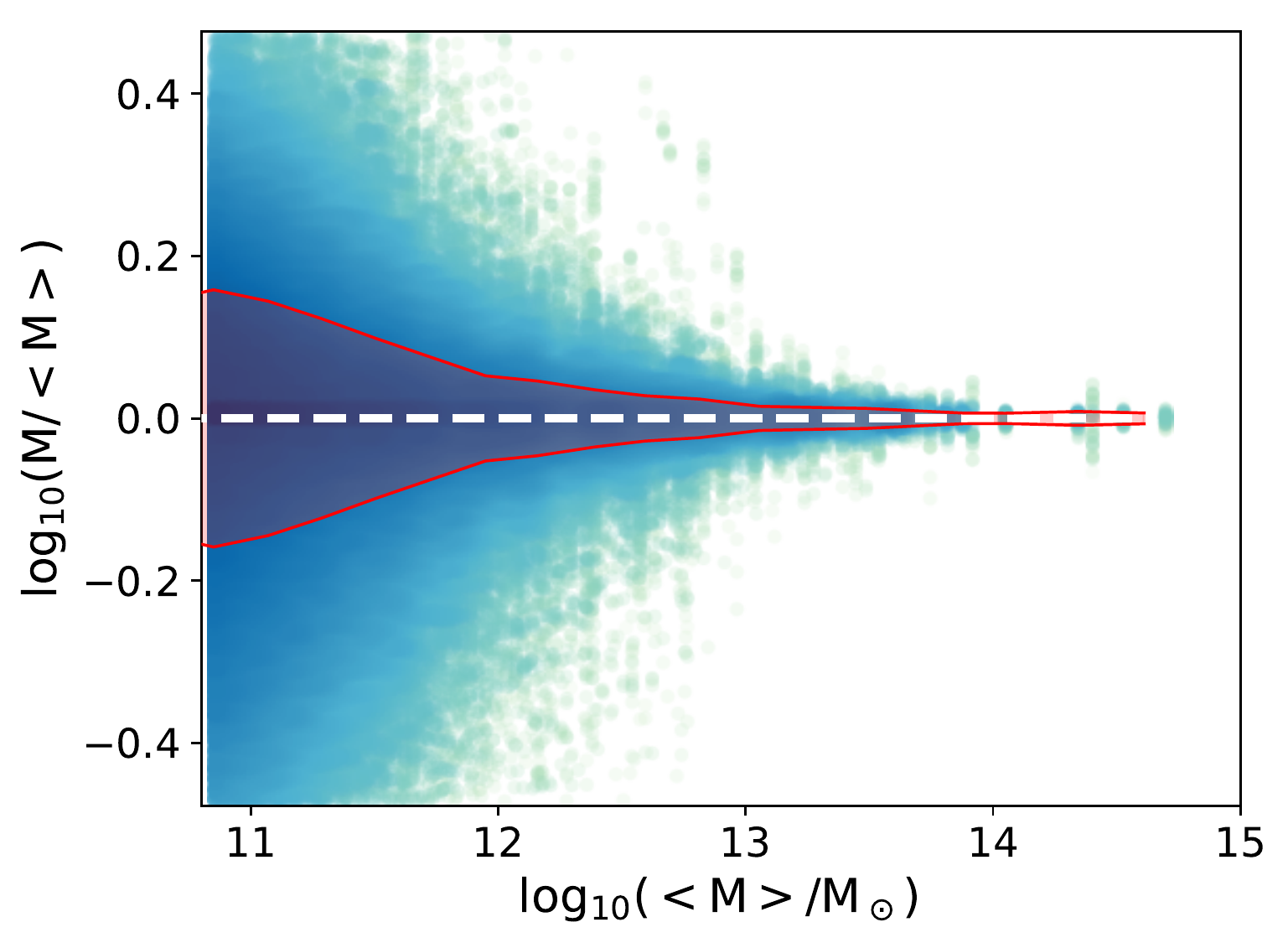}
    \put(55,65){\makebox[0pt] {\textcolor{black}{$\LL=20$}}} 
    \end{overpic} &
    \begin{overpic}[width=0.33\textwidth]{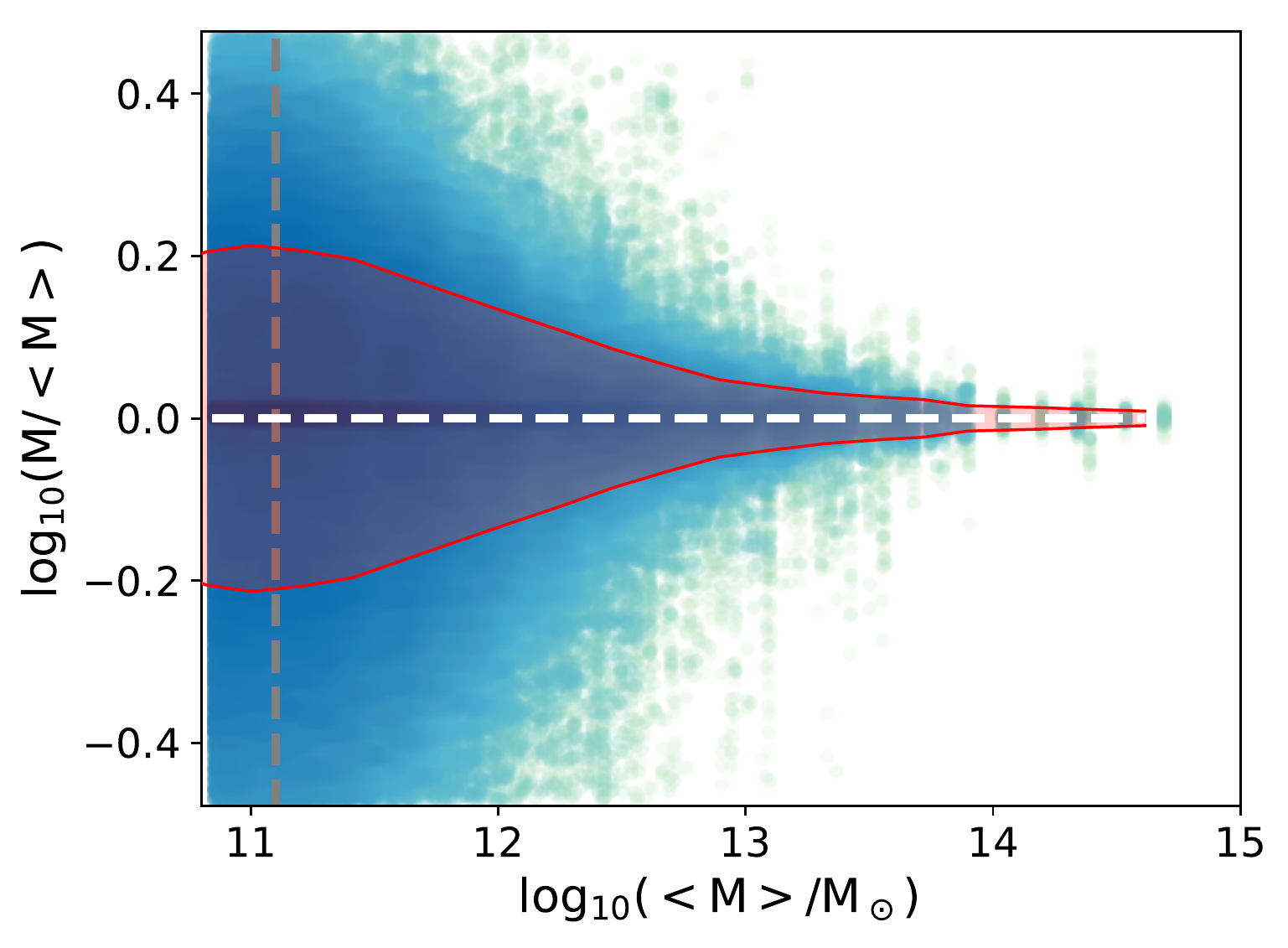}
    \put(55,65){\makebox[0pt] {\textcolor{black}{$\LL=19$}}} 
    \end{overpic} &
    \begin{overpic}[width=0.33\textwidth]{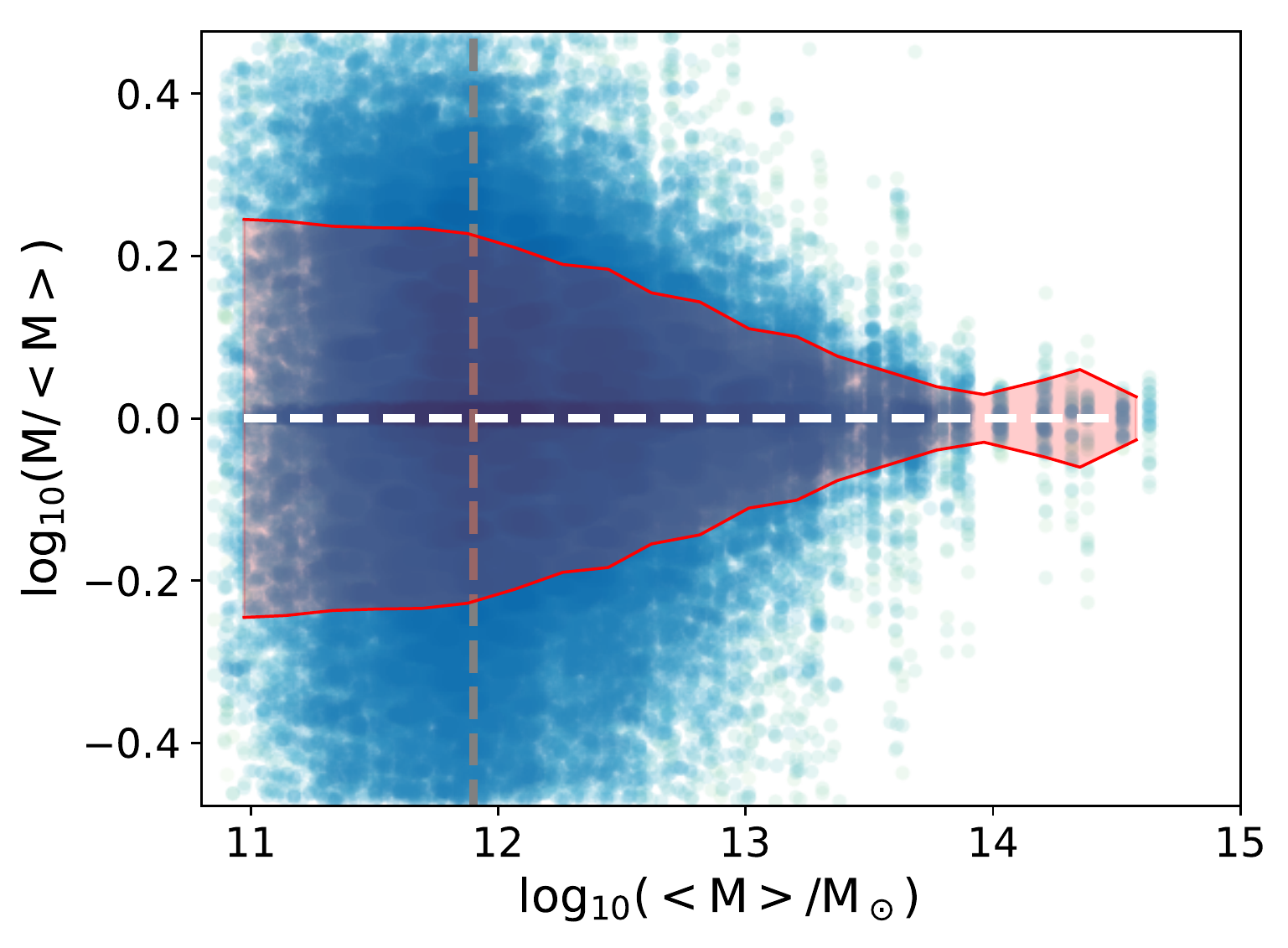}
    \put(55,65){\makebox[0pt] {\textcolor{black}{$\LL=18$}}} 
    \end{overpic} \\
  \end{tabular}
    \caption{Mass ratio between the mass of individual, matched
      central haloes in all 39 variant simulations at a given level,
      and their average mass across all 40 simulations, as a function
      of average mass. The red band shows the 1-sigma scatter. Vertical dashed lines on panels at $\LL=18$ and 19 indicate the minimum mass of haloes that {\it exist} at these level.}
    \label{fig:mass-ratios}
\end{figure*}

\begin{figure}
    \begin{overpic}[width=\columnwidth]{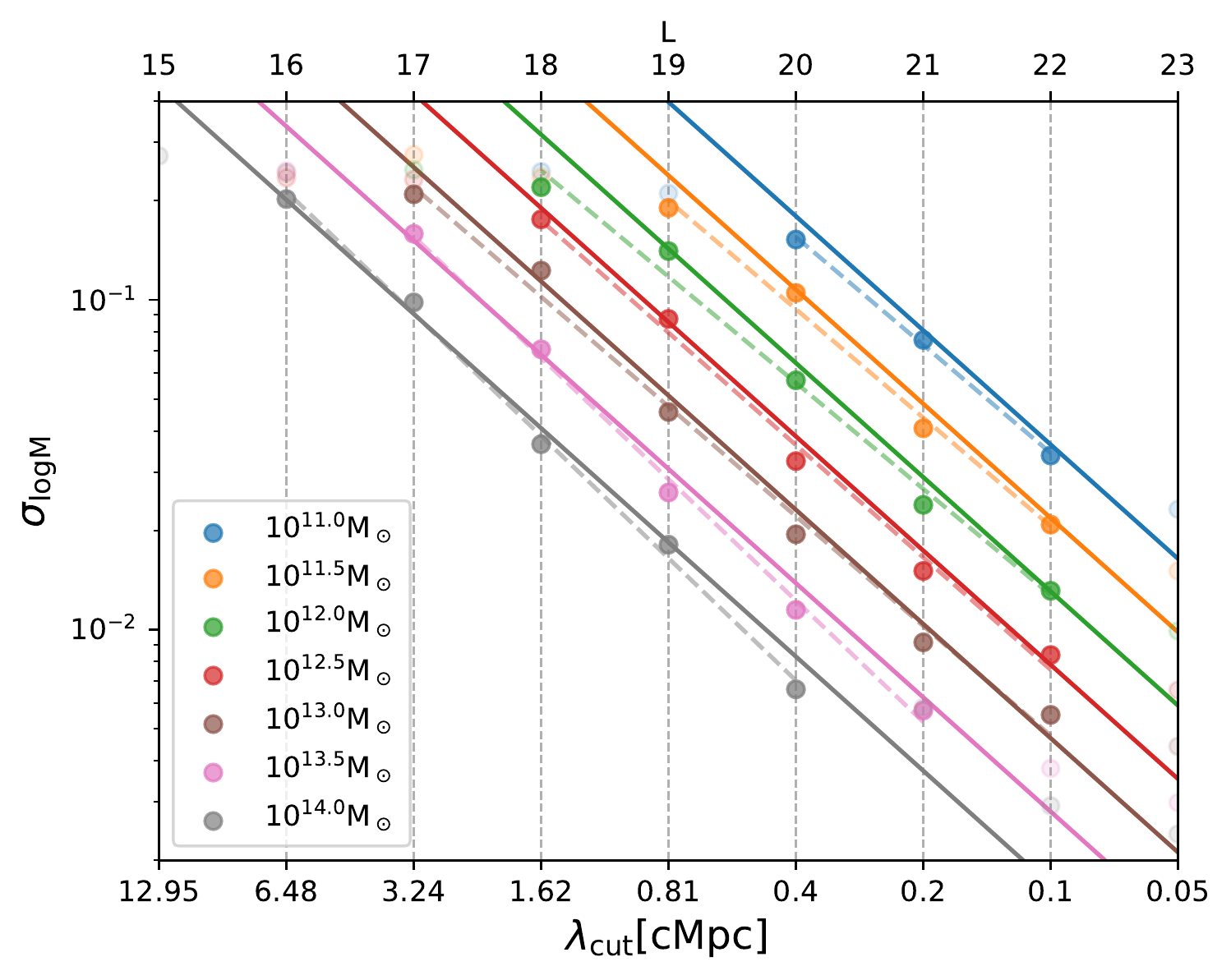}
    \end{overpic} 
    \caption{Median scatter in log halo mass, $\sigma^*_\MM$ for matched haloes of different mass (indicated by different colours), as a function of level of randomisation in the initial density field (L, upper x-axis), and corresponding wavelength $\lc$. Only bold data points are included in the fits. Dashed lines show independent, power-law fits for each mass bin, solid lines show a universal power-law fit for all masses. }
      
    \label{fig:massratio-synthesis}
\end{figure}

\begin{figure*}
\begin{tabular}{@{}ccc@{}}
    \begin{overpic}[width=0.33\textwidth]{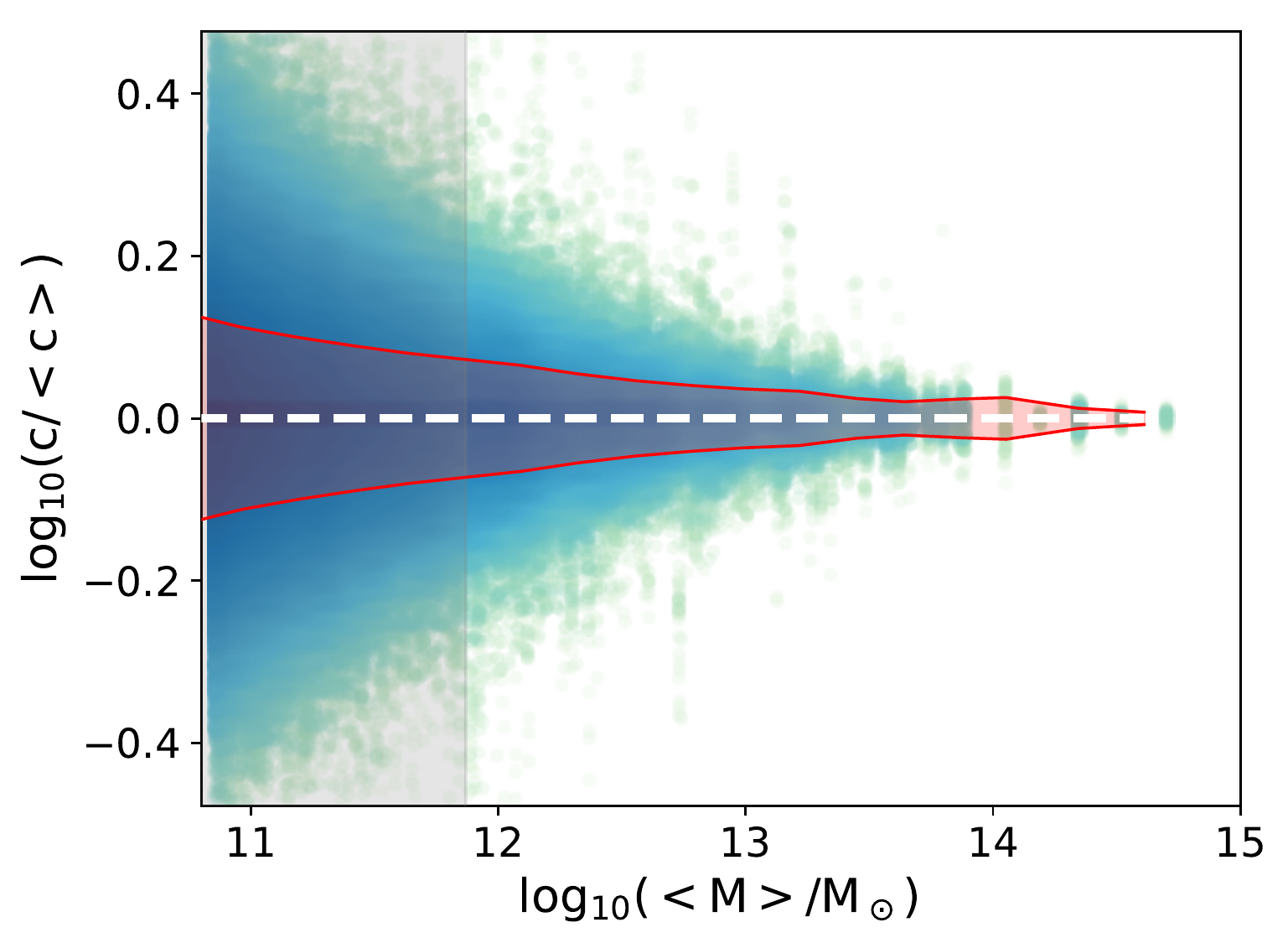}
    \put(55,65){\makebox[0pt] {\textcolor{black}{$\LL=23$}}} 
    \end{overpic} &
    \begin{overpic}[width=0.33\textwidth]{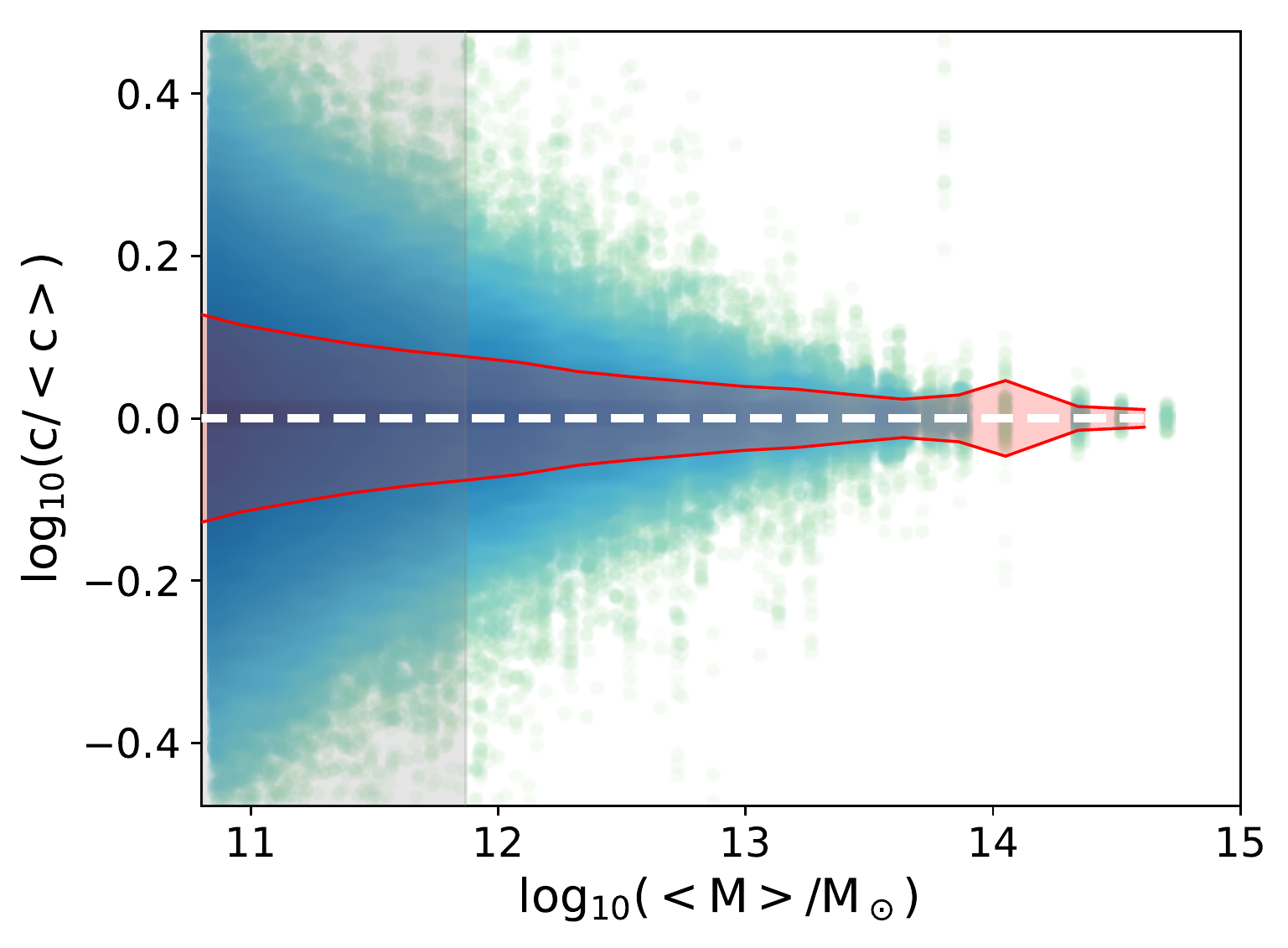}
    \put(55,65){\makebox[0pt] {\textcolor{black}{$\LL=22$}}} 
    \end{overpic} &
    \begin{overpic}[width=0.33\textwidth]{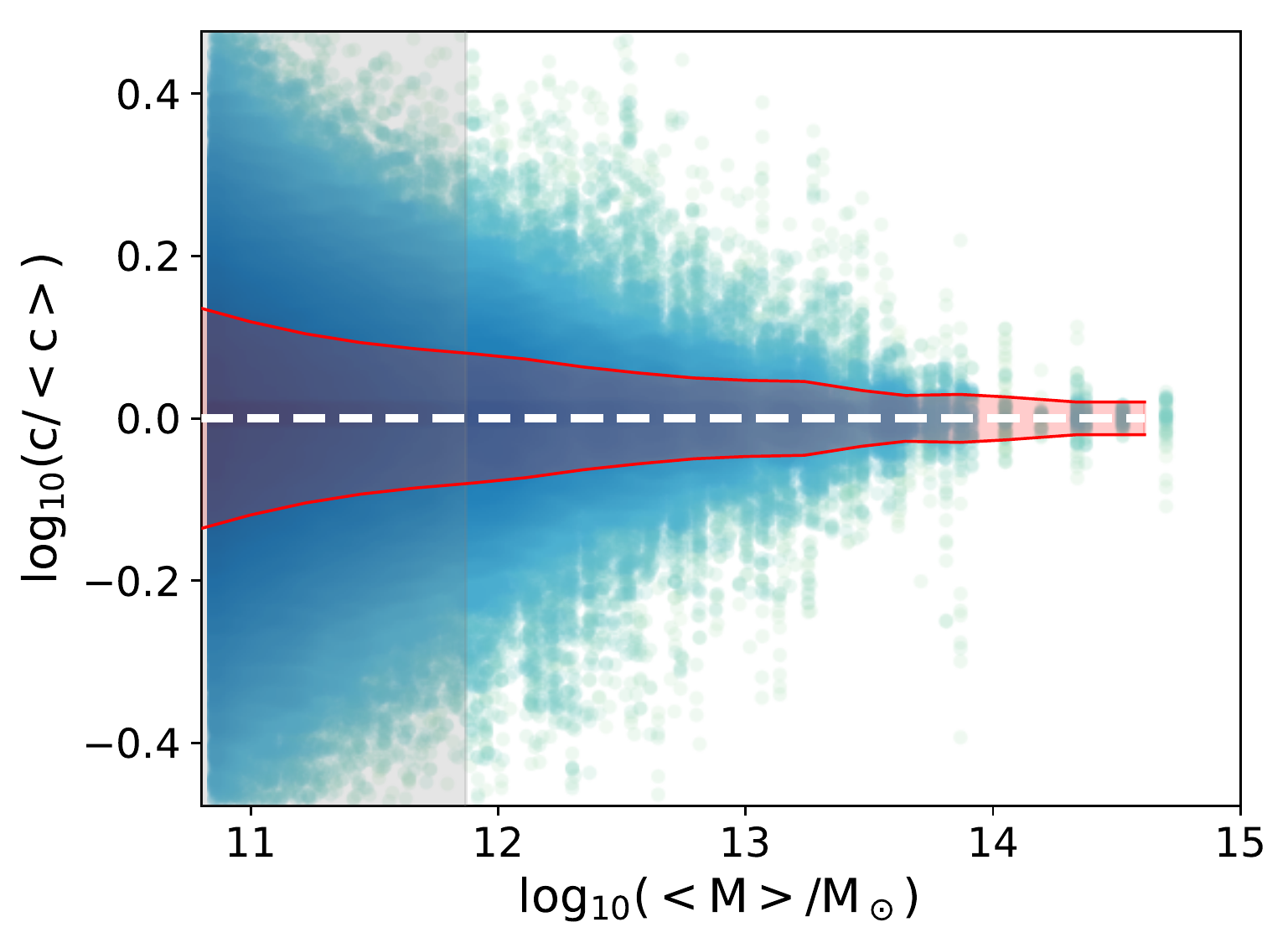}
    \put(55,65){\makebox[0pt] {\textcolor{black}{$\LL=21$}}} 
    \end{overpic} \\
    \begin{overpic}[width=0.33\textwidth]{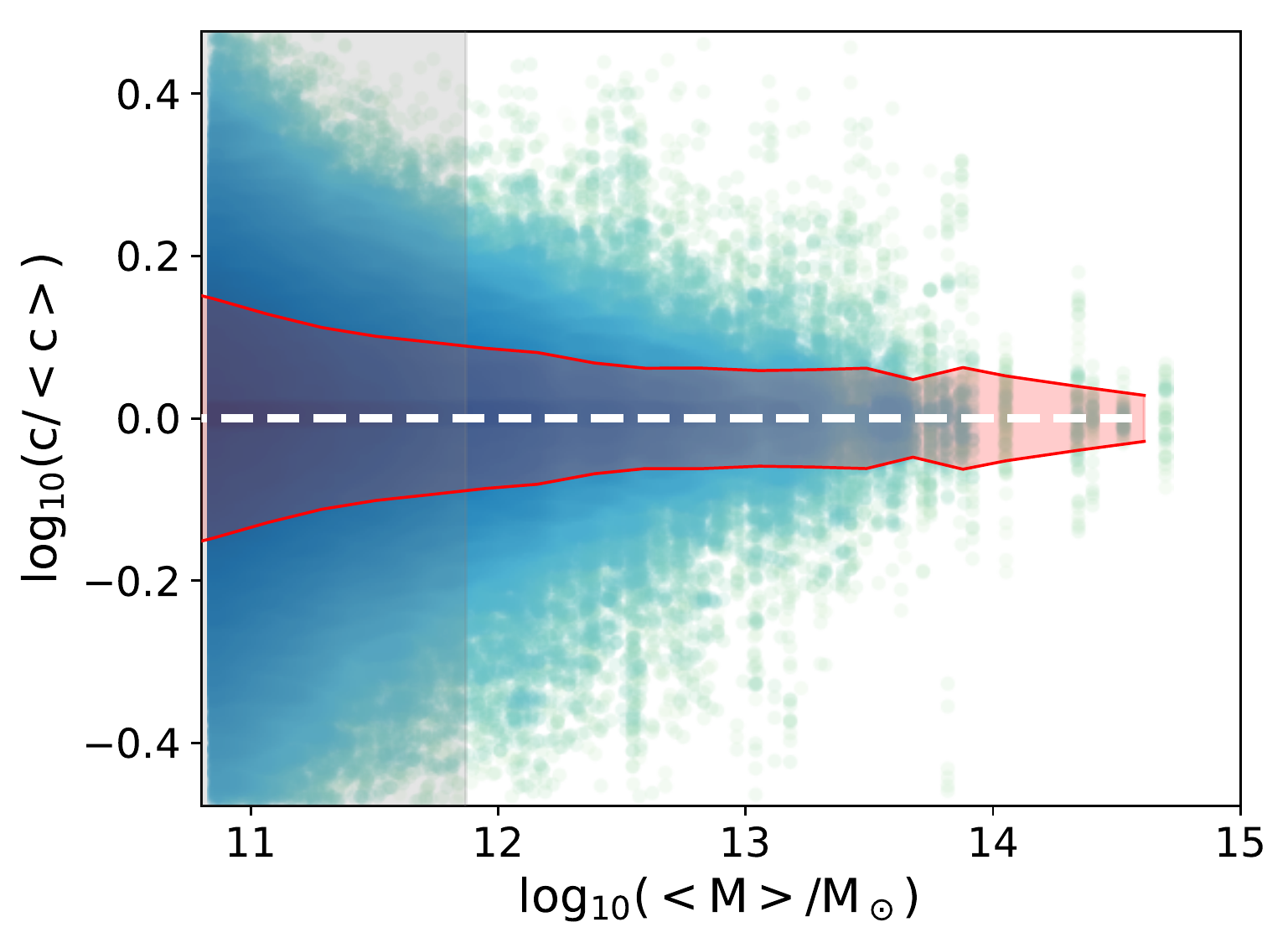}
    \put(55,65){\makebox[0pt] {\textcolor{black}{$\LL=20$}}} 
    \end{overpic} &
    \begin{overpic}[width=0.33\textwidth]{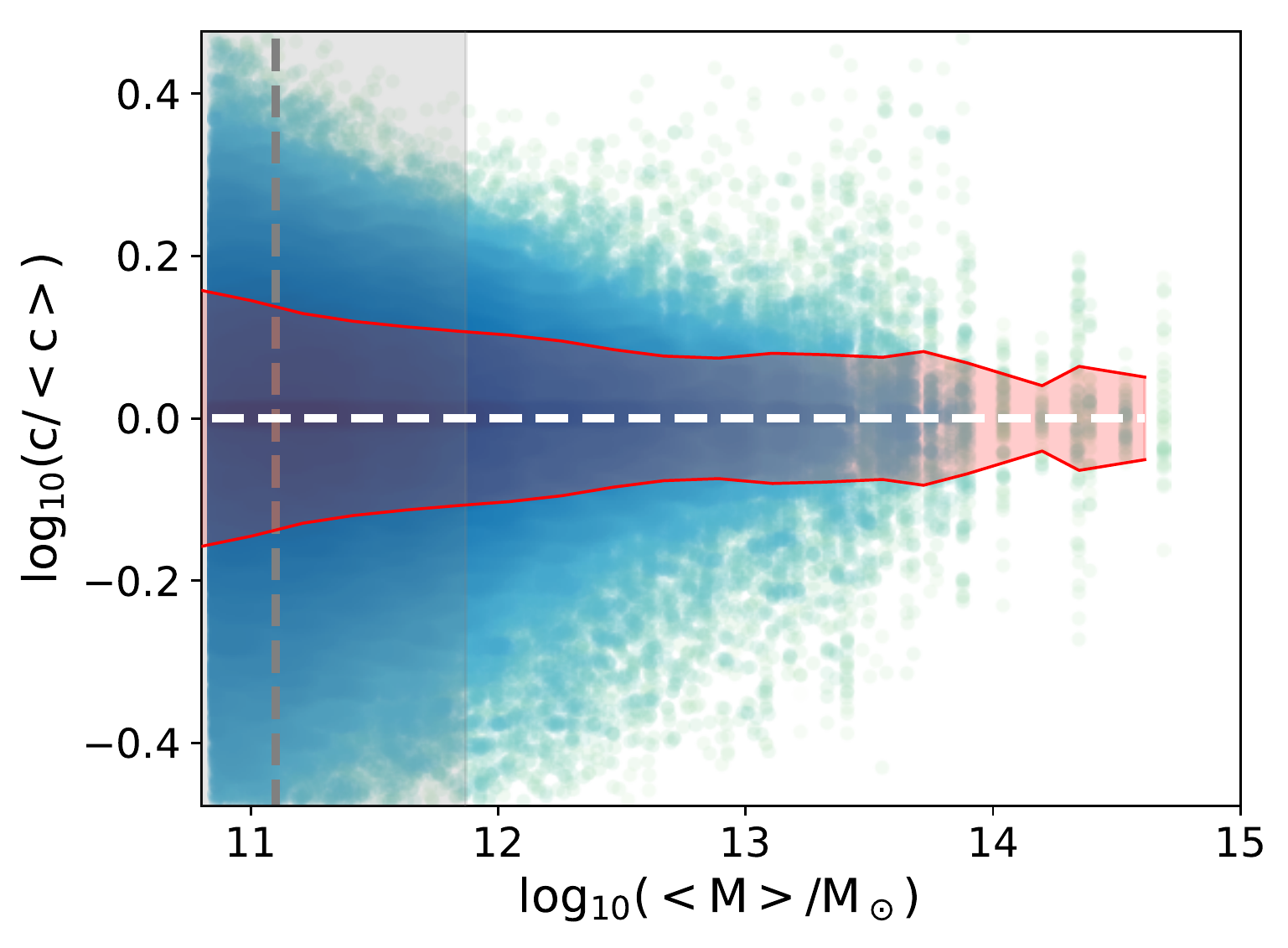}
    \put(55,65){\makebox[0pt] {\textcolor{black}{$\LL=19$}}} 
    \end{overpic} &
    \begin{overpic}[width=0.33\textwidth]{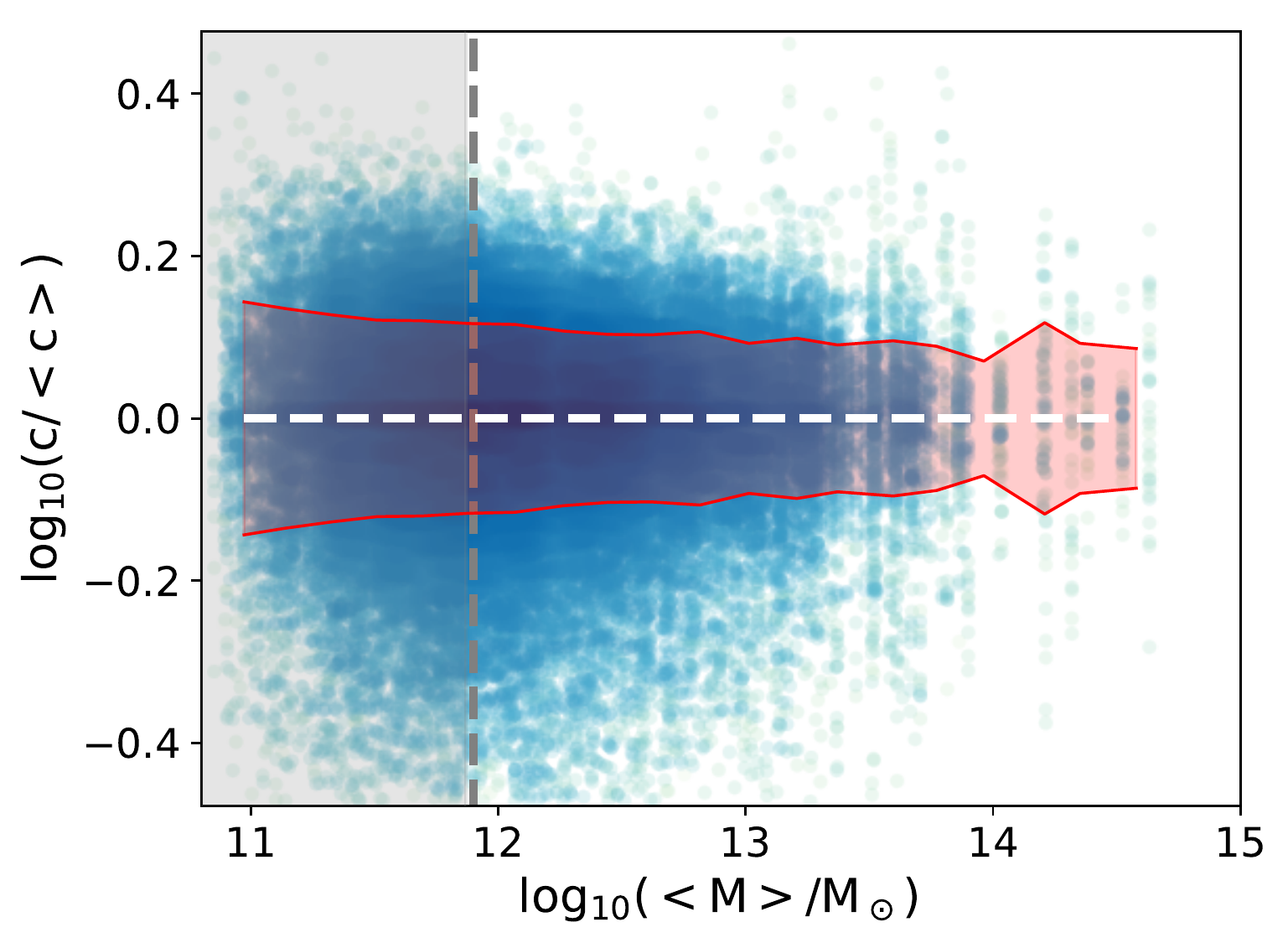}
    \put(55,65){\makebox[0pt] {\textcolor{black}{$\LL=18$}}} 
    \end{overpic} 
  \end{tabular}
    \caption{Ratio between concentration of matched central haloes in the variant simulations, and the median concentration for each halo, as a function of median halo mass, for different levels (indicated on each panel). The red regions show the standard deviation among variants. Grey regions denote haloes with fewer than 1000 particles, the convergence limit determined by \protect\cite{Neto-2007}; vertical dashed lines on panels at $\LL=18$ and 19 indicate the minimum mass of haloes that {\it exist} at these levels. \label{fig:plots/concentration_scatter} }
\end{figure*}

\begin{figure}
    \begin{overpic}[width=\columnwidth]{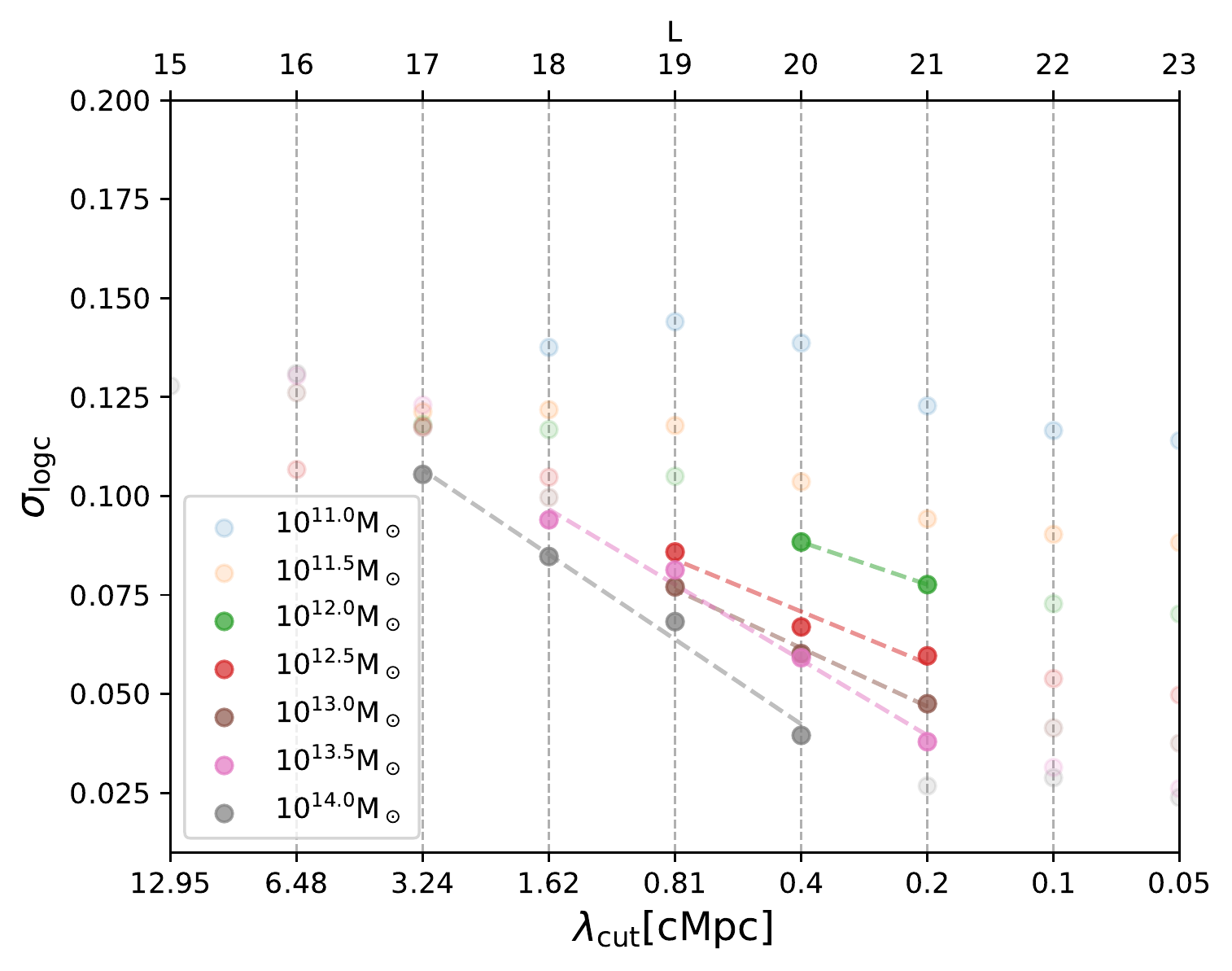}
    \end{overpic}
    \caption{Median standard deviation of halo positions, for haloes of     different mass (indicated by different colours), and as a function of level of randomisation in the initial density field (L, upper x-axis), and corresponding cut-off wavelength $\lc$. Only bold data points are included in the fits. Dashed lines show independent, power-law fits for each mass bin.}
    \label{fig:concentration-synthesis}
\end{figure}

\begin{figure*}
\begin{tabular}{@{}ccc@{}}
    \begin{overpic}[width=0.33\textwidth]{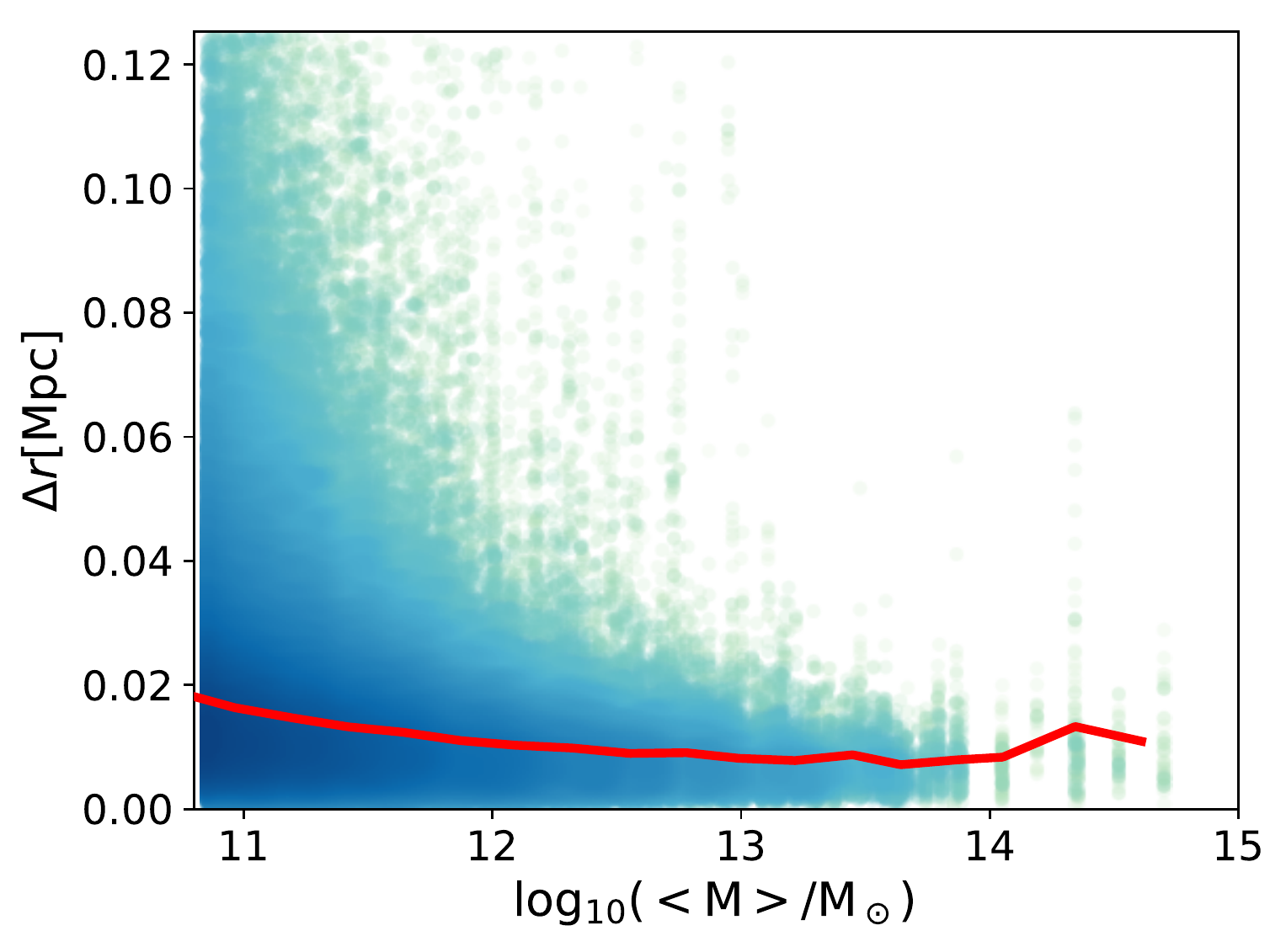}
    \put(55,65){\makebox[0pt] {\textcolor{black}{$\LL=23$}}} 
    \end{overpic} &
    \begin{overpic}[width=0.33\textwidth]{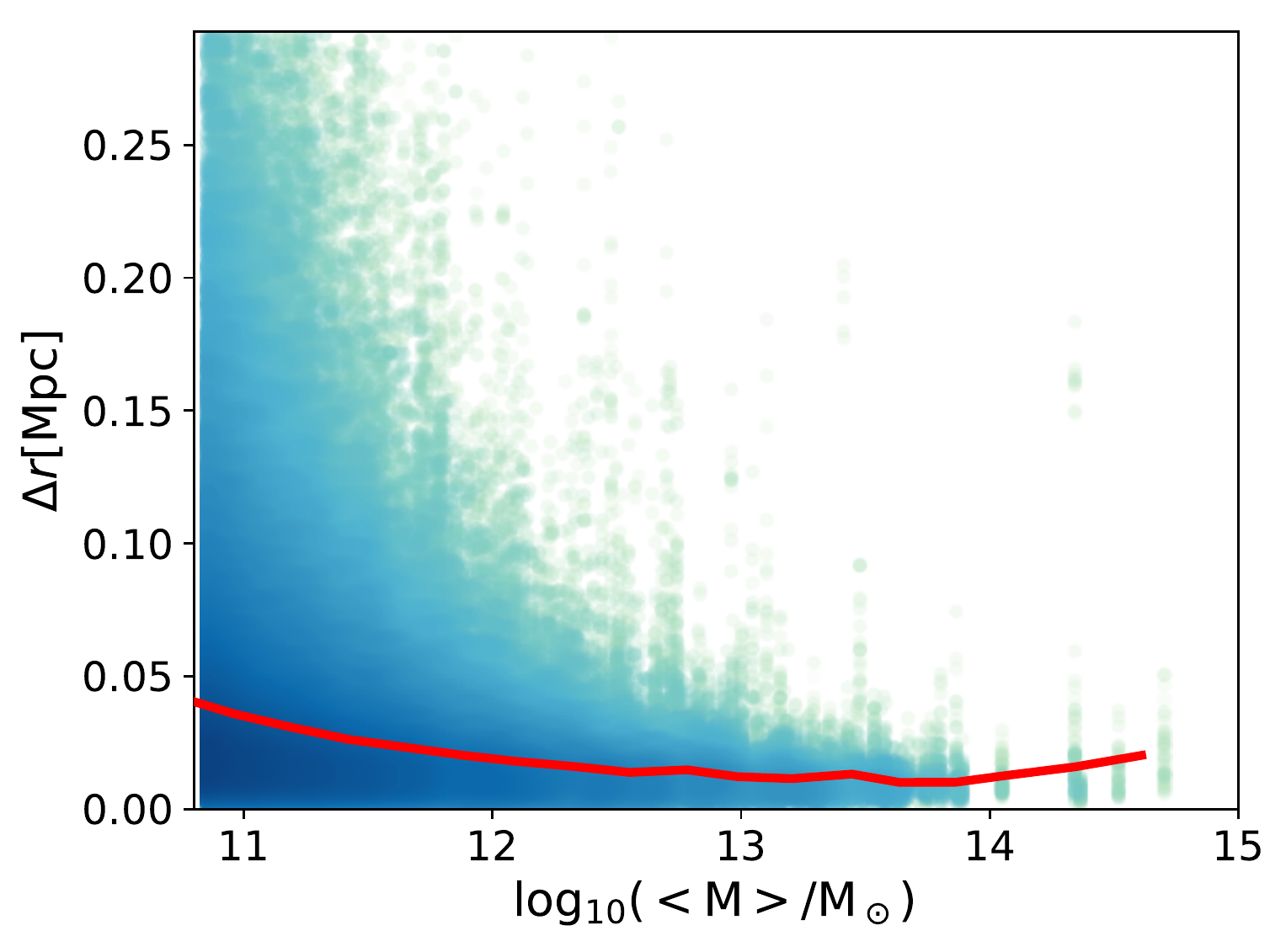}
    \put(55,65){\makebox[0pt] {\textcolor{black}{$\LL=22$}}} 
    \end{overpic} &
    \begin{overpic}[width=0.33\textwidth]{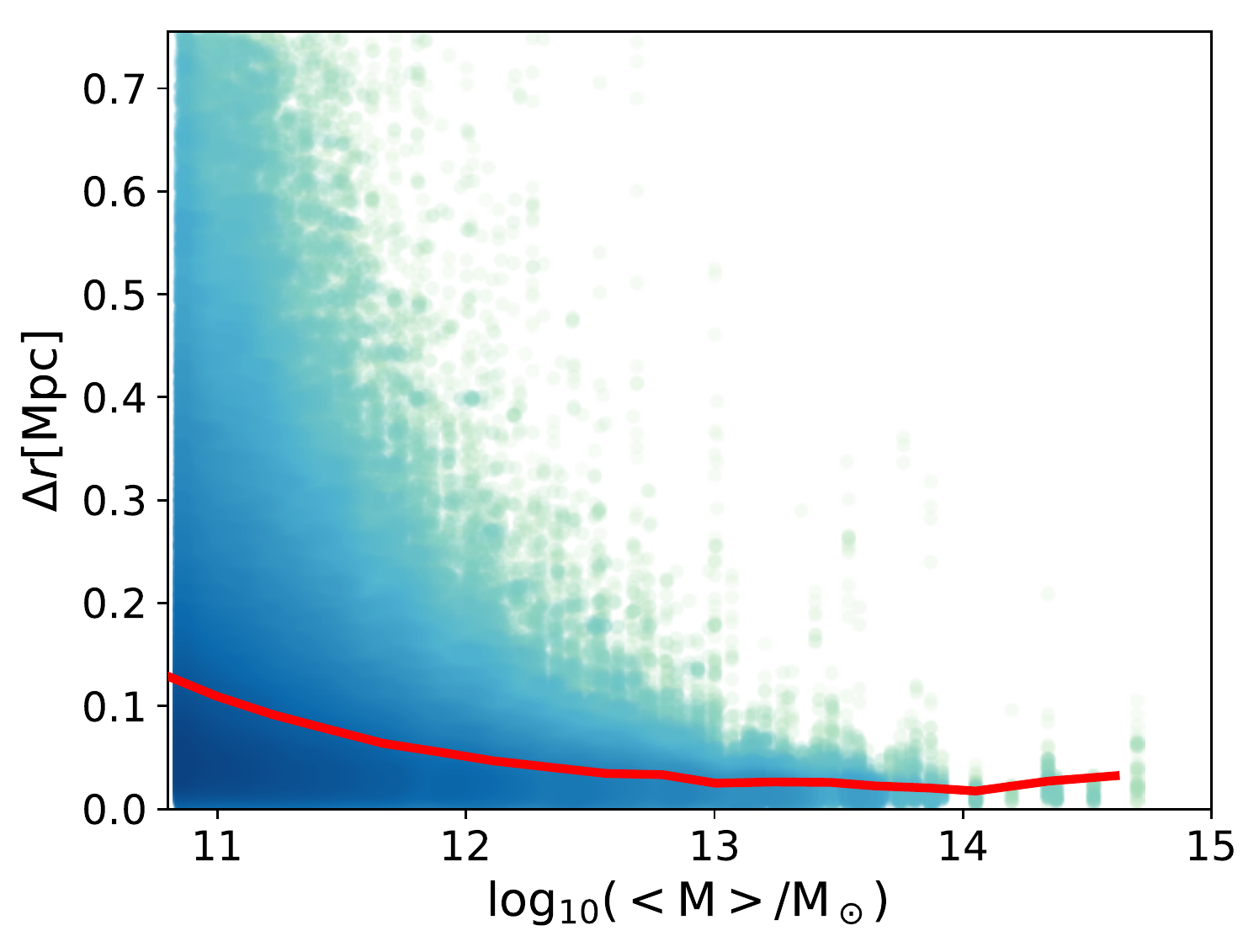}
    \put(55,65){\makebox[0pt] {\textcolor{black}{$\LL=21$}}} 
    \end{overpic} \\
    \begin{overpic}[width=0.33\textwidth]{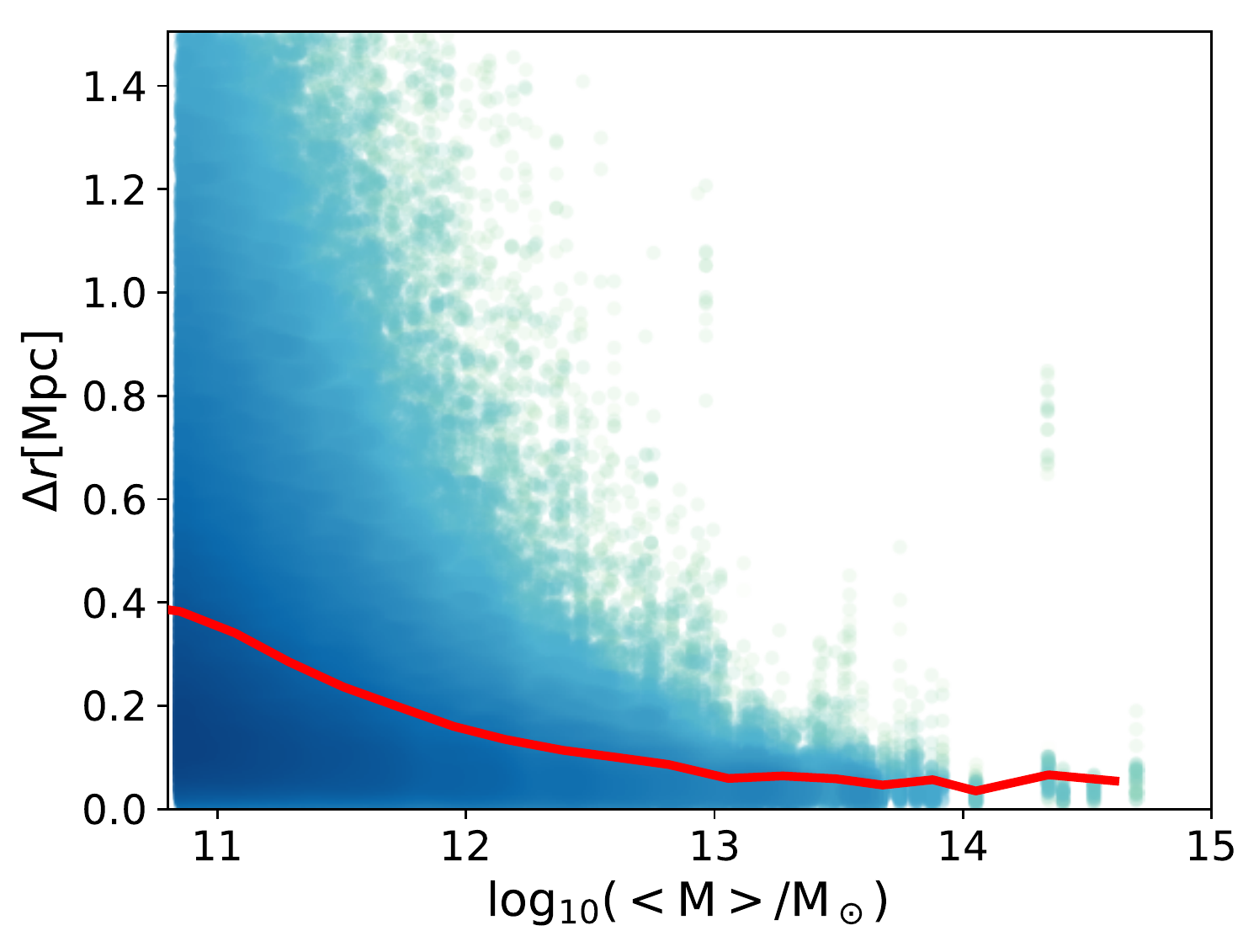}
    \put(55,65){\makebox[0pt] {\textcolor{black}{$\LL=20$}}} 
    \end{overpic} &
    \begin{overpic}[width=0.33\textwidth]{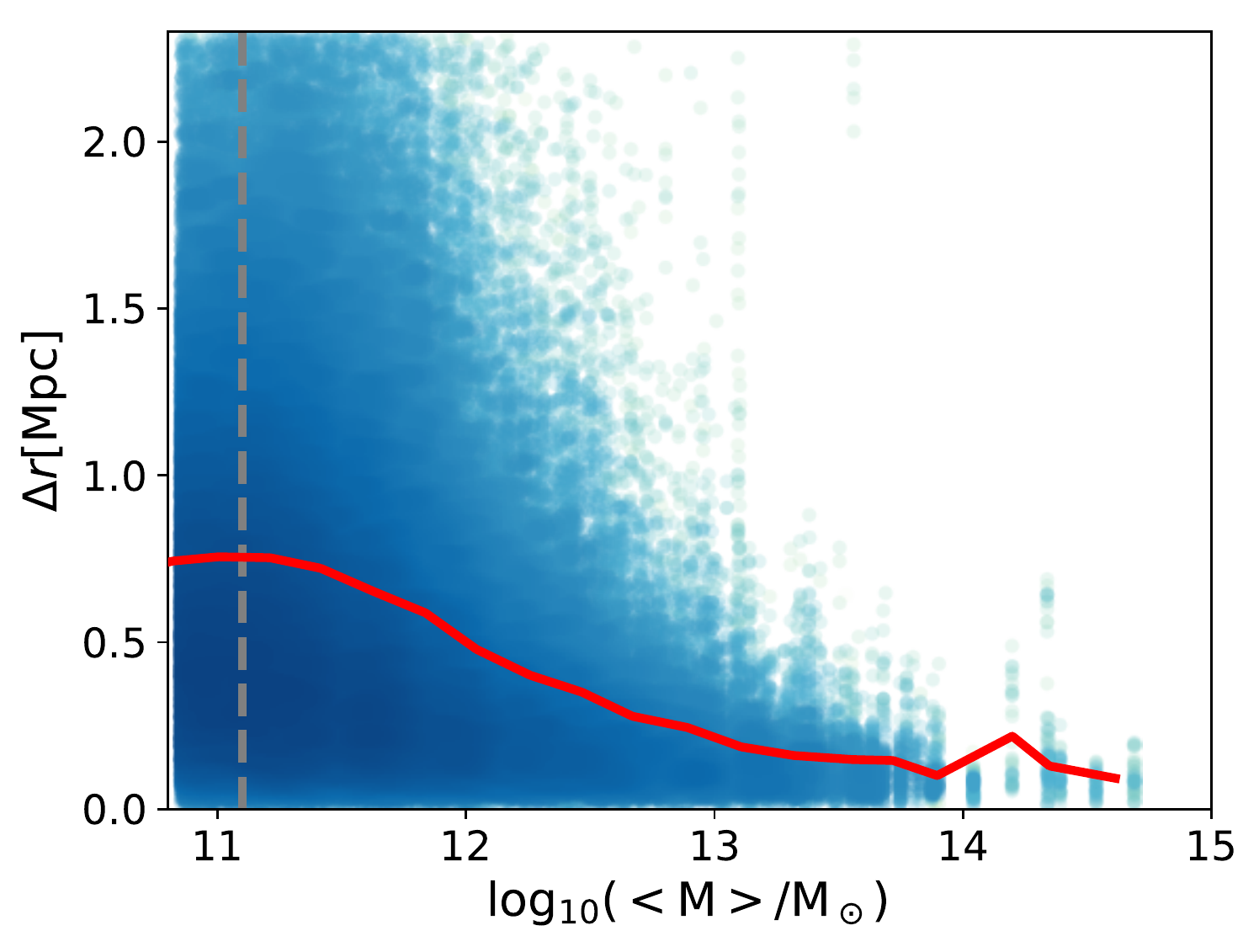}
    \put(55,65){\makebox[0pt] {\textcolor{black}{$\LL=19$}}} 
    \end{overpic} &
    \begin{overpic}[width=0.33\textwidth]{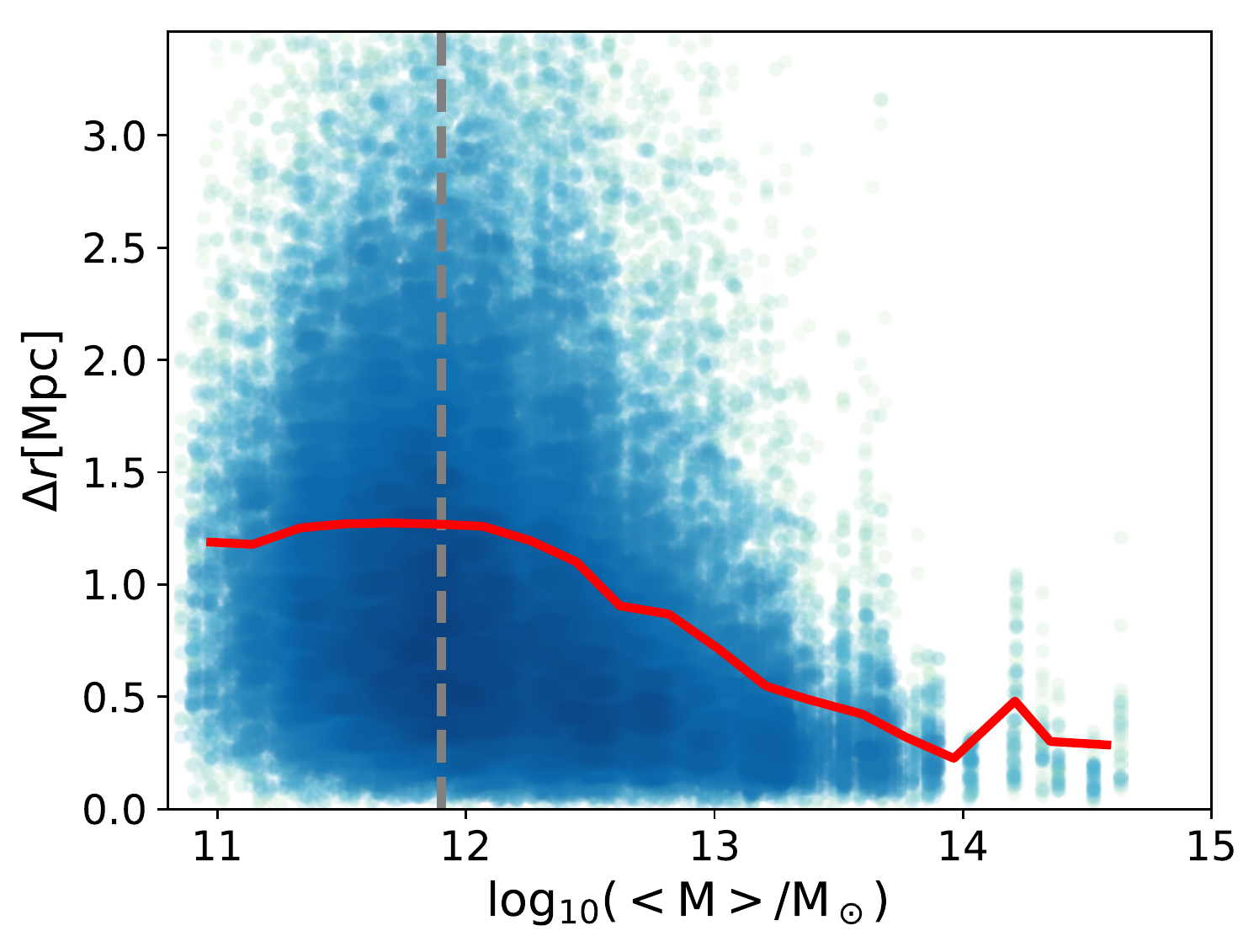}
    \put(55,65){\makebox[0pt] {\textcolor{black}{$\LL=18$}}} 
    \end{overpic} \\
  \end{tabular}
    \caption{Displacement of matched central haloes in the
      variant simulation relative to the median position for each halo, as a  function of median halo mass, for different levels (indicated on
      each panel). The red lines show one standard deviation in position among variants. Vertical dashed lines on panels at $\LL=18$ and 19 indicate the minimum mass of haloes that {\it exist} at these level. On each panel, the range extends to the 99th percentile of all haloes shown.}
    \label{fig:displacements}
\end{figure*}

\begin{figure}
    \begin{overpic}[width=\columnwidth]{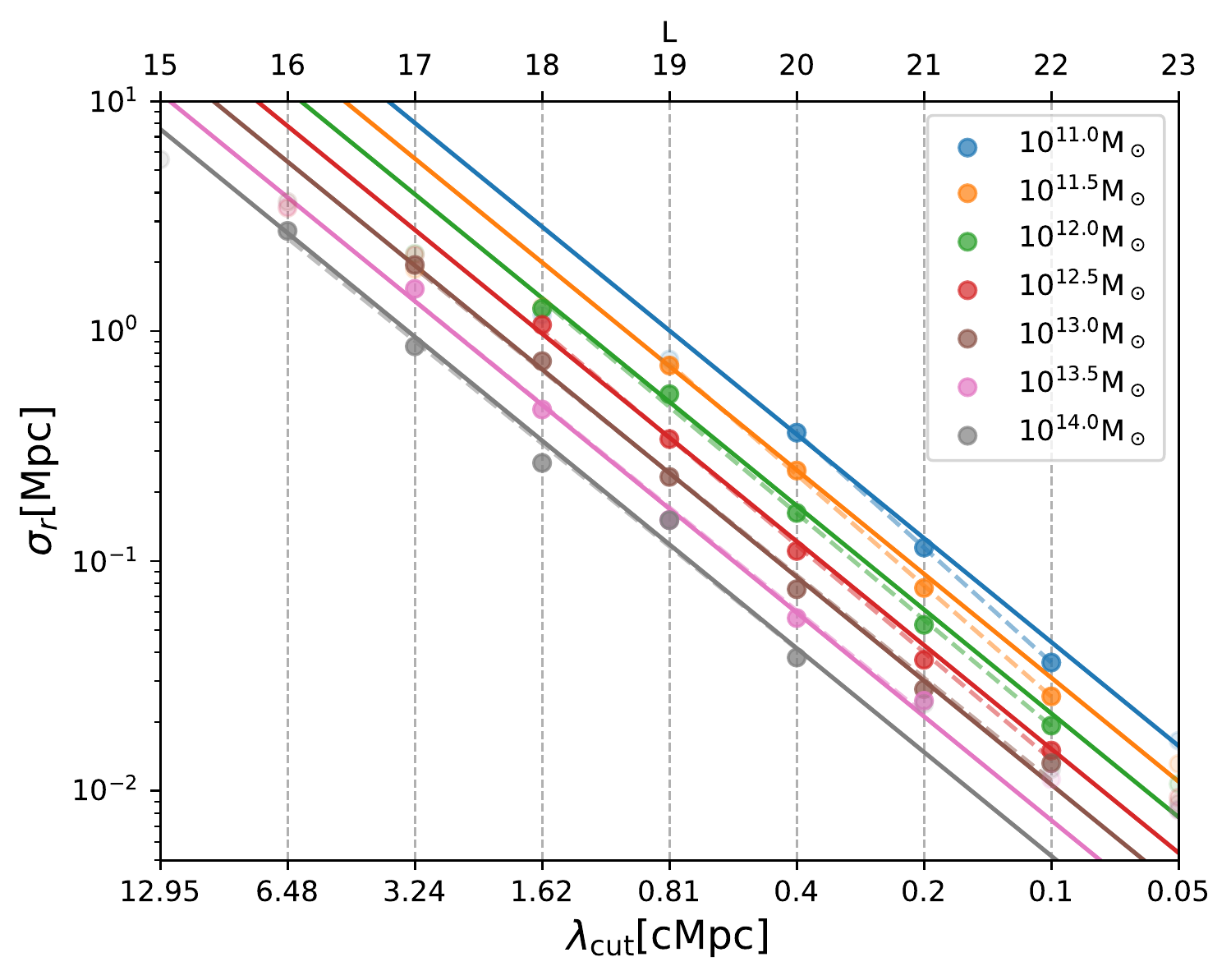}
    \end{overpic}
    \caption{Scatter of halo positions, $\sigma_r$ for central haloes of different mass (indicated by different colours), and as a function of level of randomisation in the initial density field (L, upper x-axis), and corresponding cut-off wavelength ($\lc$, lower x-axis). Only bold data points are included in the fits. Dashed lines show independent, power-law fits for each mass bin, solid lines show a universal power-law fit for all masses.}
    \label{fig:displacement-synthesis}
\end{figure}

\section{The origin of halo properties}\label{sec:results:properties}
Having defined the scales that determine the existence of particular haloes, we now turn to the changes seen in the properties of individual haloes that are matched across simulations due to variations on smaller scales. In particular, we will examine the mass and concentration of individual haloes, and their position and velocities.

At this stage, it is important to remind ourselves that, while the Reference simulation plays a special role in identifying matches, it is only one of many possible realisations. At every level, it shares the same amount of phase information with any of the variants, as they share with one another. Furthermore, any halo identified in the Reference simulation is only one possible realisation of that halo. If we consider that the possible random variations on scales below $\LE$ define the space of possible halo properties, we can consider the halo of the Reference simulation as one random sample of this space, not guaranteed to be at its centre. In measuring the variation in properties of individual haloes, we therefore do no compare the different realisations of a halo to the Reference simulation. Instead, we compute a median value from all realisations at a given level, and analyse the scatter among individual samples.

\begin{figure*}
\begin{tabular}{@{}ccc@{}}
    \begin{overpic}[width=0.33\textwidth]{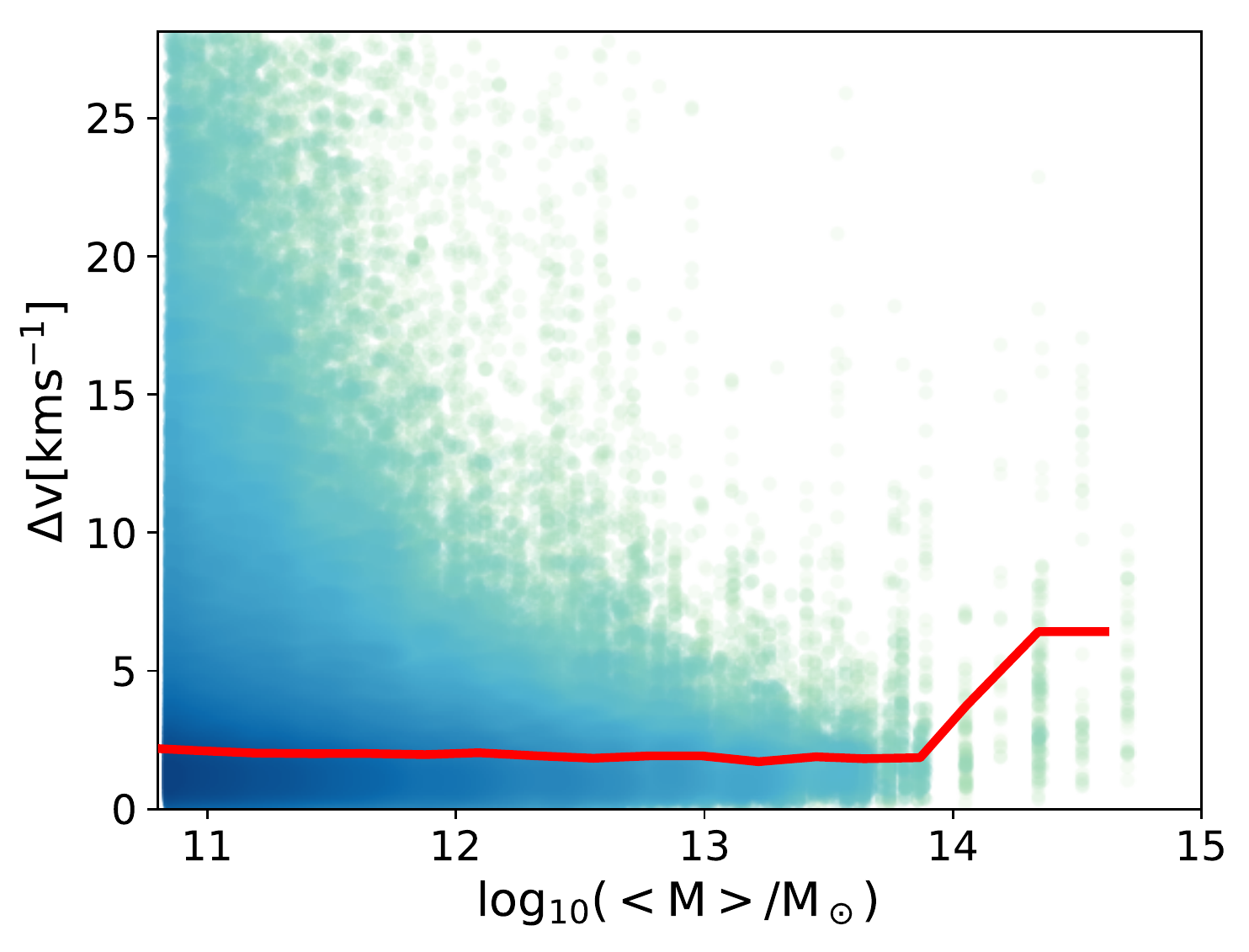}
    \put(55,65){\makebox[0pt] {\textcolor{black}{$\LL=23$}}} 
    \end{overpic} &
    \begin{overpic}[width=0.33\textwidth]{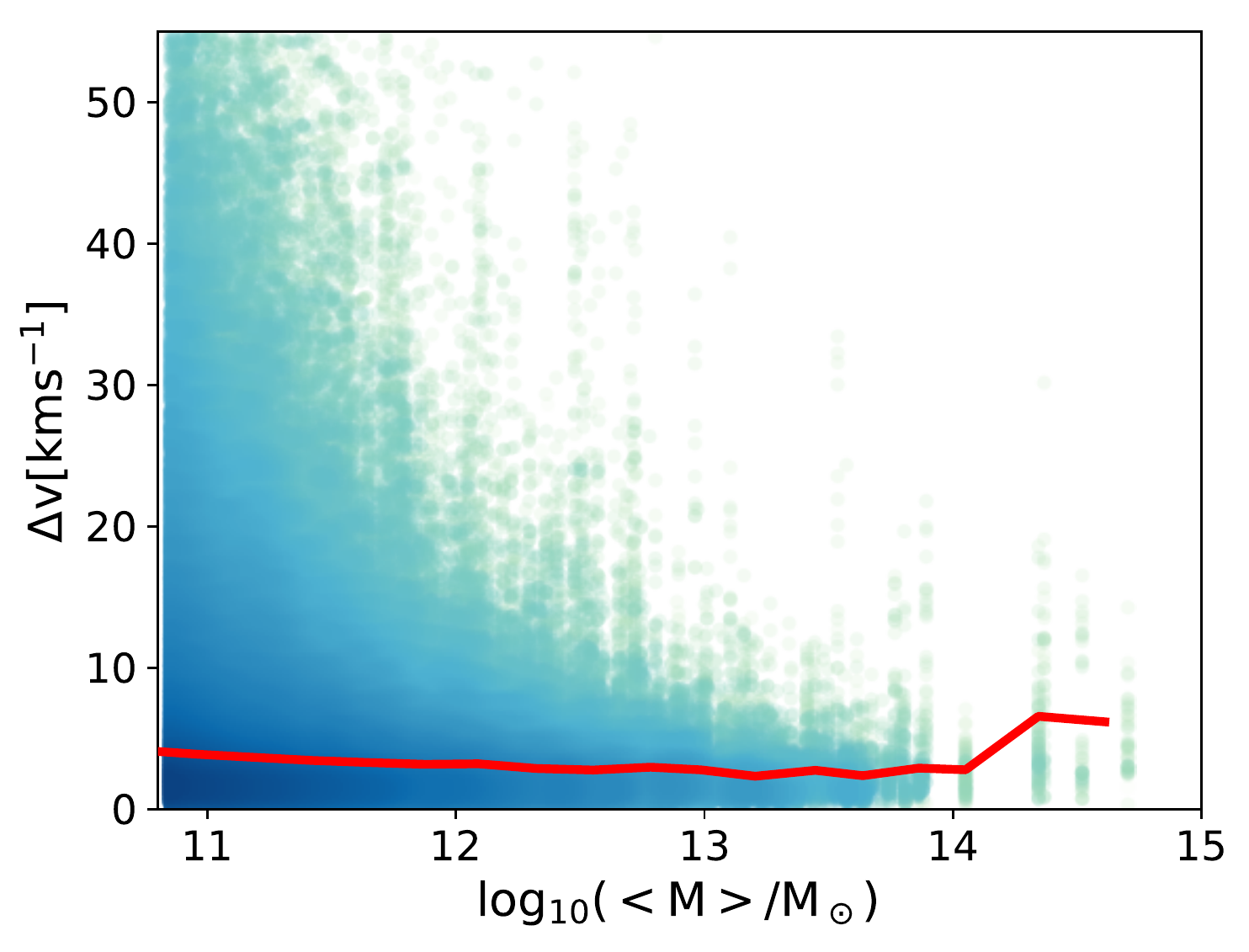}
    \put(55,65){\makebox[0pt] {\textcolor{black}{$\LL=22$}}} 
    \end{overpic} &
    \begin{overpic}[width=0.33\textwidth]{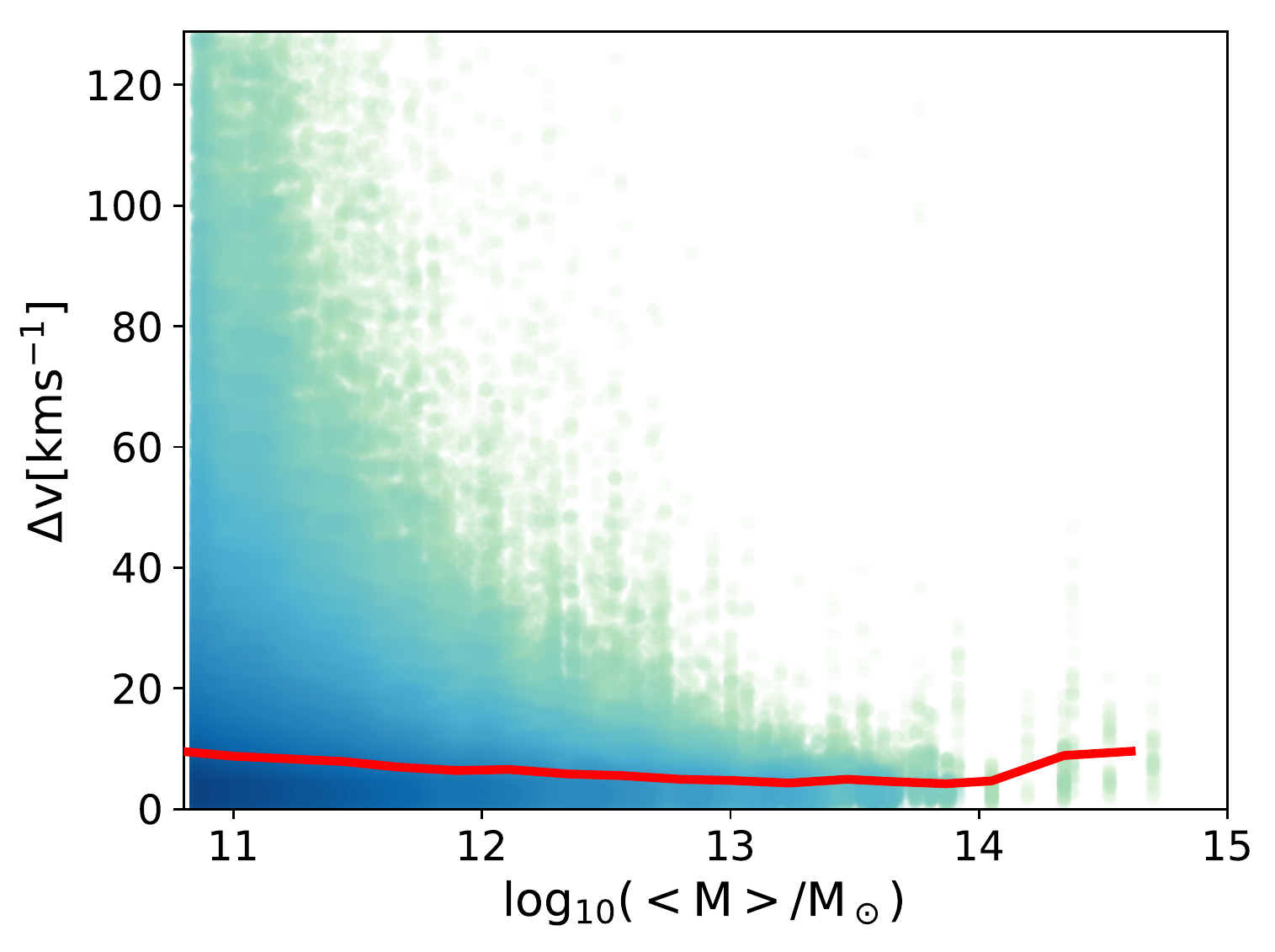}
    \put(55,65){\makebox[0pt] {\textcolor{black}{$\LL=21$}}} 
    \end{overpic} \\
    \begin{overpic}[width=0.33\textwidth]{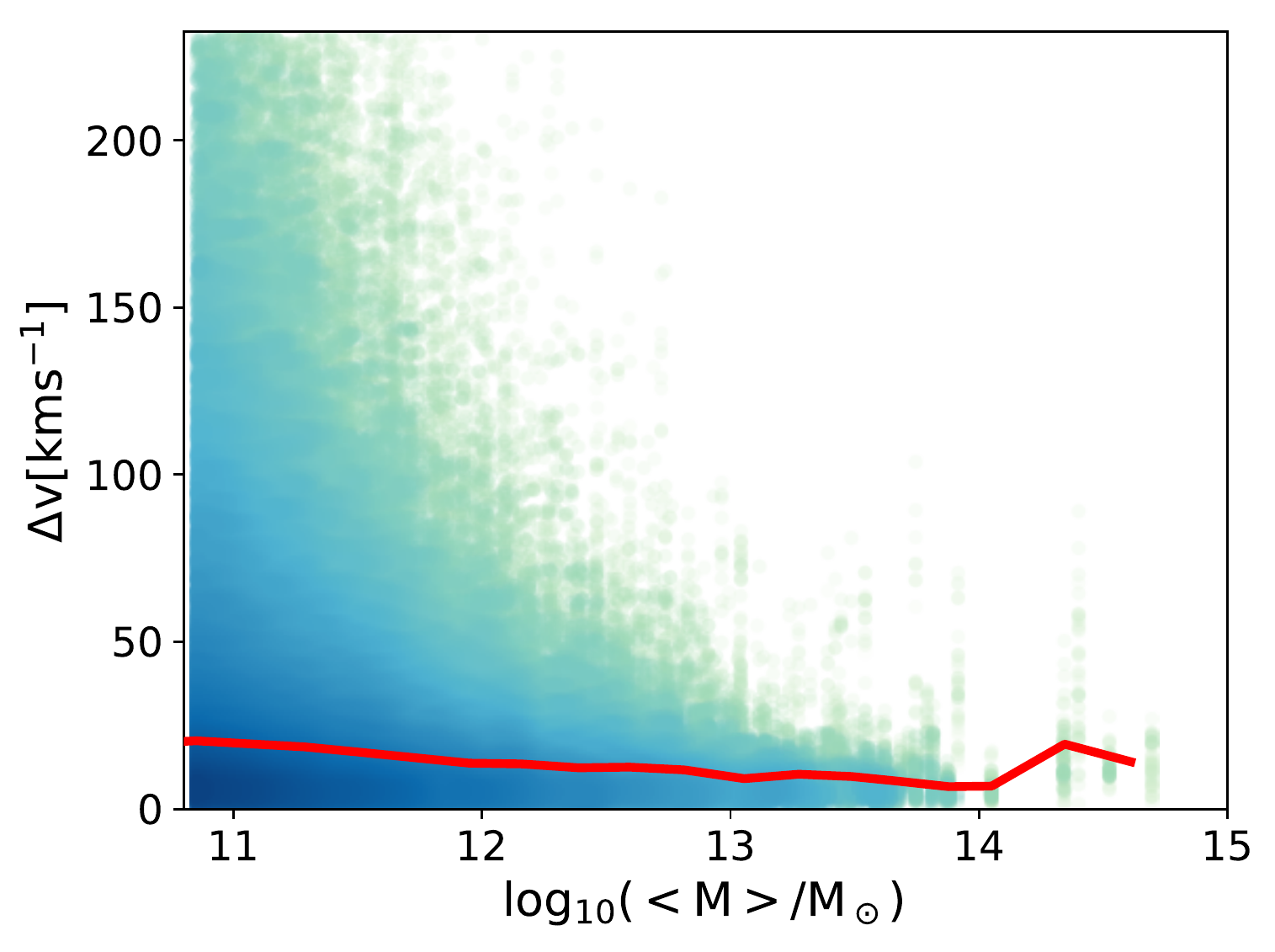}
    \put(55,65){\makebox[0pt] {\textcolor{black}{$\LL=20$}}} 
    \end{overpic} &
    \begin{overpic}[width=0.33\textwidth]{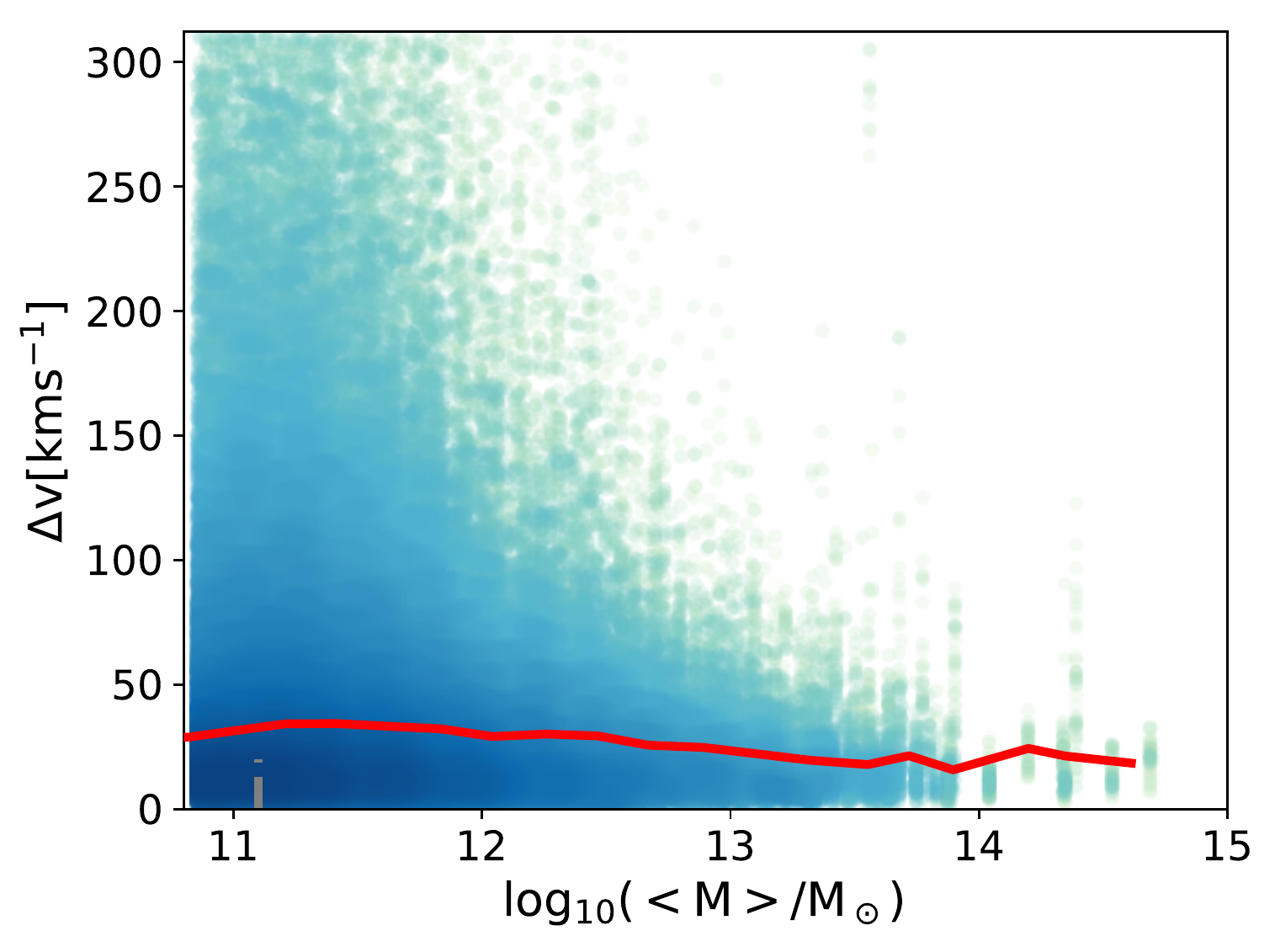}
    \put(55,65){\makebox[0pt] {\textcolor{black}{$\LL=19$}}} 
    \end{overpic} &
    \begin{overpic}[width=0.33\textwidth]{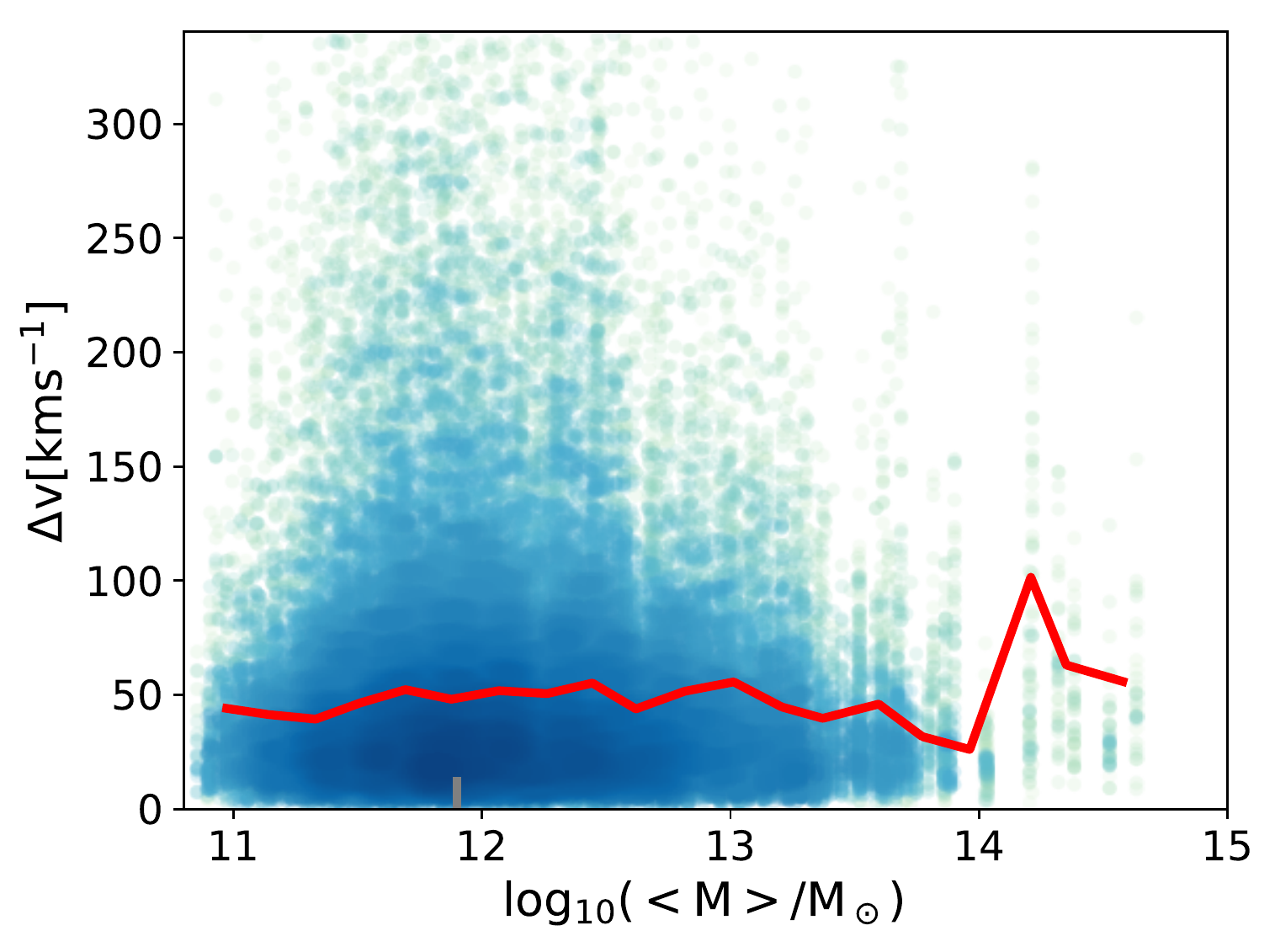}
    \put(55,65){\makebox[0pt] {\textcolor{black}{$\LL=18$}}} 
    \end{overpic} \\
  \end{tabular}
    \caption{Velocity offset of matched central haloes in the variant
      simulation relative to the average position, as a function of
      average mass (shown on the x-axis), and for different levels
      (indicated on each panel). The red lines show one standard deviation in position among variants. Vertical dashed lines on panels at $\LL=18$ and 19 indicate the minimum mass of haloes that {\it exist} at these level. On each panel, the y-axis extends to
      the 99th percentile for all matched haloes.}
    \label{fig:plots/dv_scatters}
\end{figure*}

\begin{figure}
    \begin{overpic}[width=\columnwidth]{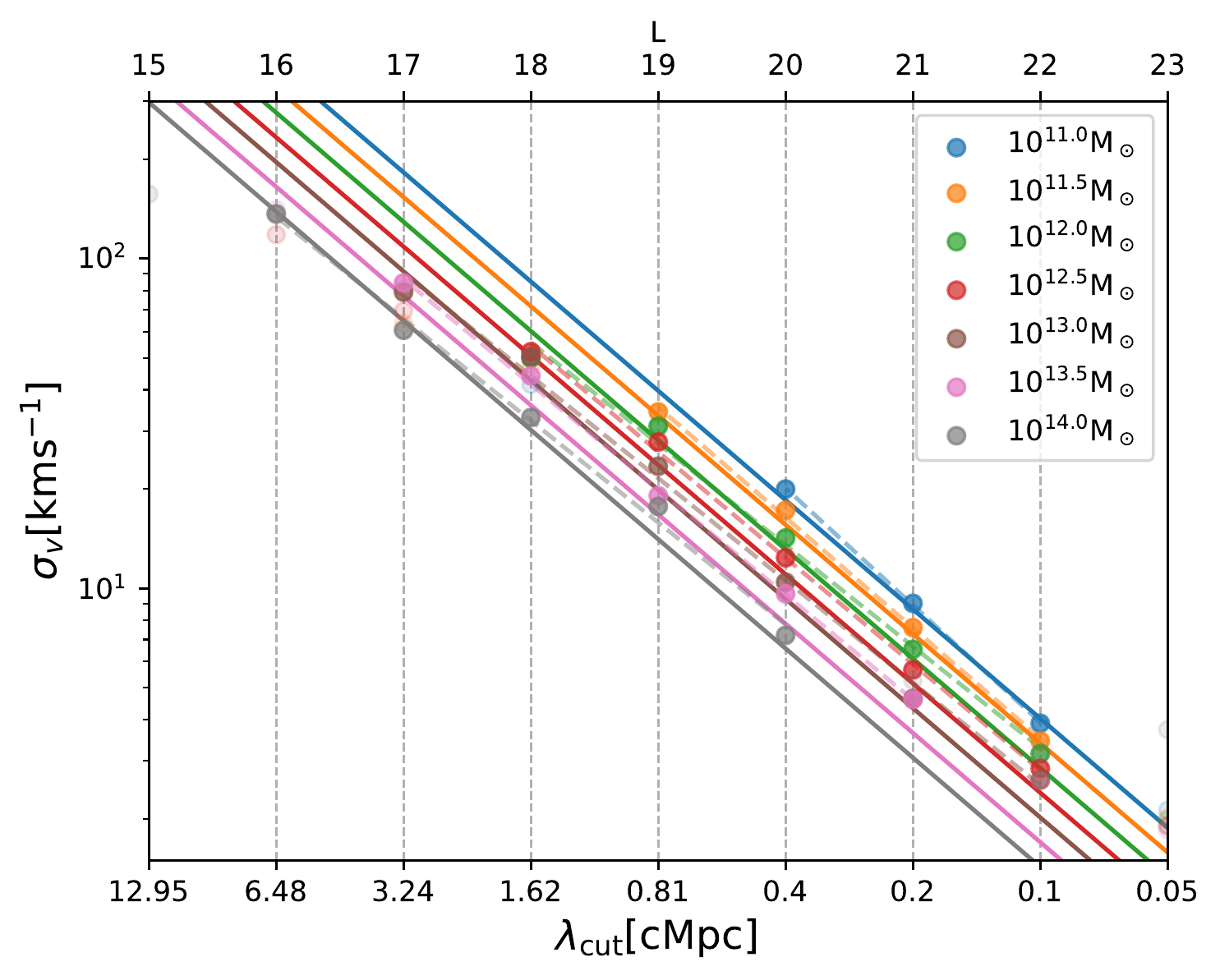}
    \end{overpic}
    \caption{Magnitude of the average velocity difference, relative to the average velocity across variants, for central haloes of different mass (as indicated by colours), and as a function of level of randomisation in the initial density field (L, upper x-axis), and corresponding cut-off wavelength ($\lc$, lower x-axis).}
    \label{fig:velocities-synthesis}
\end{figure}

\subsection{Mass}\label{sec:results:properties:mass}
Fig.~\ref{fig:mass-ratios} shows the change in mass of individual, matched haloes, relative to the median mass across variants, and as a function of median mass, for variations at levels 18 to 23. On every panel, each individual halo can appear from 20 to 40 times; haloes with fewer than 20 matches are excluded by the "majority vote" criterion described in Section~\ref{sec:results:individual}. The range on all panels covers mass ratios between $1:3$ and $3:1$. It is worth remembering that matches with mass ratios outside this range relative to the Reference simulation are excluded. While haloes can, in principle, have mass ratios of nearly $1:9$ relative to the {\it median} mass, the paucity of points approaching the limits of the range indicates that such large deviations in mass are very rare.

We find that, at a given halo mass and level of variation, the halo masses of variants are approximately log-normally distributed. Consequently, we quantify the scatter\footnote{We refer to the "scatter" of a quantity, $x$, as either the standard deviation, $\sigma_x =\sigma(x)$, or the standard deviation of its logarithm, $\sigma^*_x = \sigma\left(\mathrm{log_{10}}(x)\right)$.} in mass across the variants for each halo as
\begin{equation} \label{eqn:mass:scatter:definition}
\sigma^*_\MM = \sigma(\log_{10} \MM).
\end{equation}


The red bands on each panel of Fig.~\ref{fig:mass-ratios} indicate $\sigma*_\MM$, in bins equally spaced in $\mathrm{log(<M>})$. It can be seen that, at each level, the scatter increases with decreasing halo mass. Comparing different panels, it can also be seen that the relative variance in mass increases as the scale of the variation in the initial density field is increased from $\LL=23$ to $\LL=18$. For haloes of a given mass, the mass varies more for lower level variations.

This behaviour is summarised across masses and levels in  Fig.~\ref{fig:massratio-synthesis}, which shows $\sigma^*_\MM$ as a function of level and $\lc$. Different coloured points show the scatter measured from our simulations for haloes of different mass, from $10^{11}\Ms$ (blue) to $10^{14}\Ms$ (grey). To each set of data points, we have fitted linear relationships of the form
\begin{equation} \label{eqn:mass}
    \sigma^*_\MM = a_{\MM} \lc + b_{\MM}
\end{equation}
which are shown in Fig.~\ref{fig:massratio-synthesis} by dashed lines of corresponding colours, with coefficients $a_{\MM}$ and $b_{\MM}$ depending on mass.

Because haloes of a given mass drop out of {\it existence} below the level $\LE(M)$, given by Eqn.~\ref{eqn:existence}, data points below $\LE(M)$ are excluded from the fit. We also exclude haloes at $\LL=23$, where we find a slight upturn in $\sigma^*_\MM(L)$ relative to the fit at all masses. We attribute this to the finite spatial resolution of our simulations. 

We also find a universal relation for the mass scatter, with a value of $a = a\MM=1.15$ for all mass bins, and a mass-dependence for $b_\MM = 4.6 - 0.445~\mathrm{log}_{10}(\MM/\Ms)$, i.e.

\begin{equation} \label{eqn:mass:universal}
    \sigma^*_\MM = 4.6 + 1.15 \lc -0.445~\mathrm{log}_{10}(\MM/\Ms).
\end{equation}
Fits to this universal relation are shown by solid lines in Fig.~\ref{fig:massratio-synthesis}.

From the observed regularity, we expect that the relations can be extrapolated both to higher mass haloes and to smaller scales. However, it should not be extrapolated to levels below the existence scale for a given halo mass.

Comparing $\sigma^*_\MM(L)$ to the existence scale, $\LE$, we find that the scatter in halo mass typically reaches $\sigma^*_\MM \sim 0.23$, or a factor of $\sim 1.7$ in mass before haloes drop out of existence. Scatters in mass of $\sim 10\%$ ($\sigma^*_\MM = 0.041$) or $\sim 1\%$ ($\sigma^*_\MM = 0.0043$) are found for variations at 2 and 5 levels above the existence level $\LE$, respectively.

\subsection{Concentration}\label{sec:results:properties:concentration}
In addition to the mass, a second parameter is required to characterise a $\Lambda$CDM halo whose density is described by an NFW profile \citep{Navarro-1996}, 
\begin{equation} \label{eqn:nfw}
    \rho = \frac{\rho_s}{ (r/r_s)(1 + r/r_s)^2},
\end{equation}
where $r_s$ and $\rho_s$ are the scale radius and characteristic density, respectively. The mass is commonly complemented by the concentration, $c$, as a second parameter, defined through the equation 
\begin{equation} \label{eqn:concentration}
\delta_c = \frac{200}{3} \frac{c^3}{ \mathrm{ln}(1+c) - c / (1+c) } ,
\end{equation}
where $\delta_c$ is the characteristic overdensity.

Following~\cite{Springel-2008}, we calculate $c$ from the measured values of the maximum circular velocity, $v_\mmax$, and its corresponding radius, $r_\mmax$, from which the overdensity inside $r_\mmax$ can be obtained via:
$$
\delta_v = \frac{\rho(r_\mmax)}{\rho_{\mathrm{crit}}} = 
2 \left(\frac{v_\mmax}{\mathrm{H}_0~r_\mmax} \right)^2
$$

For an NFW halo, the characteristic overdensity, $\delta_c$, of Eqn.\ref{eqn:concentration} is related to the overdensity inside $r_\mmax$,  $\delta_v$, via the relation
$$
\delta_c = \frac{\rho_s}{\rho_{crit}} = 7.213~\delta_v .
$$

We can thus calculate the concentration parameter, $c$, of each halo from the measured values of $r_\mmax$ and $v_\mmax$ by computing $\delta_v$ and $\delta_c$, and solving Eqn.~\ref{eqn:concentration} for $c$.

Because stripping of satellites affects $r_\mmax$ disproportionately relative to $v_\mmax$,  Eqn.~\ref{eqn:concentration} is not expected to hold for satellites, and we limit our analysis to centrals only. Furthermore, measuring the concentration of a halo requires sufficient numerical resolution to resolve the density inside $r_\mmax$. As \cite{Power-2003} showed, even in an accurate simulation, this is sensitive to a halo's particle number, and we limit our analysis of concentration to haloes containing at least 1000 particles, or $7.4\times10^{11}\Ms$.

\cite{Neto-2007} showed that concentration parameters for relaxed haloes of a given mass are approximately log-normal distributed, with $\sigma^*_c \sim 0.11$ at $10^{12}\Ms$. Fig.~\ref{fig:plots/concentration_scatter} shows the relative scatter in concentration of matched central haloes, as a function of median mass, and for variations at levels 18 to 23. The red band shows  $\sigma^*_c$.

In contrast to the scatter in halo mass, there is a noticeable scatter in concentration already at the smallest scale of variation. For example, the scatter in concentration of individual $10^{13}\Ms$ halos at $\LL=22$ is $8\%$, while the scatter in mass is $<1\%$ for the same variation. This reflects the fact that the scale that determines the concentration of a halo, $r_\mmax$ is much smaller than its total size, and we find a much greater scatter in $r_{s}$ than in $r_{200}$ for the same haloes.

While the measurable scatter in total mass is bounded only by the matching criteria, the scatter in concentration is naturally bounded from above by the narrow range of concentrations for $\Lambda CDM$ haloes. As expected, we find that for large variations, the scatter of matched haloes is similar to the scatter of independent haloes reported by \cite{Neto-2007}.

Fig.~\ref{fig:concentration-synthesis} shows $\sigma^*_c$ for different masses, and different scales of variation. As expected, we find that the scatter increases with decreasing halo mass, and with the scale of variations. As shown by dashed lines. Over the range we can resolve, and up to the maximum value of $\sim0.1$, $\sigma^*_c$ appears to scale linearly with $\lc$, and also show a regular scaling with mass. However, our simulations lack the dynamic range necessary to extrapolate universal scaling relations.

\subsection{Position} \label{sec:results:properties:displacements}
Fig.~\ref{fig:displacements} shows the relative displacement, $\Delta_r$, of individual, matched haloes at $z=0$ from their median position, as a function of their median mass, and for variations at levels 18 to 23. Unlike in Fig.~\ref{fig:mass-ratios}, the range on each panel is adjusted to include $99\%$ of points in every case.

We calculate the scatter in position as 
\begin{equation}
\sigma_r = \sqrt{\sigma_x^2 + \sigma_y^2 + \sigma_z^2} 
\end{equation}
where ${r}(x,y,z)$ are the Cartesian coordinates across the variants of the halo, and $\sigma_x, \sigma_y$ and $\sigma_z$ are the 1D-dispersions in position.

In Fig.~\ref{fig:displacements}, $\sigma_r$ is shown by a red line. At each level, the average displacement increases with decreasing halo mass: for fixed variations of the initial density field, lower mass haloes are displaced more than high mass haloes. Comparing different panels, it can also be seen that the displacement of haloes of the same mass increases as the scale of the variation in the initial density field is increased: haloes of a given mass are more displaced for larger scale variations. At $\LL=19$ and $\LL=18$, the mass-dependence appears to flatten at low masses. However, in both cases, this coincides with the limit in halo mass for unique haloes to exist, as indicated by the vertical dashed lines; haloes below this limit have a low probability to form across variants.

The dependence of average displacement on halo mass and level is illustrated in Fig.~\ref{fig:displacement-synthesis}. Different coloured symbols show the scatter measured from our simulations for haloes of different mass, from $(10^{11}\Ms)$ (blue) to $10^{14}\Ms$ (grey). We find the average displacement of haloes in each mass bin to be a power-law function of $\lc$ of the form
\begin{equation} \label{eqn:displacement}
    \mathrm{log}_{10}(\sigma_r / \cMpc) = a_\MM ~\mathrm{log}_{10}(\lc / \cMpc) + b_\MM.
\end{equation}
Dashed lines Fig.~\ref{fig:displacement-synthesis} show separate fits at each mass, when the same limits to the domain as in Section~\ref{sec:results:properties:mass} are applied. As in Eqns.~\ref{eqn:fmatch} and ~\ref{eqn:mass}, we find a self-similar behaviour, with a universal slope of $a = a_\MM = 1.5$, and a mass-dependence of $b_\MM = 3.55 - 0.31~\mathrm{log}_{10}(\MM/\Ms)$. Solid lines show fits with these parameters of $a$ and $b_\MM$ at every level.

The increase in the average displacement of haloes with the scale of the perturbations follows from the fact that the position of an object depends on the gravitational potential of the surrounding density field, and greater variation in surrounding structure leads to greater variation in the potential, and thus the position of the halo. The mass-dependence at a given scale of variation follows from the fact that less massive objects are sensitive to smaller external perturbations of the potential than more massive ones. 
Comparing the mass- and scale-dependencies of $\sigma_r$ to those of $\sigma^*_\MM$, we find a stronger scale-dependence for $\sigma_r$, and a stronger mass-dependence for $\sigma^*_\MM$.

\subsection{Velocity}\label{sec:results:properties:velocity}
By analogy to Fig.~\ref{fig:displacements}, Fig.~\ref{fig:plots/dv_scatters} shows the scatter in velocity of individual, matched haloes at $z=0$, as a function of their median mass, for variations at levels 18 to 23. The range on each panel is adjusted to include $99\%$ of points in every case.

The velocity dispersion, $\sigma_v$, of individual haloes is defined as the dispersion in peculiar velocities of its matched variants,
\begin{equation} \label{eqn:velocity-dispersion}
\sigma_v = \sqrt{\sigma(v_x)^2 + \sigma(v_y)^2 + \sigma(v_z)^2},
\end{equation}
where $\sigma(v_x)$, $\sigma(v_y)$ and $\sigma(v_z)$ are the corresponding 1D-velocity dispersions.

The median velocity scatter among matches, as a function of halo mass and level, is illustrated in Fig.~\ref{fig:velocities-synthesis}. Different coloured symbols show the scatter measured from our simulations for haloes of different mass, from $10^{10.5}\Ms$ (blue) to $10^{10.5}\Ms$ (grey). We find the familiar power-law behaviour of Eqns.~\ref{eqn:mass} and ~\ref{eqn:displacement}, and parametrise the velocity offset as

\begin{equation} \label{eqn:velocity}
    \mathrm{log}_{10}(\sigma_v / (\mathrm{kms}^{-1})) = a_\MM~\mathrm{log}_{10}(\lc/\cMpc) + b_\MM.
\end{equation}

Dashed lines show separate fits at each mass, solid lines show fits assuming a fixed slope at every level. The same cuts to the domain as in Section~\ref{sec:results:properties:mass} and Section~\ref{sec:results:properties:displacements} are applied. Solid lines assume a fit to Eqn.~\ref{eqn:velocity} with a universal slope and regular mass dependence, and we find values of $a_\MM=1.1$ and a mass-dependence of $b_\MM=3.35 - 0.15  ~\mathrm{log}_{10}(\mathrm{\MM/\Ms})$.

The velocity dispersion of halo variants increases as the scale of the variation in the initial density field is increased, and also scales inversely with halo mass. Comparing the mass- and scale-dependencies of $\sigma_v$ to those of $\sigma_r$, we find an even stronger scale-dependence and even weaker mass-dependence for $\sigma_v$. This can be explained because the velocity of a halo is set not by its internal mass (or even the density of its environment), but by larger scale tidal fields. Haloes respond to a change in environment, but the mass-dependence of the velocity on halo mass appears much weaker than for both the mass scatter and displacement.

\begin{figure}
     \begin{overpic}[width=0.156\textwidth]{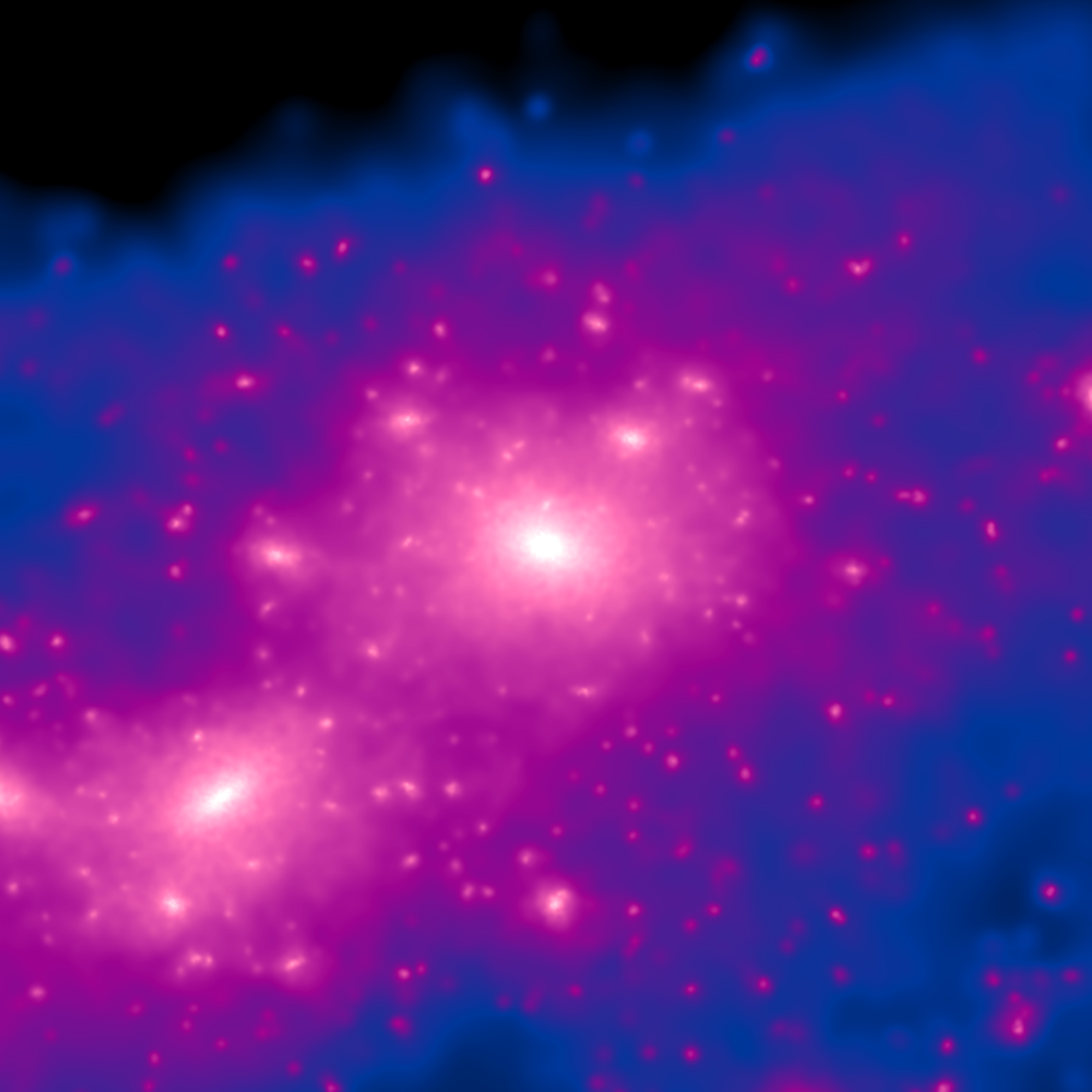}
     \put(48,5){\textcolor{white}{\bf Reference}}
     \end{overpic} 
     \begin{overpic}[width=0.156\textwidth]{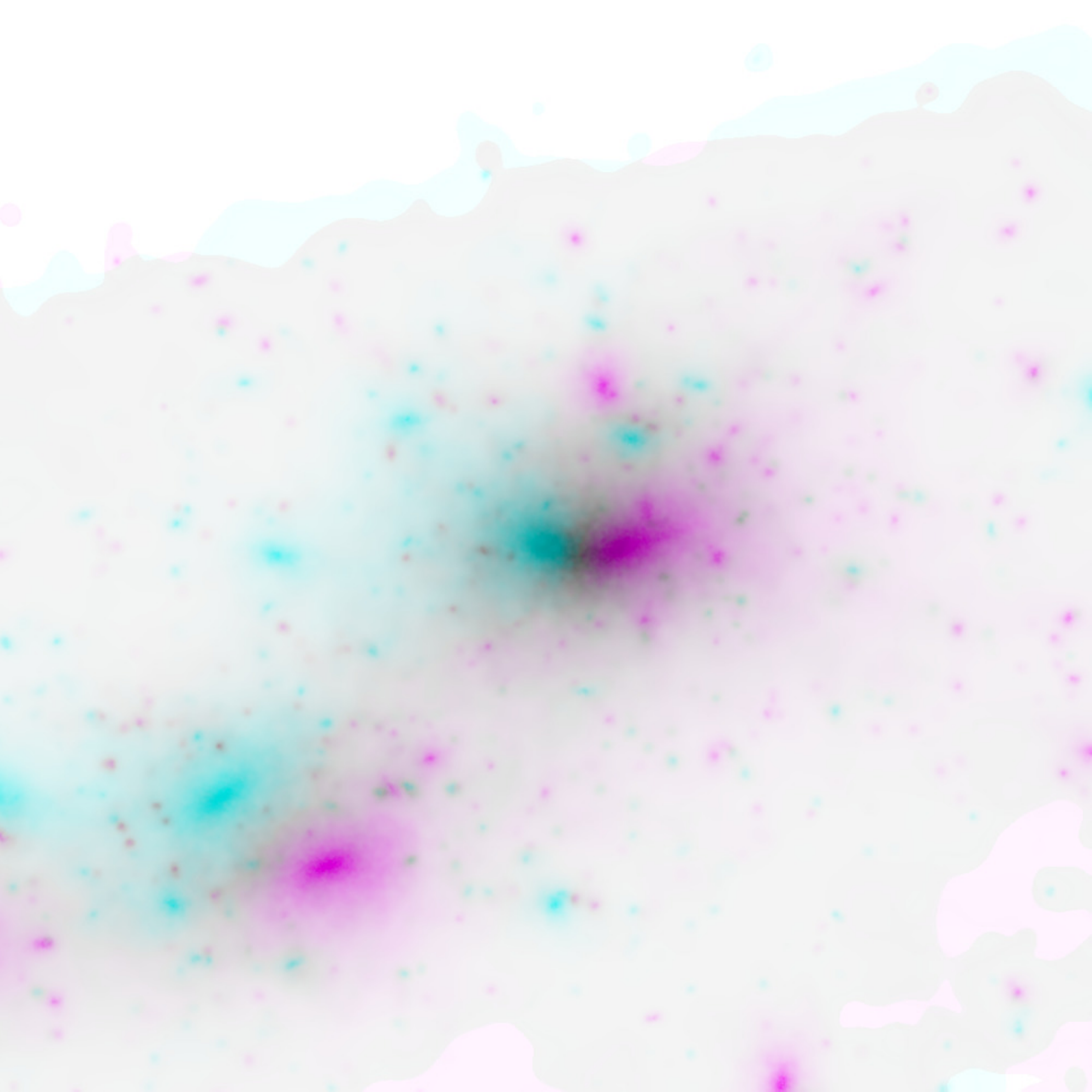}
     \put(2,5){\textcolor{black}{Reference vs. V18$_1$}}
     \end{overpic}
     \begin{overpic}[width=0.157\textwidth]{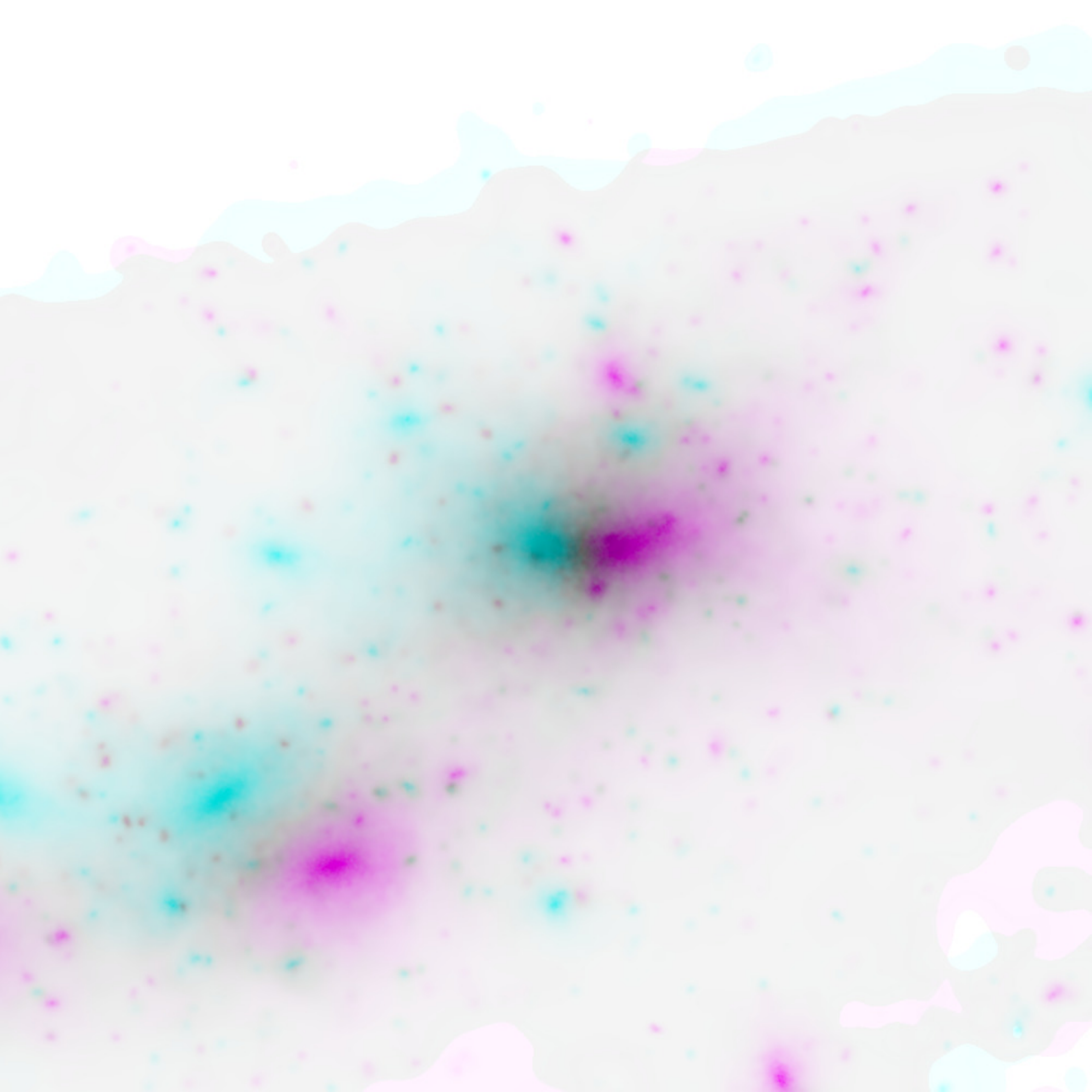}
     \put(2,5){\textcolor{black}{Reference vs. V$18_1/21_2$}}
     \end{overpic}
      \begin{overpic}[width=0.156\textwidth]{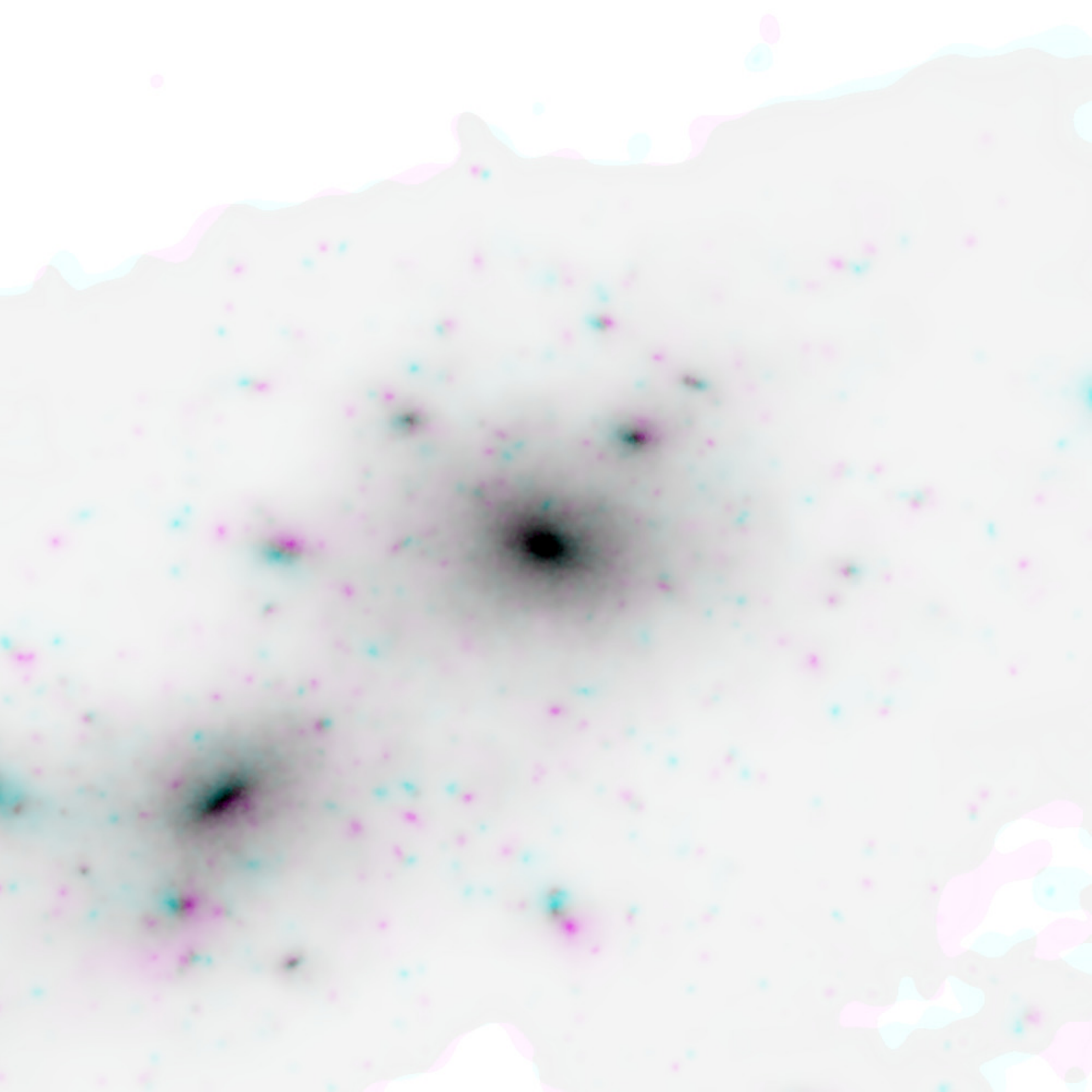}
     \put(2,5){\textcolor{black}{Reference vs. V21$_1$}}
     \end{overpic} 
     \begin{overpic}[width=0.157\textwidth]{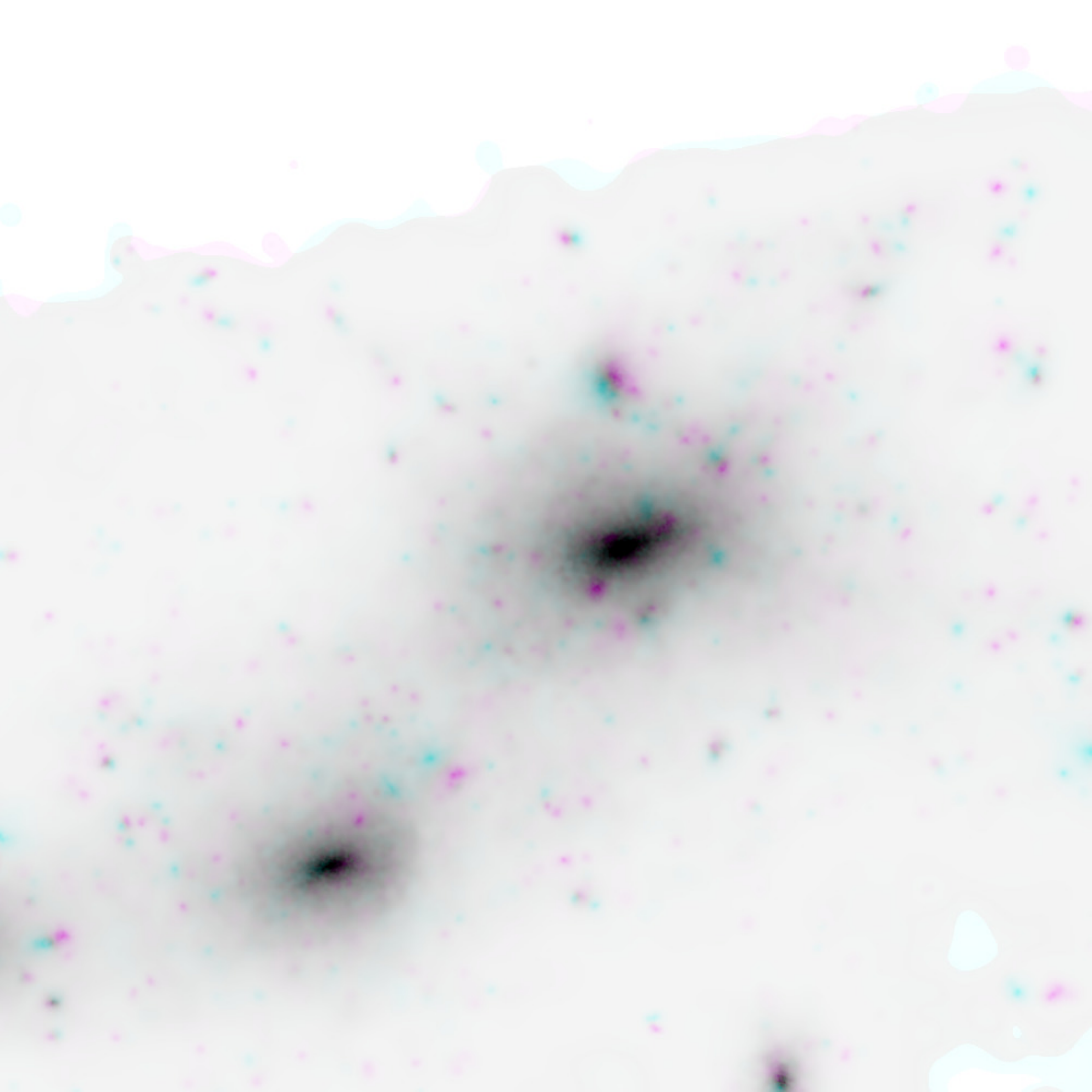}
     \put(15,5){\textcolor{black}{V18$_1$ vs. V$18_1/21_2$}}
     \end{overpic}
     \begin{overpic}[width=0.157\textwidth]{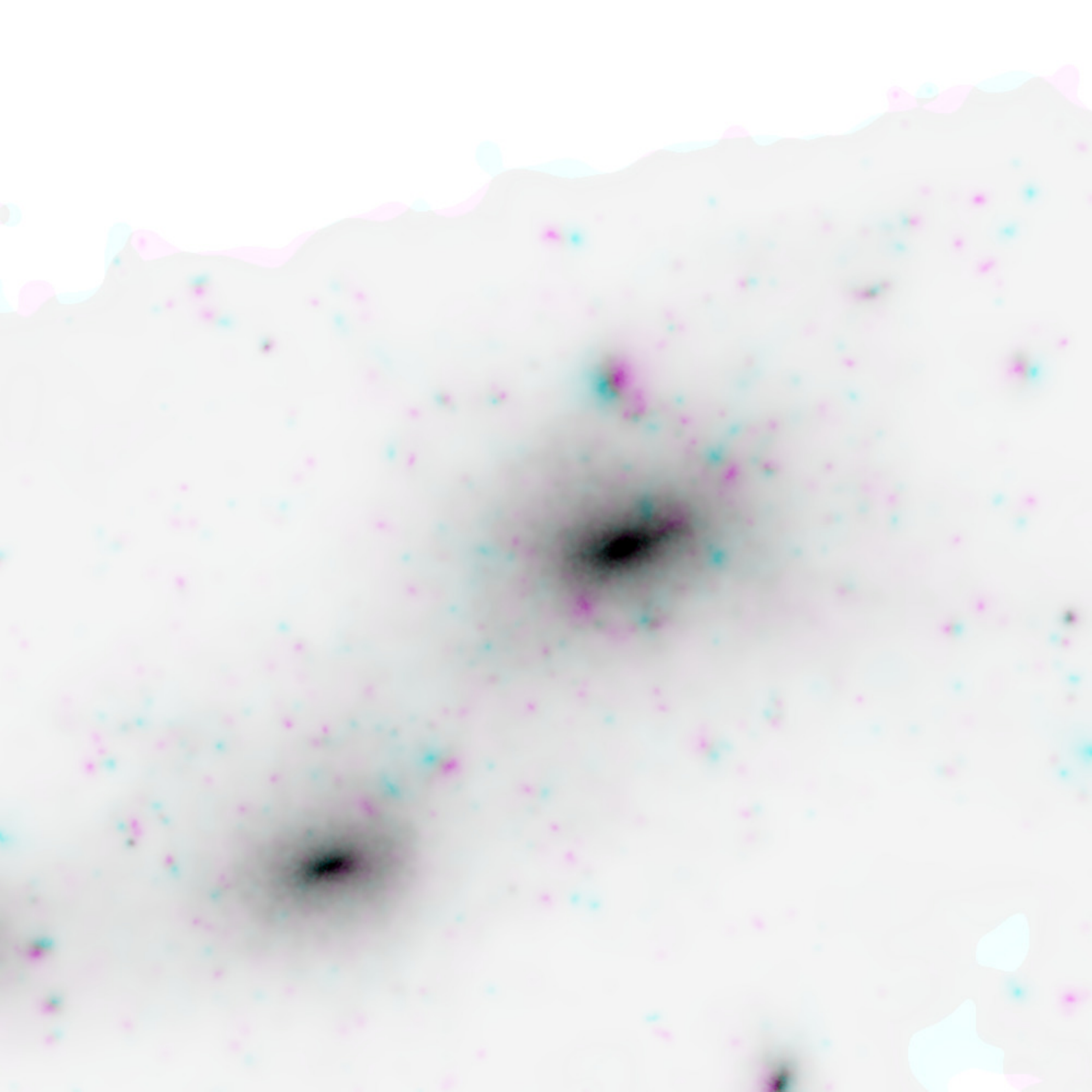}
     \put(2,5){\textcolor{black}{V18$_1$ vs. V$18_1/21_3$}}
     \end{overpic}
    \caption{Top left: dark matter density in the most massive cluster in the Reference simulation at $z=0$. Analogous to Fig.~\ref{fig:structure-levels}, the remaining panels show differences in the projected density in the same Eulerian volume. The top centre, top right, and bottom left panels show the difference relative to the reference simulation, for variants V$18_1$ and V$18_1/21_2$ (top middle and right), and V$21_1$ (bottom left). As expected, the differences in density relative to the reference simulation are similar in V$18_1$ and V$18_1/21_2$, which use the same shift on levels $\LL=18$ to $20$. By comparison, the difference between V$21_1$ and the reference simulation is much smaller. The bottom centre and bottom right panels show the density relative to V$18_1$ for variants V$18_1/21_2$ and V$18_1/21_3$, respectively. While their differences relative to the reference simulation are similar in magnitude to that of V$18_1$, their differences relative to V$18_1$ are much smaller, and similar in magnitude to that between V$21_1$ and the reference simulation.}
    \label{fig:cluster}
\end{figure}

\begin{figure}
    \begin{overpic}[width=\columnwidth]{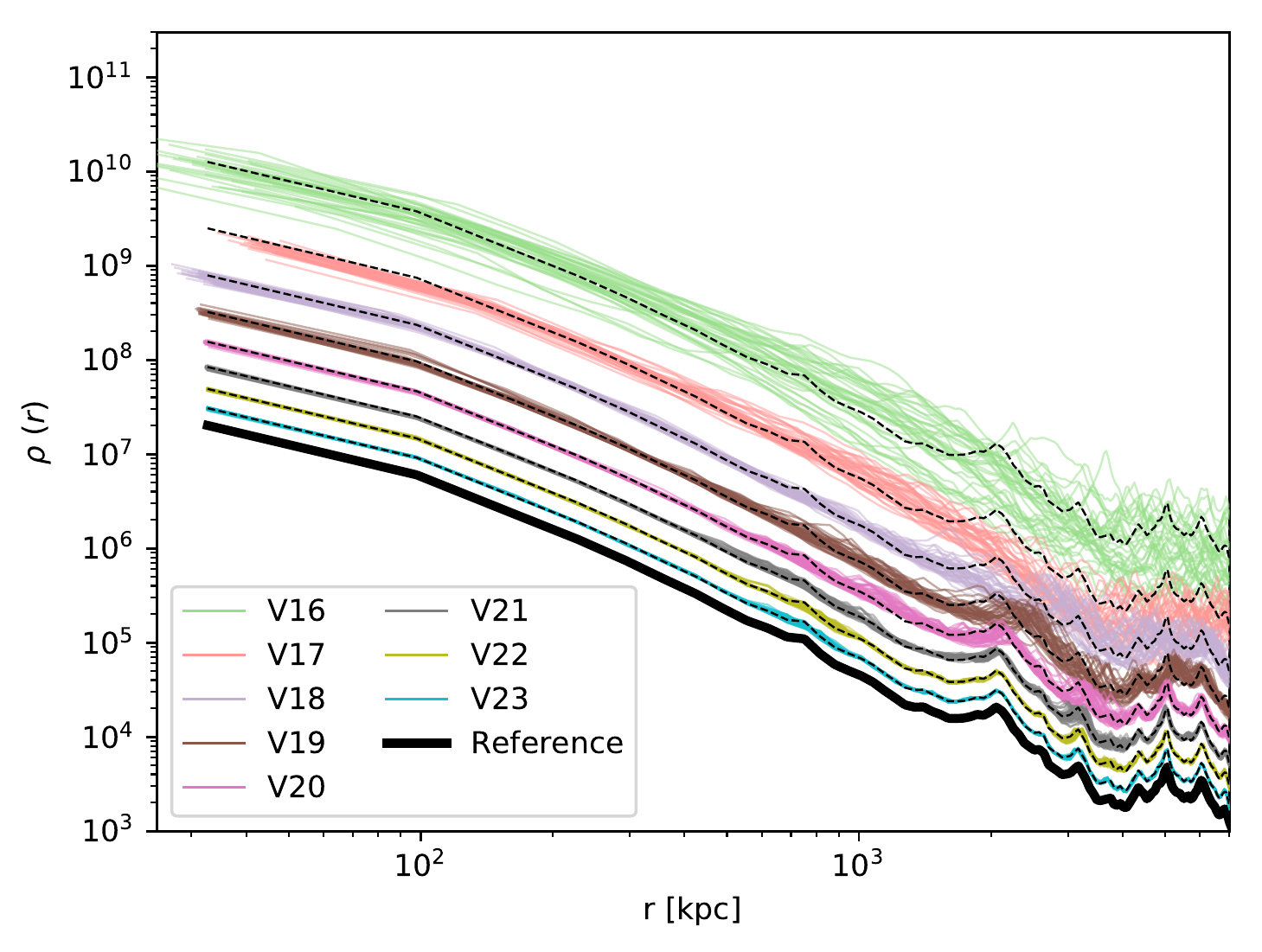}
    \end{overpic}
    \caption{Density profile of the most massive cluster at z=0. The
      thick black line shows the mass profile in the Reference
      simulation, while lines of different colour show up to 39
      matched haloes at each level from 16 to 23. For clarity, the
      lines for each level are offset, and the thin black line
      repeats the Reference simulation with the corresponding offsets.}
    \label{fig:cluster-density}
\end{figure}

\begin{figure}
    \begin{overpic}[width=\columnwidth]{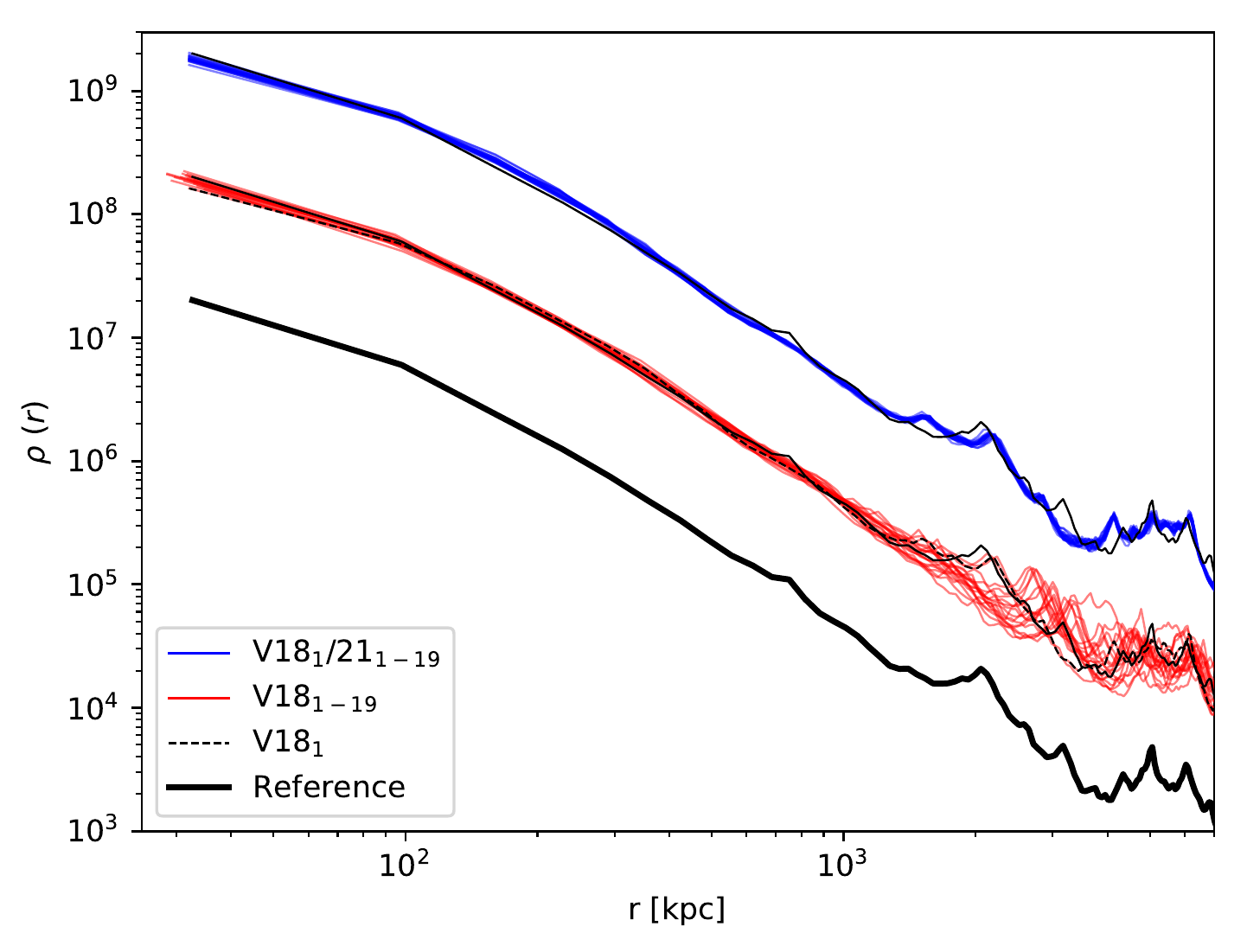}
    \end{overpic}
    \caption{Density profile of the most massive cluster at z=0, similar to Fig.~\ref{fig:cluster-density}. The
      thick black line shows the mass profile in the Reference
      simulation. Brown lines show 19 variations at level 18, offset for clarity. The Reference simulation and variant V$18_1$ are repeated as thin solid and dashed lines black, respectively. Blue lines show 19 additional variations of V$18_1$ at $\LL=21$: V$18_1/20_2$ to V$18_1/21_20$. It can be seen that these result in small perturbations, not of the cluster in the Reference simulation, but of the variation of cluster in V$18_1$.}
    \label{fig:cluster-density-variants}
\end{figure}

\section{Variations of a single halo} \label{sec:results:example}
In this section, we show how the introduction of random variations on different scales can create targeted variations of a particular object, either very similar to, or very different from the original object. In addition, we show how introducing an additional set of higher level variations can create small additional perturbations to an existing variant.

As an example, we choose the most massive halo of the Reference simulation at $z=0$, with a mass of $2.2\times 10^{14}\Ms$, whose density map is shown in the top left panel of Fig.~\ref{fig:cluster}, and which is comparable in mass to the Virgo cluster \citep{Urban-2011}.

Galaxy clusters are a frequent target for zoom-in simulations, e.g. by \cite{Eke-1998}, and more recently by \cite{Borgani-2002, Kay-2004, Nagai-2007, Martizzi-2016, Bahe-2017}. \cite{Schaller-2015c} and \cite{Barnes-2017} both analysed clusters from the same {\sc Eagle} volume that defines our reference simulation. An inherent challenge in these studies is that the largest clusters are, by definition, rare objects in any cosmological volume, limiting the predictive power of a simulation. For example, \cite{Schaller-2015c} conclude that the presence of cores in observed clusters is an outstanding problem that requires a larger samples of simulation counterparts, which our method of generating variants of existing objects may be able to provide.

\subsection{Simple Variations}
In Fig.~\ref{fig:cluster-density}, we show the density profile of the same cluster in the Reference simulation (black line) and, offset for clarity, for up to 39 matches at each level from 16 to 23. At small scales (V23 and V22), there is very little change in the density profile, both for the main halo, as well as for the identifiable substructures. At levels 21 and 20, the density in the centre of the main halo remains similar, but the positions of the larger substructures change, resulting in some scatter in the outer density profile. Below level $20$, the central density and mass of the main halo change throughout.

At levels $\LL=20$, 18, and 16, the scatter in mass is $1\%$, $10\%$ and $60\%$, respectively, while the scatter in concentration is $9\%$, $20\%$, and $35\%$. Variations below level 20 yield clusters with very similar mass and concentrations, but different individual subhaloes; variations at level 18 yield clusters with very similar mass, but different concentration (and different subhaloes, apart from the largest ones), and variations at level 16 result in clusters with different total masses and concentrations, and completely different members.

\subsection{Higher Level Perturbations}
As discussed in Section~\ref{sec:methods:ICs:octree} and illustrated in the rightmost panel of Fig.~\ref{fig:shifts}, we can also combine multiple shifts at different levels in the same simulations. As an example, we have made a set of additional simulations, in which we shift the phase information by 1 at levels 18 to 20, and by 1-19 at levels 21 and above. We label this new set of simulations "V$18_1/21_{[1..19]}$" and note that V$18_1/21_1$ is identical to V$18_1$.

In Fig.~\ref{fig:cluster}, we show the changes resulting from these higher level perturbations, by comparing the projected density in higher level perturbations at $\LL=21$ to both the reference simulation, and to the variant V$18_1$  of the same cluster. As expected, from the top centre and top right panel, it can be seen that compared to the Reference simulation, V$18_1$ and V$18_1/21_2$ show the same amount of difference, while V$18_1$, shown on the bottom left, is much more similar to the Reference simulation. 

The bottom middle and right panel show the density relative to V$18_1$ for variants V$18_1/21_2$ and V$18_1/21_3$, respectively. While their differences relative to the reference simulation are similar in magnitude to that of V$18_1$, their differences relative to V$18_1$ are much smaller, and similar in magnitude to that between V$21_1$ and the reference simulation.

Similar to Fig.~\ref{fig:cluster-density}, in Fig.~\ref{fig:cluster-density-variants}, we show the density profiles of the cluster in the 19 variants V$18_{1-19}$ (red lines), and the 19 variants V$18_{1}/21_{1-19}$ (blue lines), with offsets for clarity. It can be seen that the scatter among the density profiles for V$18_{1}/21_{1-19}$ is much reduced compared to the scatter among V$18_{1-19}$, and furthermore, that the individual variants of  V$18_{1}/21_{1-19}$ scatter around the variation V$18_{1}$ (dashed black line), rather than around the Reference simulation (solid black line).

Variations of a single object by random variations of the initial density field at scales below the existence scale can be used deliberately, to facilitate the study of rare objects. Variations just below the existence scale could be used to create a diverse set of massive clusters using only zoom-in simulations, without requiring to first simulate a much larger simulation volume. By introducing secondary variations at level 21 to an existing variant at level 18, we can create multiple, smaller scale perturbations around existing objects. This offers many possible applications: for example, from a cluster with moderate concentration, we can first create a variant with high concentration, and then create multiple high-concentration clusters with separate satellite populations.

It is also worth noting that we have applied the shift across the entire volume. This is not necessary, the real-space localisation of our basis allows us to independently vary spatial sub-volumes of the random density field. In a forthcoming paper, we will show how this technique can be used to create a faithful reproduction of the Local Group embedded within the observed constraints. I particular, it allows the generation of accurate Local Group candidates through a series of localised, small-scale perturbations, without affecting the nearby large-scale structure constraints.

\section{Summary and Discussion} \label{sec:summary}
Using cosmological N-body simulations, we have identified the scales of the initial density field that statistically determine the formation of haloes of different mass and their abundance. We have also determined the scales responsible for the existence of particular haloes, and explored how the properties of individual haloes change, as a result of variations on smaller scales.

We have defined a criterion for the existence of unique haloes across simulation volumes: a halo is said to exist at a certain scale, $\LE$, if the information in the primordial density field is sufficient for more than half of the variations at smaller scales still to result in the formation of the same halo. In Eqn.~\ref{eqn:existence}, we have parametrised the existence scale as a function of halo mass.

Beyond the mere existence of haloes, we have also quantified changes in their properties such as mass, concentration, position and velocity, and find power power-law relationships between the scale of the initial variation and the scatter in each of these properties. Furthermore, over the range $10^{11}-10^{14}\Ms$ in halo mass, we find that for each property, a single relationship can be used, when scaled by halo mass, reflecting the self-similarity of hierarchical growth from $\Lambda$CDM initial conditions.

The change in mass shows the strongest mass-dependence. By contrast, there is only a very weak mass dependence to the change in velocity. We attribute this to the fact that, while the mass accretion is set on the same scale that determines the total halo mass, the velocities of haloes are set by the tidal field on larger scales. The small remaining mass dependence can be attributed to 
the degree to which the halo itself influences its own tidal environment.

Taking a cluster-mass halo as an example, we also show that, by varying different scales in the initial conditions, variants of the same object can be produced. Small-scale variations will result in the same cluster, but substructures of different mass, different orbits, or in different phases of their orbit. Larger scale variations will change the overall mass and concentration of the cluster itself. In zoom simulations of particular objects, this can be used to estimate the expected, random change in properties, arising from the fact that the initial density field is more finely sampled. Moreover, changes to the white noise field can also be introduced deliberately to explore regions of the parameter space, which can be useful in the study of rare objects. 

Our method bares similarity to the "genetically modified haloes" approach of \cite{Roth-2016}, which extends the recursive Hoffman-Ribak algorithm \citep{Hoffman-1991} to compute the changes to the initial density field required for the formation of a structure with desired properties, along with the necessary corrections to preserve the nature of the random field. Extending the biological metaphor, our method relies on evolution through fully random mutations, rather than genetic modification. However, because it allows independent control of the scale of variations and of their spatial extent, it can be used to explore the vast space of possibilities very efficiently, while preserving the Gaussian nature of the initial random field at all times.

Our method can be naturally extended in several ways: multiple levels of the white noise field can be varied independently; different spatial regions of the white noise field can be varied independently, and variations smaller in amplitude than a full substitution can be used, all while preserving the Gaussian nature of the initial conditions. In a forthcoming paper, we will use this method to produce a faithful representation of the Local Group embedded in the observed large scale constraints of the Local Universe.

\section*{Acknowledgements}
TS is an Academy of Finland Research fellow. TS and SM are supported by the Academy of Finland grant 314238. PHJ acknowledges the support by the European Research Council via ERC Consolidator Grant (no. 818930).  
CSF acknowledges support from European Research Council (ERC) Advanced 
Investigator grant DMIDAS (GA 786910). This work was also supported by the Consolidated Grant for Astronomy at Durham (ST/L00075X/1).

This work made use the  DiRAC Data Centric system at Durham University, operated by the Institute for Computational Cosmology on behalf of the STFC DiRAC HPC Facility (www.dirac.ac.uk). This equipment was funded by BIS National E-infrastructure capital grant ST/K00042X/1, STFC capital grants ST/H008519/1 and ST/K00087X/1, STFC DiRAC Operations grant ST/K003267/1 and Durham University. DiRAC is part of the National E-Infrastructure. Simulations were also performed on facilities hosted by the CSC - IT Centre for Science, Finland.



\bibliography{paper}

\begin{thebibliography}{}
\makeatletter
\relax
\def\mn@urlcharsother{\let\do\@makeother \do\$\do\&\do\#\do\^\do\_\do\%\do\~}
\def\mn@doi{\begingroup\mn@urlcharsother \@ifnextchar [ {\mn@doi@}
  {\mn@doi@[]}}
\def\mn@doi@[#1]#2{\def\@tempa{#1}\ifx\@tempa\@empty \href
  {http://dx.doi.org/#2} {doi:#2}\else \href {http://dx.doi.org/#2} {#1}\fi
  \endgroup}
\def\mn@eprint#1#2{\mn@eprint@#1:#2::\@nil}
\def\mn@eprint@arXiv#1{\href {http://arxiv.org/abs/#1} {{\tt arXiv:#1}}}
\def\mn@eprint@dblp#1{\href {http://dblp.uni-trier.de/rec/bibtex/#1.xml}
  {dblp:#1}}
\def\mn@eprint@#1:#2:#3:#4\@nil{\def\@tempa {#1}\def\@tempb {#2}\def\@tempc
  {#3}\ifx \@tempc \@empty \let \@tempc \@tempb \let \@tempb \@tempa \fi \ifx
  \@tempb \@empty \def\@tempb {arXiv}\fi \@ifundefined
  {mn@eprint@\@tempb}{\@tempb:\@tempc}{\expandafter \expandafter \csname
  mn@eprint@\@tempb\endcsname \expandafter{\@tempc}}}

\bibitem[\protect\citeauthoryear{{Angulo}, {Springel}, {White}, {Jenkins},
  {Baugh}  \& {Frenk}}{{Angulo} et~al.}{2012}]{Angulo-2012}
{Angulo} R.~E.,  {Springel} V.,  {White} S.~D.~M.,  {Jenkins} A.,  {Baugh}
  C.~M.,   {Frenk} C.~S.,  2012, \mn@doi [\mnras]
  {10.1111/j.1365-2966.2012.21830.x}, \href
  {https://ui.adsabs.harvard.edu/abs/2012MNRAS.426.2046A} {426, 2046}

\bibitem[\protect\citeauthoryear{{Aragon-Calvo}}{{Aragon-Calvo}}{2016}]{Aragon-Calvo-2016}
{Aragon-Calvo} M.~A.,  2016, \mn@doi [\mnras] {10.1093/mnras/stv2301}, \href
  {https://ui.adsabs.harvard.edu/abs/2016MNRAS.455..438A} {455, 438}

\bibitem[\protect\citeauthoryear{{Bah{\'e}} et~al.,}{{Bah{\'e}}
  et~al.}{2017}]{Bahe-2017}
{Bah{\'e}} Y.~M.,  et~al., 2017, \mn@doi [\mnras] {10.1093/mnras/stx1403},
  \href {https://ui.adsabs.harvard.edu/abs/2017MNRAS.470.4186B} {470, 4186}

\bibitem[\protect\citeauthoryear{{Bardeen}, {Bond}, {Kaiser}  \&
  {Szalay}}{{Bardeen} et~al.}{1986}]{Bardeen-1986}
{Bardeen} J.~M.,  {Bond} J.~R.,  {Kaiser} N.,   {Szalay} A.~S.,  1986, \mn@doi
  [\apj] {10.1086/164143}, \href
  {https://ui.adsabs.harvard.edu/abs/1986ApJ...304...15B} {304, 15}

\bibitem[\protect\citeauthoryear{{Barnes} et~al.,}{{Barnes}
  et~al.}{2017}]{Barnes-2017}
{Barnes} D.~J.,  et~al., 2017, \mn@doi [\mnras] {10.1093/mnras/stx1647}, \href
  {https://ui.adsabs.harvard.edu/abs/2017MNRAS.471.1088B} {471, 1088}

\bibitem[\protect\citeauthoryear{{Bertschinger}}{{Bertschinger}}{2001}]{Bertschinger-2001}
{Bertschinger} E.,  2001, \mn@doi [\apjs] {10.1086/322526}, \href
  {https://ui.adsabs.harvard.edu/abs/2001ApJS..137....1B} {137, 1}

\bibitem[\protect\citeauthoryear{Bistolas \& Hoffman}{Bistolas \&
  Hoffman}{1998}]{Bistolas-1998}
Bistolas V.,  Hoffman Y.,  1998, \mn@doi [The Astrophysical Journal]
  {10.1086/305080}, 492, 439–451

\bibitem[\protect\citeauthoryear{{Borgani}, {Governato}, {Wadsley}, {Menci},
  {Tozzi}, {Quinn}, {Stadel}  \& {Lake}}{{Borgani} et~al.}{2002}]{Borgani-2002}
{Borgani} S.,  {Governato} F.,  {Wadsley} J.,  {Menci} N.,  {Tozzi} P.,
  {Quinn} T.,  {Stadel} J.,   {Lake} G.,  2002, \mn@doi [\mnras]
  {10.1046/j.1365-8711.2002.05746.x}, \href
  {https://ui.adsabs.harvard.edu/abs/2002MNRAS.336..409B} {336, 409}

\bibitem[\protect\citeauthoryear{{Bouchet}, {Strauss}, {Davis}, {Fisher},
  {Yahil}  \& {Huchra}}{{Bouchet} et~al.}{1993}]{Bouchet-1993}
{Bouchet} F.~R.,  {Strauss} M.~A.,  {Davis} M.,  {Fisher} K.~B.,  {Yahil} A.,
  {Huchra} J.~P.,  1993, \mn@doi [\apj] {10.1086/173289}, \href
  {https://ui.adsabs.harvard.edu/abs/1993ApJ...417...36B} {417, 36}

\bibitem[\protect\citeauthoryear{Carlesi et~al.,}{Carlesi
  et~al.}{2016}]{Carlesi-2016}
Carlesi E.,  et~al., 2016, \mn@doi [Monthly Notices of the Royal Astronomical
  Society] {10.1093/mnras/stw357}, 458, 900–911

\bibitem[\protect\citeauthoryear{{Cole} \& {Kaiser}}{{Cole} \&
  {Kaiser}}{1989}]{Cole-1989}
{Cole} S.,  {Kaiser} N.,  1989, \mn@doi [\mnras] {10.1093/mnras/237.4.1127},
  \href {https://ui.adsabs.harvard.edu/abs/1989MNRAS.237.1127C} {237, 1127}

\bibitem[\protect\citeauthoryear{{Colombi}, {Szapudi}, {Jenkins}  \&
  {Colberg}}{{Colombi} et~al.}{2000}]{Colombi-2000}
{Colombi} S.,  {Szapudi} I.,  {Jenkins} A.,   {Colberg} J.,  2000, \mn@doi
  [\mnras] {10.1046/j.1365-8711.2000.03255.x}, \href
  {https://ui.adsabs.harvard.edu/abs/2000MNRAS.313..711C} {313, 711}

\bibitem[\protect\citeauthoryear{{Davis}, {Efstathiou}, {Frenk}  \&
  {White}}{{Davis} et~al.}{1985}]{Davis-1985}
{Davis} M.,  {Efstathiou} G.,  {Frenk} C.~S.,   {White} S.~D.~M.,  1985,
  \mn@doi [\apj] {10.1086/163168}, \href
  {http://adsabs.harvard.edu/abs/1985ApJ...292..371D} {292, 371}

\bibitem[\protect\citeauthoryear{Doumler, Gottlöber, Hoffman  \&
  Courtois}{Doumler et~al.}{2013}]{Doumler-2013}
Doumler T.,  Gottlöber S.,  Hoffman Y.,   Courtois H.,  2013, \mn@doi [Monthly
  Notices of the Royal Astronomical Society] {10.1093/mnras/sts614}, 430,
  912–923

\bibitem[\protect\citeauthoryear{{Efstathiou}, {Davis}, {White}  \&
  {Frenk}}{{Efstathiou} et~al.}{1985}]{Efstathiou-1985}
{Efstathiou} G.,  {Davis} M.,  {White} S.~D.~M.,   {Frenk} C.~S.,  1985,
  \mn@doi [\apjs] {10.1086/191003}, \href
  {https://ui.adsabs.harvard.edu/abs/1985ApJS...57..241E} {57, 241}

\bibitem[\protect\citeauthoryear{{Eke}, {Navarro}  \& {Frenk}}{{Eke}
  et~al.}{1998}]{Eke-1998}
{Eke} V.~R.,  {Navarro} J.~F.,   {Frenk} C.~S.,  1998, \mn@doi [\apj]
  {10.1086/306008}, \href
  {https://ui.adsabs.harvard.edu/abs/1998ApJ...503..569E} {503, 569}

\bibitem[\protect\citeauthoryear{{Grand} et~al.,}{{Grand}
  et~al.}{2017}]{Grand-2017}
{Grand} R. J.~J.,  et~al., 2017, \mn@doi [\mnras] {10.1093/mnras/stx071}, \href
  {https://ui.adsabs.harvard.edu/abs/2017MNRAS.467..179G} {467, 179}

\bibitem[\protect\citeauthoryear{{Hahn} \& {Abel}}{{Hahn} \&
  {Abel}}{2011}]{Hahn-2011}
{Hahn} O.,  {Abel} T.,  2011, \mn@doi [\mnras]
  {10.1111/j.1365-2966.2011.18820.x}, \href
  {https://ui.adsabs.harvard.edu/abs/2011MNRAS.415.2101H} {415, 2101}

\bibitem[\protect\citeauthoryear{{Hoffman} \& {Ribak}}{{Hoffman} \&
  {Ribak}}{1991}]{Hoffman-1991}
{Hoffman} Y.,  {Ribak} E.,  1991, \mn@doi [\apjl] {10.1086/186160}, \href
  {https://ui.adsabs.harvard.edu/abs/1991ApJ...380L...5H} {380, L5}

\bibitem[\protect\citeauthoryear{{Hoffman}, {Courtois}  \& {Tully}}{{Hoffman}
  et~al.}{2015}]{Hoffman-2015}
{Hoffman} Y.,  {Courtois} H.~M.,   {Tully} R.~B.,  2015, \mn@doi [\mnras]
  {10.1093/mnras/stv615}, \href
  {http://adsabs.harvard.edu/abs/2015MNRAS.449.4494H} {449, 4494}

\bibitem[\protect\citeauthoryear{{Jasche} \& {Wandelt}}{{Jasche} \&
  {Wandelt}}{2013}]{Jasche-2013}
{Jasche} J.,  {Wandelt} B.~D.,  2013, \mn@doi [\mnras] {10.1093/mnras/stt449},
  \href {https://ui.adsabs.harvard.edu/abs/2013MNRAS.432..894J} {432, 894}

\bibitem[\protect\citeauthoryear{{Jasche}, {Leclercq}  \& {Wandelt}}{{Jasche}
  et~al.}{2015}]{Jasche-2015}
{Jasche} J.,  {Leclercq} F.,   {Wandelt} B.~D.,  2015, \mn@doi [\jcap]
  {10.1088/1475-7516/2015/01/036}, \href
  {https://ui.adsabs.harvard.edu/abs/2015JCAP...01..036J} {2015, 036}

\bibitem[\protect\citeauthoryear{{Jenkins}}{{Jenkins}}{2010}]{Jenkins-2010}
{Jenkins} A.,  2010, \mn@doi [\mnras] {10.1111/j.1365-2966.2010.16259.x}, \href
  {http://adsabs.harvard.edu/abs/2010MNRAS.403.1859J} {403, 1859}

\bibitem[\protect\citeauthoryear{{Jenkins}}{{Jenkins}}{2013}]{Jenkins-2013}
{Jenkins} A.,  2013, \mn@doi [\mnras] {10.1093/mnras/stt1154}, \href
  {http://adsabs.harvard.edu/abs/2013MNRAS.434.2094J} {434, 2094}

\bibitem[\protect\citeauthoryear{Jenkins \& Booth}{Jenkins \&
  Booth}{2013}]{Jenkins-Booth-2013}
Jenkins A.,  Booth S.,  2013, Panphasia: a user guide (\mn@eprint {arXiv}
  {1306.5771})

\bibitem[\protect\citeauthoryear{{Jenkins}, {Frenk}, {White}, {Colberg},
  {Cole}, {Evrard}, {Couchman}  \& {Yoshida}}{{Jenkins}
  et~al.}{2001}]{Jenkins-2001}
{Jenkins} A.,  {Frenk} C.~S.,  {White} S.~D.~M.,  {Colberg} J.~M.,  {Cole} S.,
  {Evrard} A.~E.,  {Couchman} H.~M.~P.,   {Yoshida} N.,  2001, \mn@doi [\mnras]
  {10.1046/j.1365-8711.2001.04029.x}, \href
  {https://ui.adsabs.harvard.edu/abs/2001MNRAS.321..372J} {321, 372}

\bibitem[\protect\citeauthoryear{{Katz}, {Quinn}, {Bertschinger}  \&
  {Gelb}}{{Katz} et~al.}{1994}]{Katz-1994}
{Katz} N.,  {Quinn} T.,  {Bertschinger} E.,   {Gelb} J.~M.,  1994, \mn@doi
  [\mnras] {10.1093/mnras/270.1.L71}, \href
  {https://ui.adsabs.harvard.edu/abs/1994MNRAS.270L..71K} {270, L71}

\bibitem[\protect\citeauthoryear{{Kay}, {Thomas}, {Jenkins}  \& {Pearce}}{{Kay}
  et~al.}{2004}]{Kay-2004}
{Kay} S.~T.,  {Thomas} P.~A.,  {Jenkins} A.,   {Pearce} F.~R.,  2004, \mn@doi
  [\mnras] {10.1111/j.1365-2966.2004.08383.x}, \href
  {https://ui.adsabs.harvard.edu/abs/2004MNRAS.355.1091K} {355, 1091}

\bibitem[\protect\citeauthoryear{{Klypin}, {Trujillo-Gomez}  \&
  {Primack}}{{Klypin} et~al.}{2011}]{Klypin-2011}
{Klypin} A.~A.,  {Trujillo-Gomez} S.,   {Primack} J.,  2011, \mn@doi [\apj]
  {10.1088/0004-637X/740/2/102}, \href
  {https://ui.adsabs.harvard.edu/abs/2011ApJ...740..102K} {740, 102}

\bibitem[\protect\citeauthoryear{{Lavaux} \& {Jasche}}{{Lavaux} \&
  {Jasche}}{2016}]{Lavaux-2016}
{Lavaux} G.,  {Jasche} J.,  2016, \mn@doi [\mnras] {10.1093/mnras/stv2499},
  \href {https://ui.adsabs.harvard.edu/abs/2016MNRAS.455.3169L} {455, 3169}

\bibitem[\protect\citeauthoryear{{Linde}}{{Linde}}{2005}]{Linde-2005}
{Linde} A.,  2005, arXiv e-prints, \href
  {https://ui.adsabs.harvard.edu/abs/2005hep.th....3203L} {pp hep--th/0503203}

\bibitem[\protect\citeauthoryear{{Martizzi}, {Hahn}, {Wu}, {Evrard}, {Teyssier}
   \& {Wechsler}}{{Martizzi} et~al.}{2016}]{Martizzi-2016}
{Martizzi} D.,  {Hahn} O.,  {Wu} H.-Y.,  {Evrard} A.~E.,  {Teyssier} R.,
  {Wechsler} R.~H.,  2016, \mn@doi [\mnras] {10.1093/mnras/stw897}, \href
  {https://ui.adsabs.harvard.edu/abs/2016MNRAS.459.4408M} {459, 4408}

\bibitem[\protect\citeauthoryear{{Nagai}, {Vikhlinin}  \& {Kravtsov}}{{Nagai}
  et~al.}{2007}]{Nagai-2007}
{Nagai} D.,  {Vikhlinin} A.,   {Kravtsov} A.~V.,  2007, \mn@doi [\apj]
  {10.1086/509868}, \href
  {https://ui.adsabs.harvard.edu/abs/2007ApJ...655...98N} {655, 98}

\bibitem[\protect\citeauthoryear{{Navarro}, {Eke}  \& {Frenk}}{{Navarro}
  et~al.}{1996}]{Navarro-1996}
{Navarro} J.~F.,  {Eke} V.~R.,   {Frenk} C.~S.,  1996, \mnras, \href
  {http://adsabs.harvard.edu/abs/1996MNRAS.283L..72N} {283, L72}

\bibitem[\protect\citeauthoryear{{Neto} et~al.,}{{Neto}
  et~al.}{2007}]{Neto-2007}
{Neto} A.~F.,  et~al., 2007, \mn@doi [\mnras]
  {10.1111/j.1365-2966.2007.12381.x}, \href
  {https://ui.adsabs.harvard.edu/abs/2007MNRAS.381.1450N} {381, 1450}

\bibitem[\protect\citeauthoryear{{Nusser}, {Dekel}  \& {Yahil}}{{Nusser}
  et~al.}{1995}]{Nusser-1995}
{Nusser} A.,  {Dekel} A.,   {Yahil} A.,  1995, \mn@doi [\apj] {10.1086/176069},
  \href {https://ui.adsabs.harvard.edu/abs/1995ApJ...449..439N} {449, 439}

\bibitem[\protect\citeauthoryear{{O'Leary} \& {McQuinn}}{{O'Leary} \&
  {McQuinn}}{2012}]{Oleary-2012}
{O'Leary} R.~M.,  {McQuinn} M.,  2012, \mn@doi [\apj]
  {10.1088/0004-637X/760/1/4}, \href
  {https://ui.adsabs.harvard.edu/abs/2012ApJ...760....4O} {760, 4}

\bibitem[\protect\citeauthoryear{{Peebles}}{{Peebles}}{1980}]{Peebles-1980}
{Peebles} P.~J.~E.,  1980, {The large-scale structure of the universe}.
Princeton University Press

\bibitem[\protect\citeauthoryear{{Planck Collaboration}}{{Planck
  Collaboration}}{2016}]{Planck-2016}
{Planck Collaboration} 2016, \mn@doi [\aap] {10.1051/0004-6361/201525836},
  \href {https://ui.adsabs.harvard.edu/abs/2016A&A...594A..17P} {594, A17}

\bibitem[\protect\citeauthoryear{{Planck Collaboration}}{{Planck
  Collaboration}}{2019}]{Planck-2019}
{Planck Collaboration} 2019, arXiv e-prints, \href
  {https://ui.adsabs.harvard.edu/abs/2019arXiv190505697P} {p. arXiv:1905.05697}

\bibitem[\protect\citeauthoryear{{Power}, {Navarro}, {Jenkins}, {Frenk},
  {White}, {Springel}, {Stadel}  \& {Quinn}}{{Power} et~al.}{2003}]{Power-2003}
{Power} C.,  {Navarro} J.~F.,  {Jenkins} A.,  {Frenk} C.~S.,  {White} S.~D.~M.,
   {Springel} V.,  {Stadel} J.,   {Quinn} T.,  2003, \mn@doi [\mnras]
  {10.1046/j.1365-8711.2003.05925.x}, \href
  {https://ui.adsabs.harvard.edu/abs/2003MNRAS.338...14P} {338, 14}

\bibitem[\protect\citeauthoryear{{Roth}, {Pontzen}  \& {Peiris}}{{Roth}
  et~al.}{2016}]{Roth-2016}
{Roth} N.,  {Pontzen} A.,   {Peiris} H.~V.,  2016, \mn@doi [\mnras]
  {10.1093/mnras/stv2375}, \href
  {https://ui.adsabs.harvard.edu/abs/2016MNRAS.455..974R} {455, 974}

\bibitem[\protect\citeauthoryear{{Salmon}}{{Salmon}}{1996}]{Salmon-1996}
{Salmon} J.,  1996, \mn@doi [\apj] {10.1086/176952}, \href
  {https://ui.adsabs.harvard.edu/abs/1996ApJ...460...59S} {460, 59}

\bibitem[\protect\citeauthoryear{{Sawala} et~al.,}{{Sawala}
  et~al.}{2016}]{Sawala-2016a}
{Sawala} T.,  et~al., 2016, \mn@doi [\mnras] {10.1093/mnras/stw145}, \href
  {http://adsabs.harvard.edu/abs/2016MNRAS.457.1931S} {457, 1931}

\bibitem[\protect\citeauthoryear{{Schaller} et~al.,}{{Schaller}
  et~al.}{2015}]{Schaller-2015c}
{Schaller} M.,  et~al., 2015, \mn@doi [\mnras] {10.1093/mnras/stv1341}, \href
  {https://ui.adsabs.harvard.edu/abs/2015MNRAS.452..343S} {452, 343}

\bibitem[\protect\citeauthoryear{{Schaye} et~al.}{{Schaye}
  et~al.}{2015}]{Schaye-2014}
{Schaye} J.,  et~al., 2015, \mn@doi [\mnras] {10.1093/mnras/stu2058}, \href
  {http://adsabs.harvard.edu/abs/2015MNRAS.446..521S} {446, 521}

\bibitem[\protect\citeauthoryear{{Springel}}{{Springel}}{2005}]{Springel-2005-gadget}
{Springel} V.,  2005, \mn@doi [\mnras] {10.1111/j.1365-2966.2005.09655.x},
  \href {http://adsabs.harvard.edu/abs/2005MNRAS.364.1105S} {364, 1105}

\bibitem[\protect\citeauthoryear{{Springel}, {White}, {Tormen}  \&
  {Kauffmann}}{{Springel} et~al.}{2001}]{Springel-2001-subfind}
{Springel} V.,  {White} S.~D.~M.,  {Tormen} G.,   {Kauffmann} G.,  2001,
  \mn@doi [\mnras] {10.1046/j.1365-8711.2001.04912.x}, \href
  {https://ui.adsabs.harvard.edu/abs/2001MNRAS.328..726S} {328, 726}

\bibitem[\protect\citeauthoryear{{Springel} et~al.,}{{Springel}
  et~al.}{2005}]{Springel-2005-millennium}
{Springel} V.,  et~al., 2005, \mn@doi [\nat] {10.1038/nature03597}, \href
  {http://adsabs.harvard.edu/abs/2005Natur.435..629S} {435, 629}

\bibitem[\protect\citeauthoryear{{Springel} et~al.,}{{Springel}
  et~al.}{2008}]{Springel-2008}
{Springel} V.,  et~al., 2008, \mn@doi [\mnras]
  {10.1111/j.1365-2966.2008.14066.x}, \href
  {http://adsabs.harvard.edu/abs/2008MNRAS.391.1685S} {391, 1685}

\bibitem[\protect\citeauthoryear{{Urban}, {Werner}, {Simionescu}, {Allen}  \&
  {B{\"o}hringer}}{{Urban} et~al.}{2011}]{Urban-2011}
{Urban} O.,  {Werner} N.,  {Simionescu} A.,  {Allen} S.~W.,   {B{\"o}hringer}
  H.,  2011, \mn@doi [\mnras] {10.1111/j.1365-2966.2011.18526.x}, \href
  {https://ui.adsabs.harvard.edu/abs/2011MNRAS.414.2101U} {414, 2101}

\bibitem[\protect\citeauthoryear{{Wang}, {Bose}, {Frenk}, {Gao}, {Jenkins},
  {Springel}  \& {White}}{{Wang} et~al.}{2019}]{Wang-2019}
{Wang} J.,  {Bose} S.,  {Frenk} C.~S.,  {Gao} L.,  {Jenkins} A.,  {Springel}
  V.,   {White} S. D.~M.,  2019, arXiv e-prints, \href
  {https://ui.adsabs.harvard.edu/abs/2019arXiv191109720W} {p. arXiv:1911.09720}

\bibitem[\protect\citeauthoryear{{White} \& {Rees}}{{White} \&
  {Rees}}{1978}]{White-1987}
{White} S.~D.~M.,  {Rees} M.~J.,  1978, \mn@doi [\mnras]
  {10.1093/mnras/183.3.341}, \href
  {https://ui.adsabs.harvard.edu/abs/1978MNRAS.183..341W} {183, 341}

\bibitem[\protect\citeauthoryear{{Yepes}, {Gottl{\"o}ber}  \&
  {Hoffman}}{{Yepes} et~al.}{2014}]{Yepes-2013}
{Yepes} G.,  {Gottl{\"o}ber} S.,   {Hoffman} Y.,  2014, \mn@doi [New Astronomy
  Reviews] {10.1016/j.newar.2013.11.001}, \href
  {http://adsabs.harvard.edu/abs/2014NewAR..58....1Y} {58, 1}

\makeatother
\end{thebibliography}



\appendix

\section{{\sc Panphasia} Descriptors used}

The Reference simulation in this paper uses the same phase information as the 100~cMpc {\sc Eagle} simulation \citep{Schaye-2014}, whose {\sc Panphasia} phase descriptor is: 
{\scriptsize
\begin{verbatim} 
[Panph1,L16,(31250,23438,39063),S12,CH1050187043,EAGLE_L0100_VOL1]
\end{verbatim}
}
The reference phase information is defined using a $12 \times 12 \times 12$\ root cell at level $\LL=16$.
At level 16 for example the reference phases in the x-direction are taken
from the range of twelve cells 31250-31261, and similarly 23438-23449 and
39063-39074 in $y$ and $z$ respectively.  

To create variants of the phase information we extract the phase
information from neighouring regions of the Panphasia field. Because
a white noise field is uncorrelated it is sufficient to use regions
that are shifted by multiples of 12 cells at level 16, in the 
$x$,$y$ or $z$ directions.  

From this, the phase information for volumes $\mathrm{V}\LL_j$ is constructed by applying an integer spatial shift of $(\Delta x, \Delta y, \Delta z)$, in units of the box size, equivalent to 12 cells in the positive x-direction at level 16 of the octree, or 100 Mpc, and similarly
for $y$ and $z$

\begin{align*}
\Delta x &= (j ~\mathrm{mod}~10), \\
\Delta y &= ((j - \Delta x)~\mathrm{mod}~100) / 10, \\
\Delta z &= ((j - 10 \Delta y - \Delta x)~\mathrm{mod}~1000) / 100. \\
\end{align*}
For example, $\mathrm{V}21_{24}$ is constructed using shifts of $\Delta x=4$ and $\Delta y=2$.

For every variant, we shift up to 10 levels at $\LL_{min}$ and above by multiples of the box size. The cell size at each level is the ratio between the box size and the root cell, $l_\mathrm{cell} = 100 / 12 / 2^{\LL - 16}$cMpc. In order to shift the phase information by the box size, at each level, a shift by $\Delta N_i = \Delta_i \times 12 \times 2^{\LL - 16}$ is required. For example, $\mathrm{V}21_{24}$ corresponds to the following cell shifts in {\sc Panphasia}:

\begin{align*}
&\begin{rcases}
    \Delta N_x &= 4 \times 12 \times 2^{21-16} = 1536~\\
    \Delta N_y &= 2 \times 12 \times 2^{21-16} = 768~\\
    \Delta N_z &= 0 \\
\end{rcases}
\LL = 21\\
&\begin{rcases}
    \Delta N_x &= 4 \times 12 \times 2^{22-16} = 3072~\\
    \Delta N_y &= 2 \times 12 \times 2^{22-16} = 1536~\\
    \Delta N_z &= 0 \\
\end{rcases}
\LL = 22\\
&\begin{rcases}
    \Delta N_x &= 4 \times 12 \times 2^{23-16} = 6144~ \\
    \Delta N_y &= 2 \times 12 \times 2^{23-16} = 1536~ \\
    \Delta N_z &= 0 \\
\end{rcases}
\LL = 23
\end{align*}

\bsp	
\label{lastpage}
\end{document}